\documentclass{aa}  

\usepackage{graphicx}
\usepackage{txfonts}
\usepackage{threeparttablex}
\usepackage{longtable}
\usepackage{hyperref}
\usepackage{multicol}
\usepackage[switch]{lineno}

\usepackage{xcolor}

\newcommand{\SIfourfiveeightnine}{[\ion{S}{i}]\,$\lambda$\,4589\xspace}
\newcommand{\SIsevenseventwofive}{[\ion{S}{i}]\,$\lambda$\,7725\xspace}
\newcommand{\SIoneoeighttwo}{[\ion{S}{i}]\,$\lambda$\,1.082\xspace}
\newcommand{\SIoneonethreeone}{[\ion{S}{i}]\,$\lambda$\,1.131\xspace}

\newcommand{\SIIoptical}{[\ion{S}{ii}]\,$\lambda\lambda$4068,4076\xspace}
\newcommand{\SIIsixseven}{[\ion{S}{ii}]\,$\lambda\lambda$6716,6731\xspace}
\newcommand{\SIINIR}{[\ion{S}{ii}]\,NIR\xspace}

\newcommand{\SIIInineosixnine}{[\ion{S}{iii}]\,$\lambda$\,9069\xspace}
\newcommand{\SIIIninefivethreeone}{[\ion{S}{iii}]\,$\lambda$\,9531\xspace}

\newcommand{\SiIfouroneothree}{\ion{Si}{I}\,$\lambda$\,4103\xspace}
\newcommand{\SiIoneoninenine}{[\ion{Si}{I}]\,$\lambda$\,1.099\xspace}
\newcommand{\SiIonesixosix}{[\ion{Si}{I}]\,$\lambda$\,1.606\xspace}
\newcommand{\SiIonesixfourfive}{[\ion{Si}{I}]\,$\lambda$\,1.645\xspace}

\newcommand{\OIdoublet}{[\ion{O}{I}]\,$\lambda\lambda$6300,6364\xspace}
\newcommand{\Msun}{$M_{\odot}$}
\newcommand{\sumocodename}{\texttt{SUMO}\xspace}
\newcommand{\el}[2]{$^{#1}$#2}
\newcommand{\necrit}{$n_{e,\mathrm{crit}}$\xspace}

\newcommand{\CaIIdoublet}{[\ion{Ca}{II}]\,$\lambda\lambda$7291,7323\xspace}
\newcommand{\CaNIR}{[\ion{Ca}{ii}]\,NIR\xspace}

\newcommand{\orcid}[1]{\href{https://orcid.org/#1}{\includegraphics[scale=0.05]{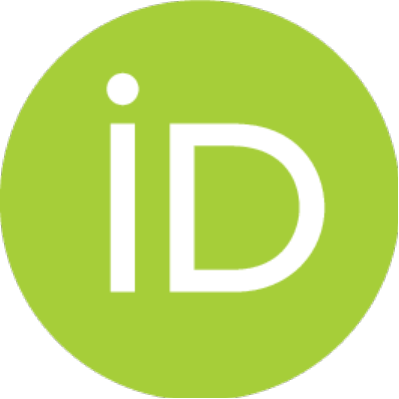}}}

\begin{document}

   \title{SN~2024afyu interpreted as a Pair Instability Supernova}

    \titlerunning{SN~2024afyu}
    
   \author{P.~J.~Pessi\inst{1}\fnmsep\thanks{\email{priscila.pessi@ncbj.gov.pl}}\orcid{0000-0002-8041-8559}
    \and  S. Barmentloo\inst{2} \orcid{0000-0003-4800-2737}
    \and {\L}. Wyrzykowski\inst{3,1}\orcid{0000-0002-9658-6151}
    \and S. Schulze\inst{4}\orcid{0000-0001-6797-1889}
    \and J. Sollerman\inst{2}\orcid{0000-0003-1546-6615}
    \and A. Gangopadhyay\inst{2}\orcid{0000-0002-3884-5637}
    \and P.~J. Miko{\l}ajczyk\inst{5,1}\orcid{0000-0001-8916-8050}
    \and S. Rose\inst{6}\orcid{0000-0003-4725-4481}
    \and K. Kotysz\inst{3,5}
    \and C. Fremling\inst{7,6}\orcid{0000-0002-4223-103X}
    \and A. Jerkstrand\inst{2}\orcid{0000-0001-8005-4030}
    \and R. Lunnan\inst{2}
    \and L. Yan\inst{7}\orcid{0000-0003-1710-9339}
    \and C. Fransson\inst{2}\orcid{0000-0001-8532-3594}
    \and J.~L. Wise\inst{8}\orcid{0000-0003-0733-2916}
    \and M. Dubey\inst{9}\orcid{0009-0002-2621-6611} 
    \and K. Misra\inst{9}\orcid{0000-0003-1637-267X} 
    \and A. P. Cotter\inst{1}\orcid{0009-0004-1910-4990}
    \and D. Athanasopoulos\inst{10}\orcid{0000-0002-2655-6459}
    \and A. Bochenek\inst{8}\orcid{0009-0008-2714-2507}
    \and M. Bronikowski\inst{3,11}\orcid{0000-0002-1537-6911}
    \and U. Burgaz\inst{12}\orcid{0000-0003-0126-3999}
    \and W. Burzynski\inst{13}\orcid{0009-0008-2110-4892}
    \and J.~M. Carrasco\inst{14,15,16}\orcid{0000-0002-3029-5853}
    \and E. Concepcion\inst{11}\orcid{0009-0001-6354-2494}
    \and C. Galdies\inst{17}\orcid{0000-0002-8908-0785}
    \and A.~F. Gillan\inst{1}\orcid{0000-0003-4094-9408}
    \and A. Gkini\inst{2}\orcid{0009-0000-9383-2305}
    \and E. Gkiokas\inst{10}\orcid{0009-0005-7924-5565}
    \and E. G{\l}owacki\inst{13}\orcid{0009-0000-8051-7605}
    \and F-J. Hambsch\inst{18,19,20}\orcid{0000-0003-0125-8700}
    \and J. Japelj\inst{11}
    \and N. Karaman\inst{21}\orcid{0000-0003-3910-2285}
    \and P. King\inst{22}\orcid{0009-0001-0980-4043}
    \and S. Kurowski\inst{23}\orcid{0000-0002-1557-0343}
    \and C. Liu\inst{24,25,26}\orcid{0000-0002-7866-4531}
    \and M. Maskoliunas\inst{27}\orcid{0000-0003-3432-2393}
    \and Z. McGrath\inst{8}\orcid{0009-0006-0726-1328}
    \and M. Mihel\v{c}i\v{c}\inst{28}
    \and M. Nikolajuk\inst{13}\orcid{0000-0003-4075-6745}
    \and W. Ogloza\inst{29,30}\orcid{0000-0002-6293-9940}
    \and E. Pak\v stien\.e\inst{27}\orcid{0000-0002-3326-2918}
    \and U. Pylypenko\inst{31,32}\orcid{0009-0002-7560-1903}
    \and A. Pucek\inst{23}\orcid{0009-0009-2238-6913}
    \and N. Rehemtulla\inst{33,25,26}\orcid{0000-0002-5683-2389}
    \and R.~M. Rich\inst{22}\orcid{0000-0003-0427-8387}
    \and A.~O. Simon\inst{34}\orcid{0000-0003-0404-5559}
    \and A. Singh\inst{2}
    \and C. Skoglund\inst{2}
    \and E. Stonkute\inst{27}\orcid{0000-0002-8028-8133}
    \and Y. Stsefanenka\inst{23}\orcid{0009-0007-7384-5812}
    \and P. Szober\inst{35}\orcid{0009-0008-9539-4079}
    \and K. Tsalapatas\inst{2}
    \and B.~F.~A. van Baal\inst{2}\orcid{0009-0001-3767-942X}
    \and C. Ventura\inst{36}\orcid{0009-0009-9751-9215}
    \and K. Vrontaki\inst{10}\orcid{0009-0002-7669-7425}
    \and A. Wozniak\inst{37}\orcid{0009-0000-4572-7682}
    \and J. Zdanavicius\inst{27}\orcid{0009-0000-9910-1124}
    \and S. Zola\inst{23}\orcid{0000-0003-3609-382X}
    \and E.~C. Bellm\inst{38}\orcid{0000-0001-8018-5348}
    \and J. Castaneda Jaimes\inst{6}\orcid{0000-0002-0987-3372}
    \and M.~J. Graham\inst{6}\orcid{0000-0002-3168-0139}
    \and G. Helou\inst{39}\orcid{0000-0003-3367-3415}
    \and F.~J. Masci\inst{39}\orcid{0000-0002-8532-9395}
    \and J.~N. Purdum\inst{7}\orcid{0000-0003-1227-3738}
    \and R. Riddle\inst{7}\orcid{0000-0002-0387-370X}
    }
   \institute{\centering\textit{(Affiliations can be found after the references)}}
    
   \date{}

 
  \abstract
{Pair-instability supernovae (PISNe) are the predicted explosions of very massive stars triggered by electron--positron pair production. Numerous transients have been proposed as PISN candidates, yet none has provided unambiguous confirmation of this explosion mechanism. The predicted strengths of nebular emission lines offer a powerful means of testing the PISN scenario.
}
{We investigate the nature of SN~2024afyu, a nearby (\textit{z} = 0.0085), long-lived (t$_{\mathrm{rise}} = 85 \pm 11.7$ days) SN with peculiar spectral evolution, with the aim of identifying its powering mechanism.
}
{We analyse multi-band photometry and optical and near-infrared spectroscopy from shortly after explosion to the nebular phase ($\sim$500 days past peak). Besides early appearance of [\ion{Ca}{ii}] features, we identify a number of sulfur and silicon emission lines, for which we estimate electron temperatures and elemental masses. We compare SN~2024afyu to core-collapse SNe and proposed PISN candidates and PISN models, given the identification of intermediate mass elements.
}
{SN~2024afyu has an inferred $^{56}$Ni mass of around 0.4--1.0~\Msun{} and an inferred sulfur mass of the order of 3~\Msun{}, substantially larger than expected for conventional core-collapse explosions. SN~2024afyu is photometrically similar (although fainter, M$_{\mathrm{Peak}}^{r} = -$18.9 $\pm$ 0.04~mag) but spectroscopically distinct to other proposed PISNe. Yet, existing PISN models broadly reproduce several key characteristics, including the overall spectral appearance and broad photometric evolution.  
}
{SN~2024afyu is a strong PISN candidate, since alternative scenarios would struggle to explain the combination of broad light curve, large intermediate-mass-element abundance, and general spectroscopic evolution. The discrepancies between the observations and currently available theoretical models highlight the need for new PISN calculations spanning a wider range of progenitor masses, metallicities, mixing prescriptions, and circumstellar environments. 
}


   \maketitle
%
\section{Introduction}
\label{sec:intro}

Massive stars end their lives in spectacular supernova (SN) explosions. Stars born with initial masses $\gtrsim$  8~\Msun{} are expected to undergo gravitational core-collapse after exhausting their nuclear fuel, producing core-collapse supernovae (CCSNe). Historically, CCSNe have been classified according to the presence or absence of specific spectral features, with hydrogen (H) rich explosions classified as Type~II supernovae (SNe~II), H-deficient explosions as Type~I supernovae (SNe~I), and events displaying H at early times that fades later on forming the transitional Type~IIb class \citep{1941PASP...53..224M,1993ApJ...415L.103F,2017hsn..book..195G}. In recent years, the advent of large, high-cadence, wide-field surveys has dramatically increased the number of discovered transients, while spectroscopic follow-up resources have remained comparatively limited. As a result, renewed effort has been devoted to classifying SNe from their photometric evolution alone, an idea first explored several decades ago and now enabled by modern surveys and machine-learning techniques \citep[e.g.][]{1977SvAL....3..215P,2002PASP..114..833P,2019ApJ...884...83V,2024A&A...692A.208F,2026arXiv260213036T,2026arXiv260414761R}.

If a star is even more massive at birth, $\sim$70--100~\Msun{} (or higher, depending on metallicity and mass loss), it may develop a sufficiently hot and dilute core to allow the conversion of energetic photons into electron--positron pairs. The reduction in radiation pressure destabilizes the core. Depending on the final helium-core mass of the progenitor, the instability may lead to one or more episodes of explosive oxygen burning that produce pulses that eject part of the stellar envelope before the star undergoes CC, producing a pulsational pair-instability supernova (PPISN). The instability may instead trigger a thermonuclear runaway explosion that completely disrupts the star in a pair-instability supernova \citep[PISN;][]{1967ApJ...148..803R,1967PhRvL..18..379B,1968Ap&SS...2...96F,1984ApJ...280..825B,2026enap....2..680R}
Although theorized for decades, the unambiguous identification of observed events as (P)PISN is challenging. A number of transients have been proposed as candidates, although in most cases their observed properties can also be reproduced by alternative powering mechanisms such as interaction with a dense circumstellar medium (CSM), the spin-down of a magnetar or fallback accretion onto a central black hole \citep[e.g.][]{2017hsn..book..939K}. 

Many of the proposed PISN candidates show superluminous light curves \citep[i.e. peaks $\lesssim -20$~mag; e.g.][]{2019ARA&A..57..305G,2023ApJ...943...41C,2025A&A...695A.142P}, that results in an initial superluminous SNe (SLSNe) classification. The following identification as PISN in these cases is mainly supported by the high amount of synthesized $^{56}$Ni necessary to explain the observed total radiated energy. Nonetheless, theoretical models predict that PISN arising from progenitors near the lower mass limit should be substantially less luminous. In these cases, the defining photometric signature is not an exceptionally high peak luminosity but rather a broad, slowly evolving light curve  \citep{2011ApJ...734..102K}.

Photometric evolution alone is generally insufficient to identify a PISN. Instead, intensive multi-frequency monitoring can provide more compelling candidates, in particular when spectroscopy is available at nebular phases. At such phases the ejecta become optically thin and the observed emission lines trace the products of explosive nucleosynthesis  \citep[e.g.][]{2024A&A...683A.223S}. 
In this paper, we investigate the observational characteristics of SN~2024afyu, a long-lived supernova that reached a luminosity above the typical CCSN threshold (M$_{\mathrm{Peak}}^{r} = -$18.9 $\pm$ 0.04~mag), although below that of SLSNe. The SN was initially classified as Type~II and soon after as Type~IIb. Its proximity (\textit{z} = 0.0085) and transitional classification motivated us to start an intensive follow-up campaign. We monitored its photometric evolution from the ultraviolet (UV) to the near-infrared (NIR), including the optical $coUBVgrRiIz$ bands. In addition, we obtained a total of 48 spectroscopic epochs (seven of which were observed in the NIR) using eight different telescopes. Combined with the classification spectra\footnote{Publicly available on The Transient Name Server (TNS), \url{https://www.wis-tns.org/}.} presented by \cite{2025TNSCR..53....1B} and \cite{2025TNSCR1202....1P}, the spectroscopic observations cover the evolution of SN~2024afyu from approximately $-48$ to $+468$ rest-frame days relative to the $r$-band peak. 

Given its observed properties, we explore whether SN~2024afyu can be interpreted as a PISN. We examine its nebular spectra to assess whether its explosion is consistent with the nucleosynthetic yields predicted for PISNe and compare its photometric and spectroscopic evolution with those of well-studied SNe, previously proposed PISN candidates, and publicly available PISN theoretical models.
This paper is organized as follows in Sec.~\ref{sec:observations} we present the observational data. In Sec.~\ref{sec:analysis} we present the analysis that includes the inspection of the host galaxy in Sec.~\ref{sec:host}, the characterization of the light curve in Sec.~\ref{sec:light-curve}, and the analysis of the spectral evolution in Sec.~\ref{sec:spectral-evol}. In addition, we present the comparison of SN~2024afyu's observed properties with other CCSNe. in Sec.~\ref{sec:comparison}. In Sec.~\ref{sec:sulfurANDSiliconID} we identify distinctive spectral lines, based on these, we inspect a PISN interpretation in  Sec.~\ref{sec:PISNinterpretation}. We discuss our findings in Sec.~\ref{sec:discussion} and conclude in Sec.~\ref{sec:conclusion}.

\section{Observations}
\label{sec:observations}

SN~2024afyu (also known as ZTF24abzxalv, GOTO25na, ATLAS25axv, BGEMJ171343.05+073730.9 and PS25djk) was discovered on 2024 December 30 by \cite{2024TNSTR5161....1S} and was spectroscopically classified on 2025 January 4 as a SN~II at redshift (\textit{z}) 0.01 by \cite{2025TNSCR..53....1B}. Further spectroscopic observations led to a reclassification as a SN~IIb on March 29 2025 by \cite{2025TNSCR1202....1P}. Additional spectroscopic follow up reveal narrow Balmer lines from the underlying H~II region (see Sec.~\ref{sec:spec}) that allow us to derive a \textit{z} $=$ 0.0085. At this low redshift, peculiar velocities can constitute a significant fraction of the Hubble flow thus, we correct the obtained \textit{z} for local peculiar motions using the 2M++ reconstructed velocity field \citep{2015MNRAS.450..317C}, following the methodology adopted in Pantheon+ \citep{2022PASA...39...46C,2022ApJ...938..110B}. Adopting H$_{0} = 73$ km s$^{-1}$ Mpc$^{-1}$, $\Omega _{\mathrm{Matter}} = 0.27$, $\Omega _{\mathrm{Lambda}} = 0.73$ as cosmological parameters, we obtain a corrected distance modulus $\mu \sim$ 32.88~mag, that translates to a luminosity distance D$_{\mathrm{L}} \sim$ 37.7~Mpc.

\subsection{Photometry}
\label{sec:phot} 

\begin{figure*}
   \centering
   \includegraphics[scale=0.55]{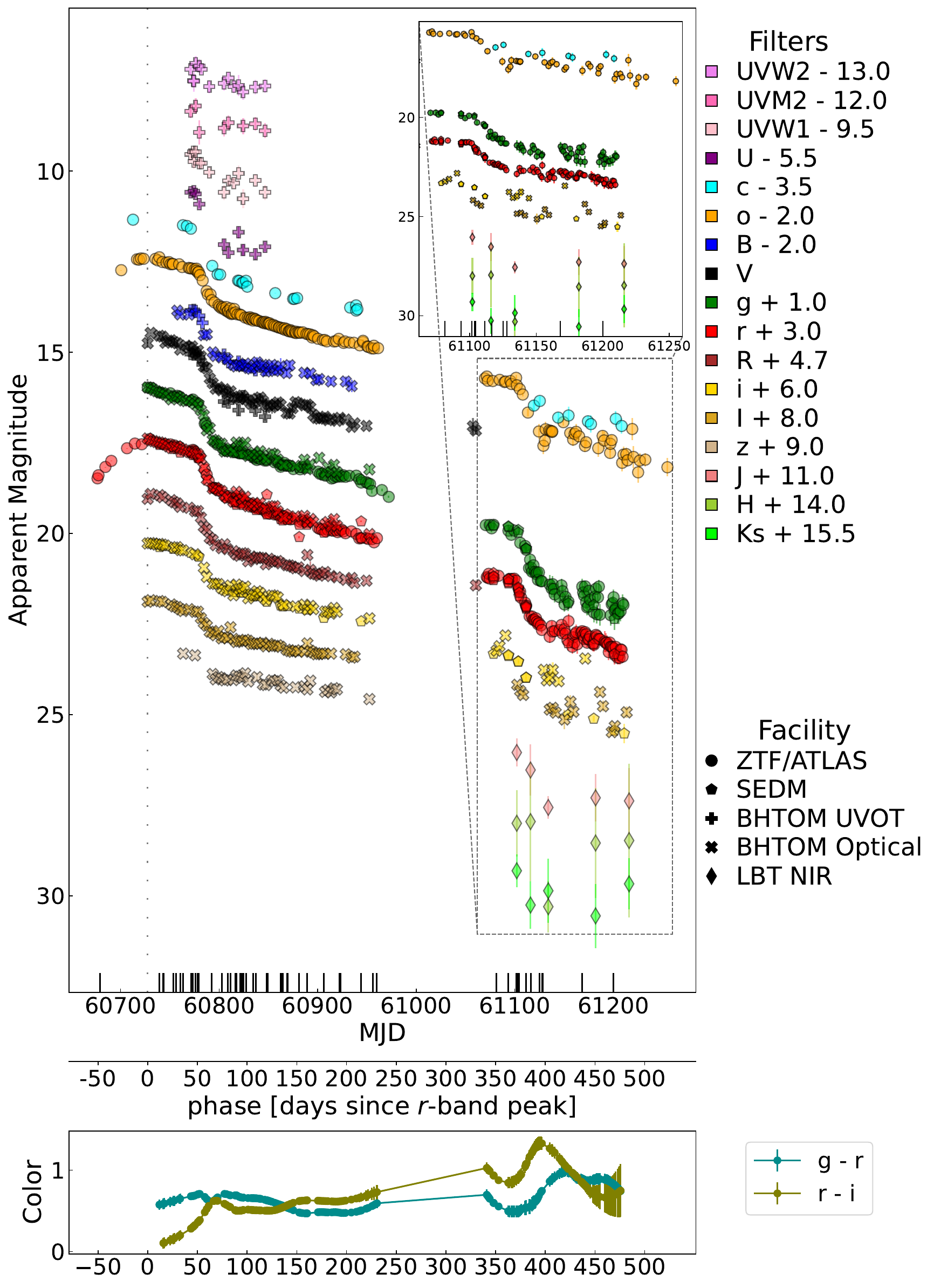}
      \caption{Top panel: Observed light curves. Different colors correspond to different filters, these have been artificially shifted for visualization, the corresponding shifts are indicated in the label. Different symbols correspond to different facilities. Vertical lines at the bottom of the plot indicate epochs of spectroscopic observations. An inset in the upper-right corner of the plot provides a zoomed-in view of the late-time phases, highlighting the small-scale ``wiggles'' visible in the light curve. Bottom panel: observed $g - r$ and $r - i$ colors.
              }
         \label{fig:obslc}
   \end{figure*}
   
SN~2024afyu was detected by the Zwicky Transient Facility \citep[ZTF;][]{2019PASP..131a8002B,2019PASP..131g8001G,2019PASP..131a8003M,2020PASP..132c8001D} and the Asteroid Terrestrial-impact Last Alert System \citep[ATLAS;][]{2018PASP..130f4505T,2020PASP..132h5002S}. Additional photometric follow up was obtained by the ZTF collaboration with the Spectral Energy Distribution Machine \citep[SEDM;][]{2018PASP..130c5003B} on the 60-inch telescope (P60) at Palomar Observatory and the Liverpool Telescope \citep[LT;][]{2004SPIE.5489..679S}; and by the Black Hole Target and Observation Manager \citep[BHTOM\footnote{\url{https://bhtom.space}.};][]{2024lsstWyrzykowski, 2025RMxACMikolajczyk} which includes optical broad-band photometry as well as 
publicly available archival observations covering a target's coordinates. The late time evolution of SN~2024afyu was also covered in the NIR with the Large Binocular Telescope (LBT), located on Mt. Graham in south eastern Arizona. The photometry presented on this work is listed in Tables~\ref{tab:ZTFSEDMATLASphot}, \ref{tab:BHTOMphot} and \ref{tab:LBTNIRphot}, which are available in their entirety as online material.

The corresponding light curves are shown in Fig.~\ref{fig:obslc}. Absolute magnitudes are obtained as $M_{\lambda} = m_{\lambda} - \mu - A_{\lambda}$, where $m_{\lambda}$ is apparent magnitude, $\mu$ is the distance modulus (see Sec.~\ref{sec:observations}), and $A_{\lambda}$ is the Milky Way (MW) extinction in the considered band, for reference $A_{V} = 0.532$ mag. MW extinction was obtained using NED's Galactic Extinction Calculator\footnote{NED's Extinction Calculator considers the recalibration of \citet{2011ApJ...737..103S} to the extinction map presented by \citet{1998ApJ...500..525S}, assuming a \citet{1999PASP..111...63F} reddening law with R$_{\mathrm{v}} = 3.1$.}, accessed through the \texttt{ned\_extinction\_calc} script\footnote{\url{https://github.com/mmechtley/ned_extinction_calc}.}. 
The effective wavelength corresponding to each filter was obtained from the Spanish Virtual Observatory Filter Information service \citep[SVO;][]{2012ivoa.rept.1015R,2020sea..confE.182R}. Host galaxy extinction was not considered (see Sec.~\ref{sec:host}). 

\subsubsection{ATLAS}

ATLAS surveys the sky in broad filters: $c$ (4200–6500~\AA), $o$ (5600–8200~\AA) and more recently $w$ (4200–7200~\AA) at their Teide Observatory unit \citep{2023sndd.confE...2L}. The survey operates with a typical cadence of about two days \citep{2020PASP..132h5002S,2021TNSAN...7....1S}. We obtained the corresponding $co$-band forced-photometry from the dedicated repository\footnote{\url{https://fallingstar-data.com/forcedphot/}.}, and processed the data using the pipeline developed for this purpose by \citet{Young_plot_atlas_fp}, adopting intra-night stacking. 

\subsubsection{BHTOM}
\label{sec:bhtom}

BHTOM coordinates a telescope network of $\sim$170 telescopes around the world (mirror diameter sizes range from 0.03 to 2.6 m), some of which are robotic, and some are manually operated by either amateur or professional observatories. The system operates through an open-source online infrastructure that delivers science-ready photometry via a uniform, automated pipeline. BHTOM is built on the Target and Observation Manager (TOMToolkit) framework developed by the Las Cumbres Observatory \citep{StreetTOM2024}. 
Observers upload reduced images (corrected for bias, dark current, and flat field) which are then automatically processed within the BHTOM pipeline. This performs calibration using established software packages, including CCDPhot \citep{2019CoSka..49..125Z,2020past.conf..190Z}, Source Extractor \citep[SExtractor;][]{1996A&ASBertin}, Software for Calibrating AstroMetry and Photometry \citep[SCAMP;][]{2006ASPCBertin}, Dominion Astrophysical Observatory Photometry \citep[DAOPHOT II;][]{1987PASPStetson}, and WCSTools\footnote{\url{http://tdc-www.harvard.edu/wcstools/}.}. Photometric calibration is performed using Gaia Synthetic Photometry \citep[GaiaSP;][]{2023A&AGaiaCollaboration}, which provides colour-corrected synthetic magnitudes suitable for point-spread function (PSF) photometry. The uncertainty assigned to each measurement corresponds to the formal PSF-fitting error returned by DAOPHOT~II \citep{1987PASPStetson}.
Astrometric solutions use several reference catalogues\footnote{Used catalogues include the U.S. Naval Observatory Robotic Astrometric Telescope catalog (URAT-1; \citealp{2015AJZacharias}), the fourth U.S. Naval Observatory CCD Astrograph Catalog (UCAC4; \citealp{2013AJZacharias}), the USNO-B1.0 catalog \citep{2003AJMonet}, and Gaia DR3 \citep{2023A&AGaiaDR3}.}. Photometric calibration is tied to magnitudes from GaiaSP and the Two Micron All Sky Survey (2MASS; \citealp{2006AJSkrutskie}). 
BHTOM then merges the newly processed observations with available archival photometry to produce the most complete photometric history possible. The photometry obtained with the BHTOM telescope network is not host-subtracted. However, the BHTOM measurements are fully consistent with the ZTF and SEDM observations in analogous filters (see below), indicating that any contribution from the host galaxy is negligible. Still, we exclude photometric points observed after 370 rest frame days after $r$-band peak, as host contamination may become increasingly significant as the SN fades, with the exception of $i/I$ band points, which remain consistent with host-subtracted SEDM observations.  
BHTOM also includes public observations from the Neil Gehrels Swift Observatory's \citep{2004ApJ...611.1005G} UV/optical Telescope \citep[UVOT;][]{2005SSRv..120...95R}. These are retrieved from the UVOT data from the HEASARC Data Archive\footnote{\url{https://heasarc.gsfc.nasa.gov/cgi-bin/W3Browse/w3browse.pl}.} and analyzed it using HEASoft\footnote{\url{https://heasarc.gsfc.nasa.gov/docs/software/lheasoft/}.} version 6.36 with UVOT CALDB version 20240201\footnote{\url{https://swift.gsfc.nasa.gov/caldb/}.}. Magnitudes are extracted using the standard UVOTSOURCE\footnote{\url{https://www.swift.ac.uk/analysis/uvot/mag.php}.} pipeline considering a 5\farcs\ circular source region centered on the source, and a background annulus with an inner radius of 6\farcs\ and an outer radius of 12\farcs\ also centered on the source. A point is considered a detection when the detection sigma is more than 3.

To reduce effects introduced by different observing conditions across the telescope network, we bin measurements in 3-day intervals. To do this, magnitudes are first converted to flux units. Each 3-day interval flux bin is assigned a time corresponding to the midpoint of the interval. For bins containing at least four measurements, we apply an iterative $3\sigma$ clipping procedure to remove outliers. The final binned flux is then calculated as the inverse-variance weighted mean of the remaining measurements. The uncertainty assigned to each bin is taken as the larger value between the formal uncertainty of the weighted mean and the empirical scatter of the flux values within the bin. This approach accounts for both the reported photometric uncertainties and for any additional dispersion between measurements.

\subsubsection{ZTF}
 
ZTF's wide-field survey (47-square-degree field of view) scans the northern sky with a cadence that goes from minutes to several days \citep[with a three day average for the public survey;][]{2019PASP..131f8003B}. Forced PSF photometry is retrieved from the Science Data System at IPAC\footnote{\url{https://www.ipac.caltech.edu}.} following ZTF guidelines\footnote{\url{https://irsa.ipac.caltech.edu/data/ZTF/docs/ztf_zfps_userguide.pdf}.}. The resulting light curves were processed using the approach described by \citet{2025A&A...695A.142P}, which includes removing measurements associated to poor quality flags and determining the zero-flux baseline for each filter. 

Additional photometry was obtained using the Palomar 60-inch telescope (P60) through the ZTF collaboration. Photometry is derived using PSF-fitting calibrated against standard stars from the Sloan Digital Sky Survey \citep[SDSS;][]{1995AAS...186.4405G}. Data reduction was performed using the \texttt{Fpipe} pipeline \citep{2016A&A...593A..68F}, employing SDSS reference images for calibration. 

Furhter $griz$ photometry was obtained with the IO:O photometer mounted on the LT. Images were processed using the \texttt{subphot\_pipe}\footnote{\url{https://github.com/kryanhinds/subphot_pipe}.} pipeline for image subtraction and PSF photometry, considering reference images from Pan-STARRS1 \citep{2016arXiv161205560C}. PSF fitting is performed relative to Pan-STARRS1 standards based on the techniques outlined in \cite{2008ApJ...680..550G} and \cite{2016A&A...593A..68F}.

\subsubsection{LBT}

Late time $JHK_s$ photometry was obtained with the LUCI (LBT Utility Camera in the Infrared; \citealt{Seifert2003}) instruments mounted on the LBT. The images were processed using the data reduction pipeline developed at INAF - Osservatorio Astronomico di Roma \citep{Fontana2014a} which includes bias subtraction and flat-fielding, bad pixel and cosmic ray masking, astrometric calibration, sky-field subtraction and coaddition. Field stars from the GAIA DR3 catalogue were used for astrometry calibration. The images were then processed by BHTOM's photometry reduction software (see Sec.~\ref{sec:bhtom} above) using 2MASS as a reference catalogue. The small number of catalogued reference stars contributes to the relatively large photometric uncertainties. Host-galaxy subtraction was not possible due to the lack of deep pre-explosion archival images.

\subsection{Spectroscopy}
\label{sec:spec} 

\begin{figure*}
   \centering
   \includegraphics[scale=0.47]{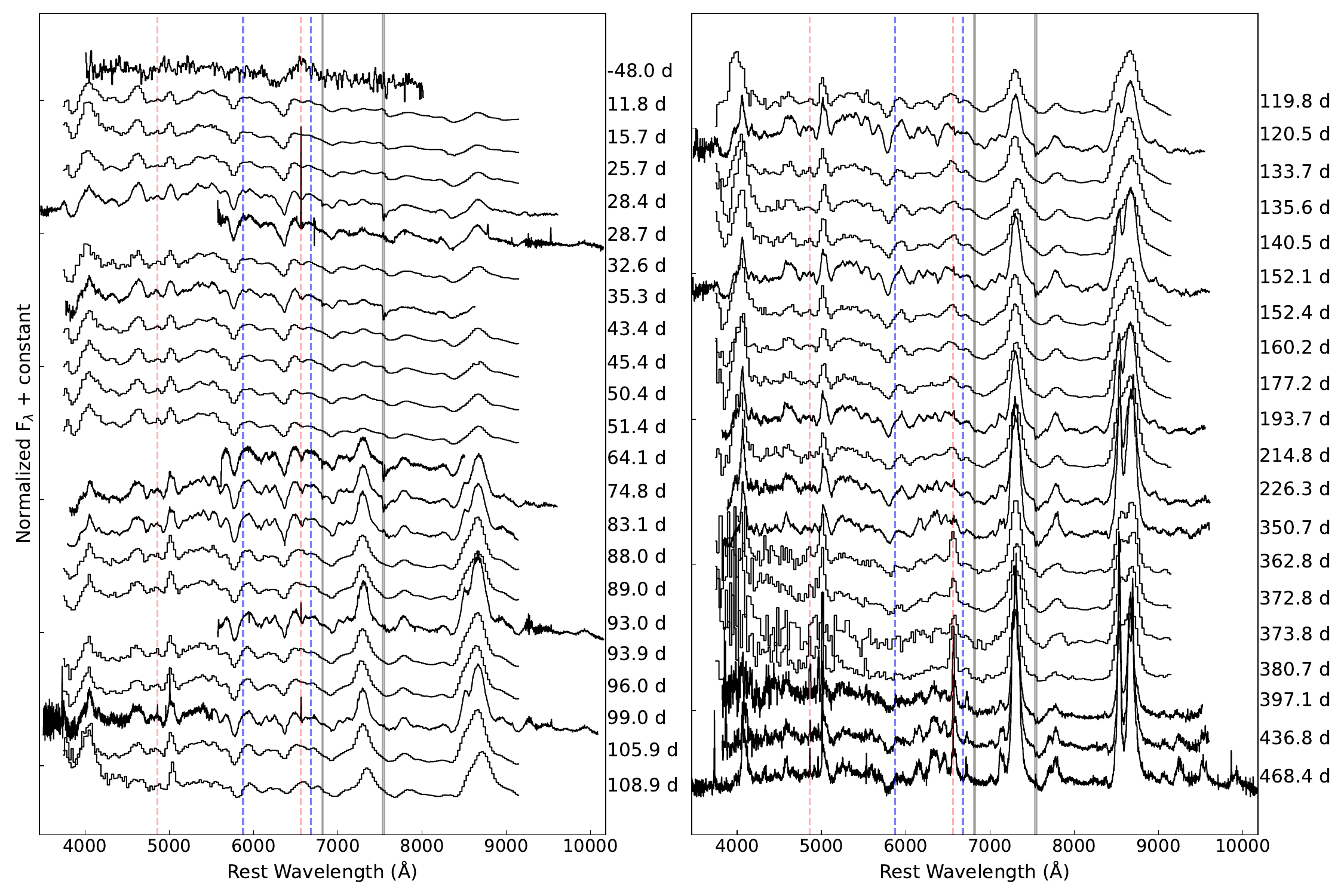}
      \caption{Optical spectral evolution. Vertical red dashed lines mark the rest wavelength of  H$\beta$ and H$\alpha$ and vertical blue dashed lines \ion{He}{i}~$\lambda$5876 and \ion{He}{i}~$\lambda$6678. Gray shaded regions indicate telluric absorption. The phase with respect to $r$-band peak is indicated at the right end of each spectrum.}
         \label{fig:obsspec}
   \end{figure*}

\begin{figure}
   \centering
   \includegraphics[width=\linewidth]{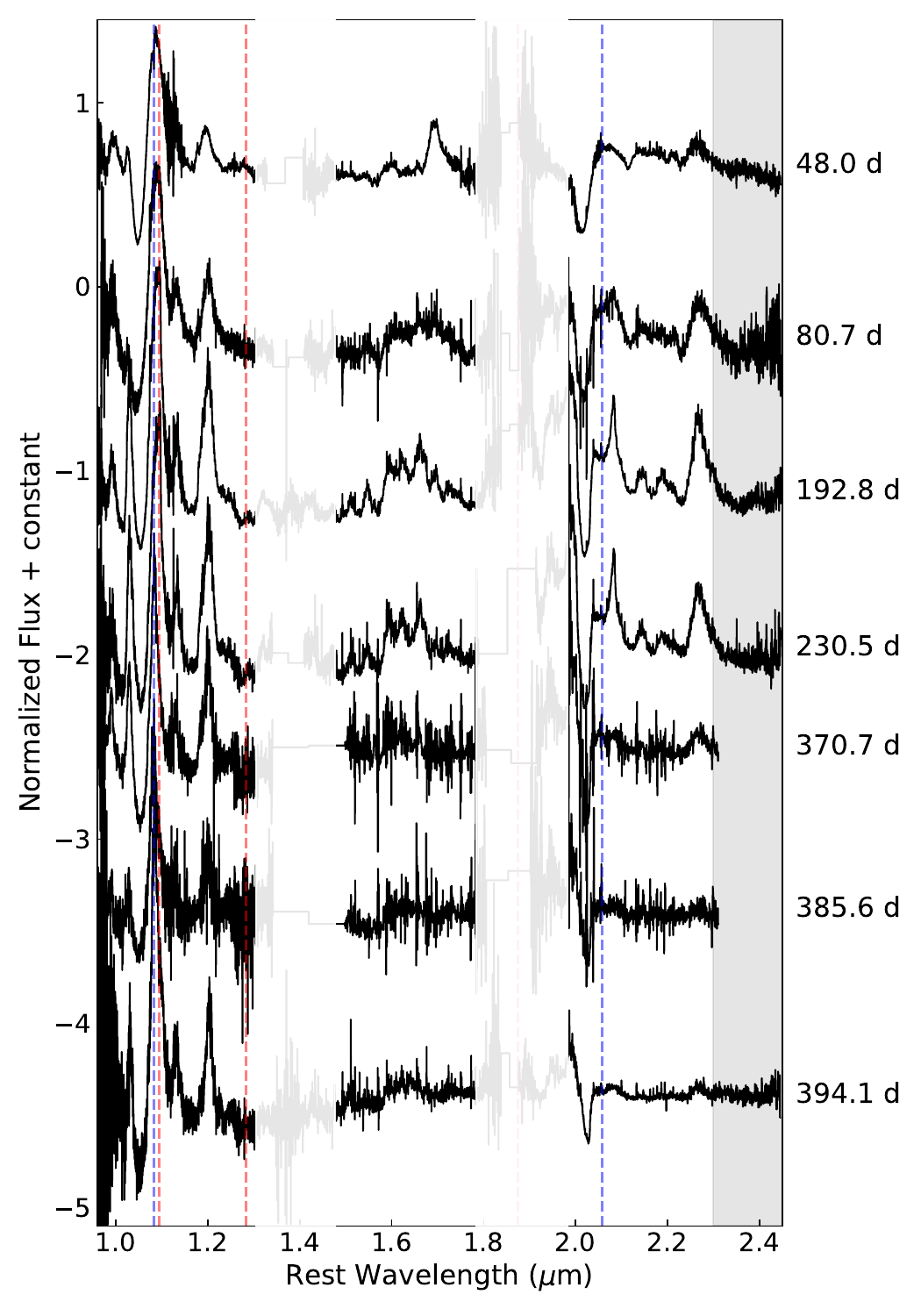}
      \caption{NIR spectral evolution. Telluric regions are shaded in white. Vertical red dashed lines mark Pa$\alpha$ (within the telluric region), Pa$\beta$ and Pa$\gamma$ lines, vertical blue dashed lines mark He\,\textsc{i}~$\lambda$1.083 and He\,\textsc{i}~$\lambda$2.0581. The grey shaded area corresponds to the region where the first CO overtone would be detected if present. The spectra were normalized by dividing each spectrum by its continuum, which was estimated using the \texttt{fit\_continuum} function from the \texttt{specutils} package, after masking regions containing strong features and telluric absorption.}
         \label{fig:obsspecNIR}
   \end{figure}
   
Spectroscopic follow-up was coordinated through the ZTF collaboration using the Fritz instance of SkyPortal \citep{2019JOSS....4.1247V, 2023ApJS..267...31C}. 
We obtained several epochs of spectroscopic observations using the SEDM spectrograph; the Alhambra Faint Object Spectrograph and Camera (ALFOSC) at the Nordic Optical Telescope; the Next Generation Palomar Spectrograph \citep[NGPS;][Fremling et al. in prep.]{2026PhDT........16D}
at the Palomar 200-inch telescope; the Binospec \citep{2019PASP..131g5004F}
spectrograph mounted on the 6.5-meter Multiple Mirror Telescope (MMT) at the MMT Observatory; the KAST Double Spectrograph on the 3.05-meter C. Donald Shane telescope at the Lick Observatory; and the Low-Resolution Imaging Spectrometer \citep[LRIS;][]{1995PASP..107..375O} on the Keck-I telescope. 
In addition, six spectral epochs were obtained in the NIR. Four with the R~2700 Near-Infrared Echelle Spectrograph \citep[NIRES;][]{2004SPIE.5492.1295W} located on the Keck-II telescope and two with the LUCI instruments at the LBT. The log of spectroscopic observations can be found in Table~\ref{tab:speclog}. The optical spectral evolution is shown in Fig.~\ref{fig:obsspec}, including the two available classification spectra; the NIR spectral evolution is shown in Fig.~\ref{fig:obsspecNIR}. 
We describe corresponding spectral reduction in Appendix.~\ref{app:spec-reduction}.

The spectra were de-reddened for MW extinction (see Sec.~\ref{sec:phot}) and absolutely flux-calibrated against the available $ri$ photometry by comparing the observed photometry with spectral synthetic photometry generated using the \texttt{synphot} package. When photometry was unavailable at the spectroscopic epoch, we used light curve interpolations to estimate the corresponding magnitudes (see Sec.~\ref{sec:light-curve}). Due to the lack of NIR photometric coverage, we flux-calibrated the NIR spectra using optical spectra obtained at similar phases, fitting a blackbody continuum to each optical spectrum and scaling the corresponding NIR spectrum to match the blackbody continuum over the first 10 percent of its wavelength range.
   
\section{Analysis}
\label{sec:analysis} 

SN~2024afyu is classified as a Type IIb supernova, but several of its observed properties differ from those typically associated with this class. Below we examine the main observational characteristics of the event to assess the validity of the proposed classification and to investigate its powering mechanism.

\subsection{Host Galaxy}
\label{sec:host}

\begin{figure}
   \centering
   \includegraphics[width=\linewidth]{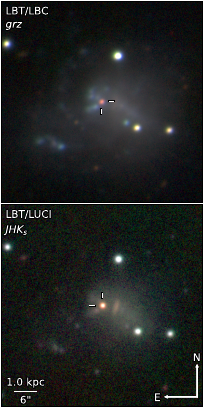}
      \caption{Colour composite images of the host galaxy of SN~2024afyu generated with images from the LBT obtained in June 2026. Both panels are centred on the SN position, marked by the crosshair. The optical composite (top panel) resolves the host into discrete blue star-forming knots, whereas the NIR composite (bottom panel) shows a smooth, centrally concentrated stellar distribution containing a short bar. The images were corrected for neither Milky Way extinction nor host attenuation.}
         \label{fig:host}
   \end{figure}

SN~2024afyu is located 10.3\farcs\ \citep[1.9~kpc, calculated as $\mathrm{D} = \mathrm{D}_{\mathrm{L}} / (1 + z)^{2}$;][]{1999astro.ph..5116H} northeast of the reported coordinates of the galaxy 2MFGC~13744. However, the reported coordinates appear to be offset from the bulk of the galaxy, and the true centre of the host may instead be closer to the observed NIR bar (see Fig.~\ref{fig:host}). The host does not have a reported spectroscopic redshift. To investigate the extinction associated with the host galaxy we searched for \ion{Na}{i} D features in the supernova spectra. Although a shallow absorption feature is visible near 5890~\AA\ at approximately 28 and 35 days, it is absent at all other epochs. In addition, the host H$\alpha$/H$\beta$ seen in our latest spectrum is close to the expected ratio for zero extinction (although the continuum around the H$\alpha$ is contaminated by a SN feature). Thus, we assume that the host-galaxy extinction is negligible.

Figure~\ref{fig:host} shows an optical and a NIR false-colour image of the host galaxy of SN~2024afyu, built with the software package \texttt{STIFF} version 2.4.0 \citep{Bertin2012a} from $grz$ and $JHK_s$ images obtained in June 2026, respectively. The host presents markedly different morphologies in the optical and the near-infrared. In the optical composite, the galaxy resolves into several discrete blue knots embedded in a faint, mottled envelope (Fig.~\ref{fig:host}, left panel), with no coherent spiral pattern or bar apparent. The $JHK_s$ false colour image instead reveals a smooth, centrally concentrated light distribution dominated by an elongated bar (Fig.~\ref{fig:host}, right panel). This morphological contrast reflects differences in the underlying stellar populations. Optical light is dominated by luminous, young massive stars and associated \ion{H}{ii} regions that contribute little to the overall stellar mass content. Near-infrared wavelengths instead trace the older red giant and asymptotic giant branch populations that account for the bulk of the stellar mass, making the $JHK_s$ image an effective tracer of the galaxy's underlying mass distribution \citep{Eskridge2000a, Eskridge2002a, MendendezDelmestre2007a}.
While bars are major drivers of secular evolution, funneling gas inward and redistributing angular momentum, their high mass concentration of older stars makes them most prominent in NIR mass maps \citep{Kormendy2013a, DiazGarcia2016a}. 
The lack of a corresponding spiral pattern in $JHK_s$ highlights the distinct dynamical origins of bars and arms. In low-mass or gas-rich discs, optical spiral features often trace localised, transient star-forming regions rather than massive stellar density waves, allowing a central bar to exist without prominent underlying stellar arms \citep{Hunter2006a, Erwin2018a}.

The SED fit of the host (Fig.~\ref{fig:host:sed} in Appendix~\ref{app:host}) yields living-star mass of $\log(M_\star/M_\odot) = 8.83^{+0.14}_{-0.09}$ and star-formation rate of 
${\rm SFR} = 0.1\,M_\odot\,{\rm yr^{-1}}$. Extrapolating the main-sequence parametrisation of \citet{Elbaz2007a} (their equation 5) to this mass at $z \simeq 0$ predicts $\sim0.19\,M_\odot\,{\rm yr^{-1}}$. The measured rate is consistent with that value. The host is a normal star-forming dwarf, not a starburst. Its stellar mass falls between those of the SMC and the LMC \citep[$460\times10^6$ and $1500\times10^6~M_\odot$, respectively;][]{McConnachie2012a}. Following \citet{Kennicutt1998a}, the H$\alpha$ flux measured from our latest spectrum translates to a star-formation rate of $(5.0\pm0.2)\times10^{-3}~M_\odot\,\rm yr^{-1}$ at the explosion site (converted to the Chabrier IMF using \citealt{Madau2014a}).
The origin of clumpy off-centre star formation in SN~2024afyu's host is unclear. Since the integrated star-formation rate is comparable to the expected SFR for a star-forming galaxy at this mass. Although there is some evidence of galaxy interaction (see Appendix~\ref{app:host}) the clumpyness may not be connected with this but with secular galaxy evolution, radio observations would be necessary to answer this question.

\subsection{Light curve}
\label{sec:light-curve}

The top panel of Fig~\ref{fig:obslc} shows the observed magnitudes in the different filters. We can see that the peak was only fully covered in $o$ and $r$ bands. 
The light curve shows a ``flattening'' after peak, visible in every band. Right before this flattening, a slightly fainter photometric point is visible in every band covering such phases (marked with a vertical dotted line in Fig.~\ref{fig:obslc}). This fainter point is most noticeable in the $V$ band. the fact that is seen in different bands observed with different instruments makes it likely real and not an instrumental effect. We note that the absence of photometry obtained earlier than this fainter point could have resulted in largely underestimated rise times.

To estimate light curve parameters, we first interpolate them using Gaussian Process \citep[GP, e.g:][]{2006gpml.book.....R}, utilizing the \texttt{GPy} Python package\footnote{\url{https://gpy.readthedocs.io/en/deploy/}.}. Using the GP median as representation of the light curve, we estimate the peak epoch and absolute magnitude of SN~2024afyu to be t$_{\mathrm{Peak}}^{r} =$ 60727.6 $\pm$ 2.9~d and M$_{\mathrm{Peak}}^{r} = -$18.9 $\pm$ 0.04~mag, respectively. All phases are reported as rest frame days since t$_{\mathrm{Peak}}^{r}$ unless stated otherwise.
The corresponding error bars represent the standard deviation of the estimated values considering 1000 posterior samples. GP was also used to estimate the explosion epoch of SN~2024afyu. To do this, we considered the ZTF $r$-band forced photometry flux (see Fig.~\ref{fig:exp_peak_epoch}). The intersection between the observed baseline (flux = 0 $\mu$Jy) and the interpolated light curve is adopted as the explosion epoch t$_{\mathrm{Exp}} =$ 60641.9 $\pm$ 11.3~d, the associated error bar represent the standard deviation of the intersection between the baseline and 1000 posterior samples of the GP. 

GP interpolation was also used to estimate rise and decline times, considering these as 10\% and 1/e flux fractions with respect to peak \citep[$\Delta$mag $=$ 2.5 and $\Delta$mag $=$ 1.09, respectively;][]{2023ApJ...943...41C,2025A&A...695A.142P}. t$_{\mathrm{rise,10\%}}$ cannot be determined due to the lack of early observations, t$_{\mathrm{rise,1/e}} = 53.6 \pm 2.9$ days, t$_{\mathrm{dec,1/e}} =63.9 \pm 2.9$ days and t$_{\mathrm{dec,10\%}} = 184.4 \pm 2.9$ days. Given the time of explosion, the rise time computed as time elapsed since explosion is t$_{\mathrm{rise}} =$ (t$_{\mathrm{Peak}}^{r} -$ t$_{\mathrm{Exp}}$) / (1$+ z$) $= 85 \pm 11.7$ days. SN~2024afyu became unobservable at $\sim$230 days after peak, and became observable again at $\sim$340 days after peak. Soon after, it started declining much more rapidly than before (see Fig.~\ref{fig:obslc}). 

We also compute the observed $g - r$ and $r - i$ colors (see bottom panel of Fig.~\ref{fig:obslc}). In this case, given the large gap in the light curve between $\sim$240 and $\sim$336 days, these interpolations were performed using the Automated Loess Regression (\texttt{ALR}) pipeline presented by \cite{2019MNRAS.483.5459R}, as this method is less sensitive to gaps than GP. We can see that the colors initially redden with time, the become slightly bluer at around the time of the first peak decline to redden again with time, until the sharp drop seen in the light curves, at which time the color becomes suddenly blue, to continue to redden after that. 

To constrain the energetics of SN~2024afyu, we construct its pseudo-bolometric light curve using only $gri$ 
observations, as these provide the most extensive follow-up coverage. The SED is estimated by interpolating these bands with respect to the adopted peak epoch and integrating over the resulting \texttt{ALR} interpolated curve. 
Our approach only provides a lower limit for the bolometric luminosity as it completely ignores UV and IR contributions to the SED.
Still, it allow us to place a lower limit on the total radiated energy, estimated to be $E_{\mathrm{rad}} \gtrsim 2.2 \times 10^{49}$~erg. 

In addition, we estimate the $^{56}$Ni mass by scaling the late-time luminosity ($\sim$200 days) to that of SN~1987A, whose distance, extinction, and radioactive-powered tail are well constrained, making it the standard reference for such measurements. We perform this comparison using the $r$-band and pseudo-bolometric light curves. To ensure consistency, we construct a $BRI$\footnote{We chose these bands as  they have good photometric coverage and similar central wavelengths to $gri$.} pseudo-bolometric light curve for SN~1987A using the same methodology adopted for SN~2024afyu. The resulting $^{56}$Ni mass depends on the comparison employed: $\sim$0.6 \Msun{}, from the $r$ versus $R$-band light curves, $\sim$0.4 \Msun{} when comparing the pseudo-bolometric light curve of SN~2024afyu with the most complete bolometric light curve of SN~1987A \citep[as presented by][]{1990AJ.....99..650S}, and $\sim$1.0 \Msun{} when comparing the two $BRI$/$gri$ pseudo-bolometric light curves. 

\subsection{Spectral evolution}
\label{sec:spectral-evol}

Figures~\ref{fig:obsspec} and \ref{fig:obsspecNIR} show the optical and NIR spectral evolution, respectively. Using the Supernova Identification code \citep[\texttt{SNID};][]{2007ApJ...666.1024B}, we find that the earliest spectrum is best matched to a SN~1987A, consistent with the reported classification. 
Our earliest observed spectrum was obtained by SEDM at 11.8 days, $\sim$23 days before the publicly reported reclassification spectrum. This SEDM spectrum already shows signs of a trough on top of the feature identified as H$\alpha$ in typical SNe~IIb. SNID's SN~IIb classification holds up to $\sim$88 days. 
Beyond this epoch, a SN~Ib becomes the preferred match up to $\sim$109 days. Unconstrained SNID fits of later spectra return much higher redshifts ($z \sim0.1-0.5$) than the observed one ($z = 0.0085$). Fixing the redshift to the observed value, the best matches remain SN~II/IIb/Ib, although at severely discrepant phases. The spectrum at $\sim$108 days, for instance, is matched to templates at $\sim$300 days. This phase inconsistency persists until $\sim$362 days, at which point the best match is again SN~1987A at comparable phases.
We find that the absorption attributed to H$\alpha$ may be formed by a blend of H and Si and/or maybe Ca (see Appendix~\ref{app:synapps}). And the associated trough on top of the emission component of the same line does not significantly evolve with time, as seen in other SN~IIb. Instead, it gradually shifts redward with time. By $\sim$350 days, the blue component gives way to an emission feature with a broad base. 

\begin{figure*}
   \centering
   \includegraphics[width=\textwidth]{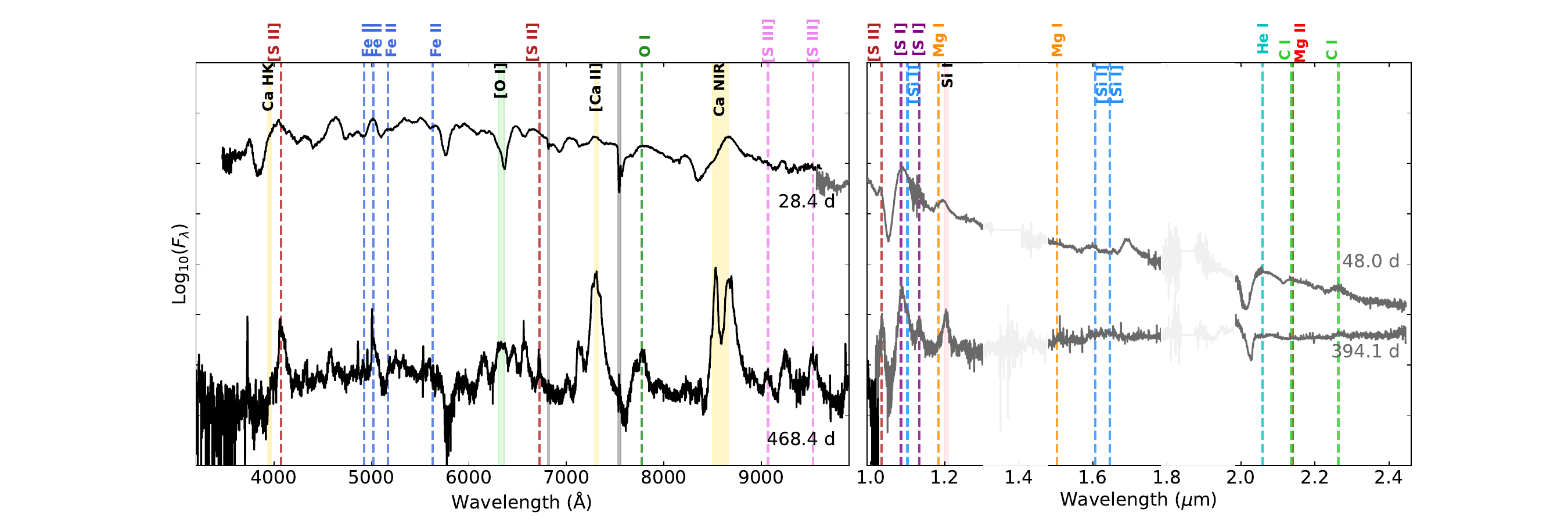}
      \caption{Identified spectral lines. Left panel: absolute flux-calibrated NIR spectra at $\sim$48 and 394 days past peak. Right panel: absolute flux-calibrated optical spectra at $\sim$28 and 468 days past peak. Telluric regions are shaded in grey in the optical region and in white in the NIR region.}
         \label{fig:speclines}
   \end{figure*}

Main H and He wavelengths are shown in Figs.~\ref{fig:obsspec} and \ref{fig:obsspecNIR}; Fig.~\ref{fig:speclines} presents the remaining identified spectral features. 
Features at $\sim$4600\AA\ and $\sim$5000\AA\ persist through the spectral evolution. 
These may be attributed to \ion{Mg}{i}]$~\lambda$4571 and \ion{He}{i}$~\lambda$5016, respectively, or alternatively to \ion{Fe}{ii} line blends. A feature at $\sim$4070\AA\ becomes prominent as the spectra evolves. We identify this as [\ion{S}{ii}] $\lambda\lambda$4068,4076 in Sec.~\ref{sec:sulfurANDSiliconID}.  
[\ion{Ca}{ii}]~$\lambda\lambda$7291,7324 becomes detectable from $\sim$28 days onward, strengthening steadily with time. The early appearance of this feature is remarkable because [\ion{Ca}{ii}] emission is more typically associated with nebular phases \citep[e.g.][]{2017MNRAS.467..369S,2022MNRAS.514.5686P}, and it is commonly accompanied by [\ion{O}{i}] $\lambda\lambda$6300,6364 \citep[e.g.][]{2012ApJ...755..161K,2022ApJ...928..151F}. In SN~2024afyu, however, [\ion{O}{i}] only becomes marginally detectable at $\sim$177 days. 

The NIR spectra also show some notable features. 
An emission feature at $\sim$1.03~$\mu$m emerges at $\sim$193 days, which we associate with [\ion{S}{ii}]~$\lambda\lambda$1.029,1.034,1.037 (see Sec.~\ref{sec:sulfurANDSiliconID}). We also see a broad line appearing at every epoch at $\sim$2.26$\mu$m that is identified as \ion{Ca}{i}~$\lambda\lambda$2.261,2.263,2.266 \citep{2018MNRAS.481..806B}. We also identify the line $\sim$2.1$\mu$m as \ion{C}{i}~$\lambda2.13$, although it could also be associated to Mg\,\textsc{ii}~$\lambda$2.14. A relatively narrow emission line is visible at $\sim$193 and 230 days at $\sim$2.08$\mu$m, these line remains unidentified, although it could be associated to \ion{Fe}{ii}~$\lambda$2.0888.

\subsection{Comparison to other events}
\label{sec:comparison}

\begin{figure}
   \includegraphics[width=\linewidth]{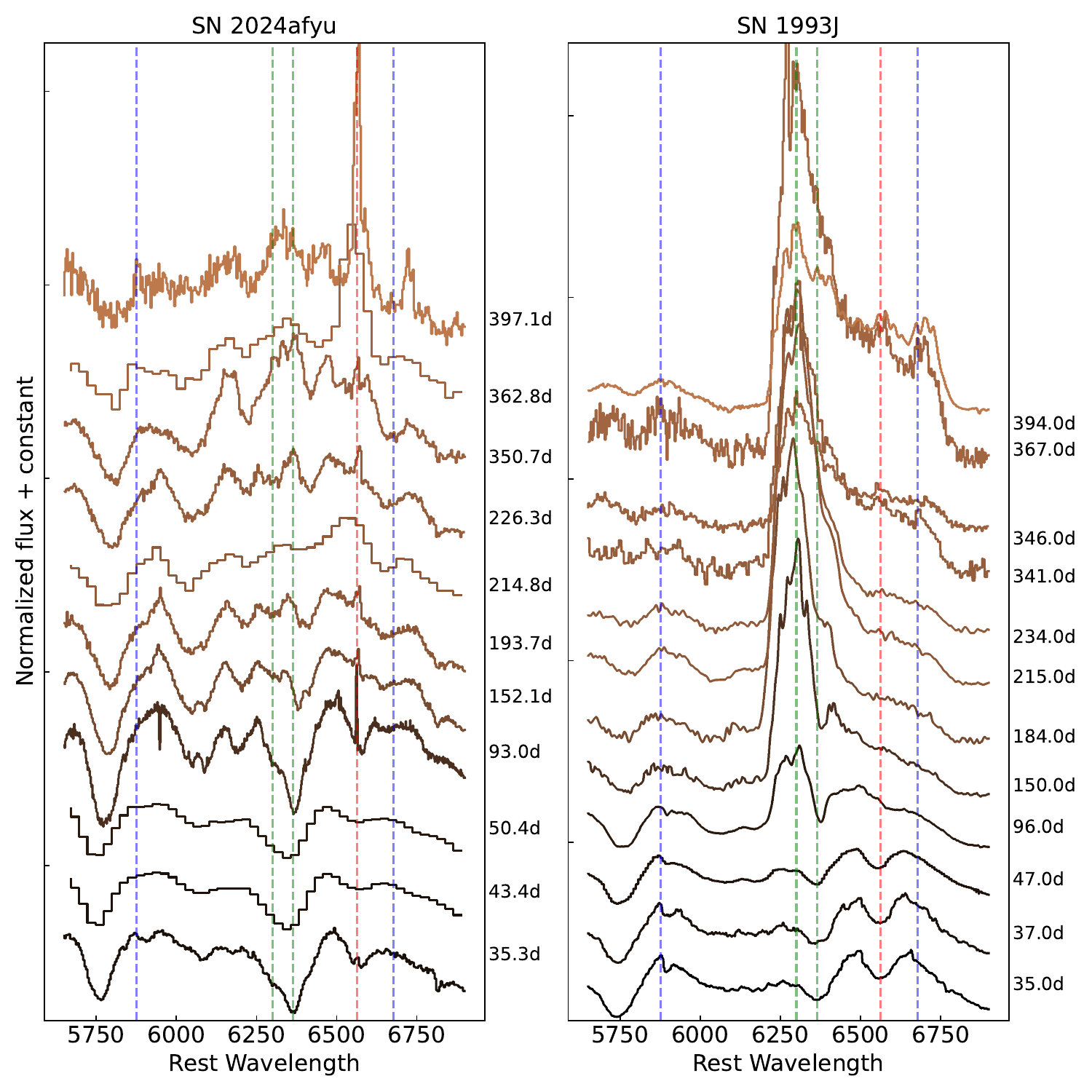}
      \caption{Evolution of the region around He\,\textsc{i}~$\lambda$5876 and He\,\textsc{i}~$\lambda$6678 for SN~2024afyu (left) and SN~1993J (right). Color coding follows the spectral phase annotated to the right of each spectrum. A vertical red dashed line marks the rest wavelength of H$\alpha$, vertical blue dashed lines mark He\,\textsc{i}~$\lambda$5876 and He\,\textsc{i}~$\lambda$6678, and the vertical green dashed lines mark [O\,\textsc{i}]$\lambda\lambda$6300,6364. The spectra of SN~1993J were published by \cite{1995A&AS..110..513B,2000AJ....120.1499M,2015A&A...573A..12J}.}
         \label{fig:comparison93J}
   \end{figure}

In this section we compare the light curve and spectra of SN~2024afyu to other well-studied CCSNe, in order to further explore its assigned Type IIb classification.
First we compare the spectral region around \ion{He}{i}$\lambda$5876 and \ion{He}{i}$\lambda$6678 to that of the canonical Type~IIb SN~1993J \citep{1993ApJ...415L.103F} in Fig.~\ref{fig:comparison93J}, as this region most clearly traces the defining H$\alpha$-to-\ion{He}{i} transition in SN~IIb. We can see that the evolution of SN~2024afyu's features differs substantially from that of SN~1993J at comparable phases, particularly after $\sim$90 days. In SN~1993J, H$\alpha$ absorption progressively gives way to [\ion{O}{i}]$\lambda\lambda$6300,6364, with the feature broadening and developing a characteristic boxy profile, while the \ion{He}{i}$\lambda$5876 absorption weakens steadily with time. In contrast, SN~2024afyu's H$\alpha$ associated feature weakens steadily with time, evolving into an emission line. The surrounding region develops a complex multi-component structure suggestive of blending from multiple species. At late phases, a relatively narrow emission feature appears on top of a broad base at the rest wavelength of H$\alpha$, we attribute the narrow component to contamination from the underlying host galaxy, while the broad base could be produced by the SN.

\begin{figure}
   \includegraphics[width=\linewidth]{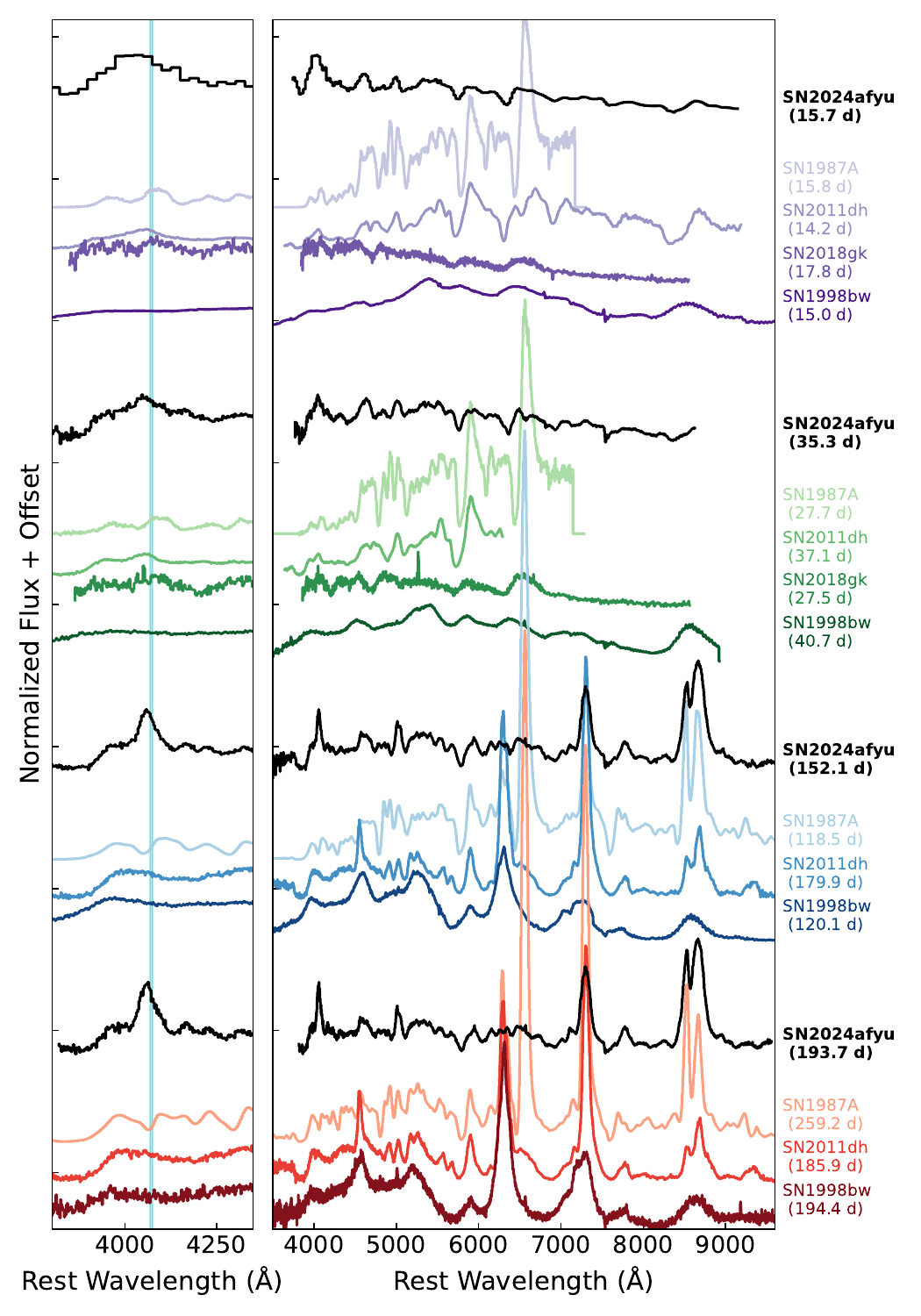}
      \caption{Optical spectral comparison of SN~2024afyu (black) with Type~II SN~1987A, Type~IIb SNe~2011dh and 2018gk, and Type~Ic-BL SN~1998bw, colour-coded by spectral phase. Left panel: zoom into the $\sim$4070~\AA\ region, with a cyan vertical line marking the rest wavelengths of \SIIoptical. Right panel: full optical wavelength range. Spectra of SN~1987A are from \citet{1995ApJS...99..223P}; SN~2011dh from \citet{2013MNRAS.436.3614S} and \citet{2014A&A...562A..17E}; SN~2018gk from \citet{2021MNRAS.503.3472B}; and SN~1998bw from \citet{2001ApJ...555..900P}. }
         \label{fig:comparisonALL}
   \end{figure}

\begin{figure}
   \includegraphics[width=\linewidth]{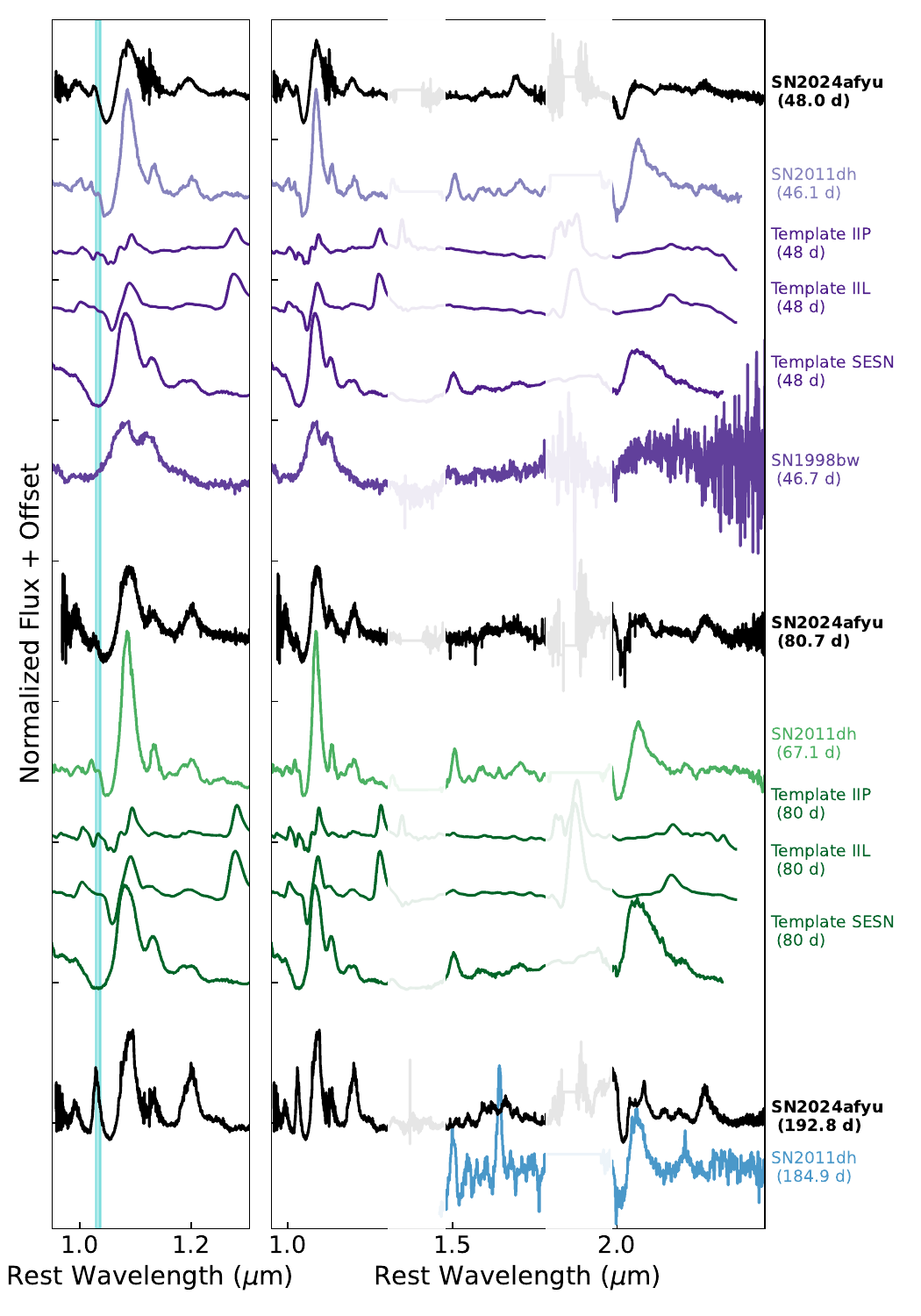}
      \caption{NIR spectral comparison of SN~2024afyu (black) with Type~IIb SN~2011dh \citep[][]{2014A&A...562A..17E,2015A&A...580A.142E}, Type~Ic-BL SN~1998bw \citep[][]{2001ApJ...555..900P}, and NIR templates for Type II SNe \citep{2019ApJ...887....4D} and He-rich SESNe \citep{2022ApJ...925..175S} colour-coded by spectral phase. Left panel: zoom into the $\sim$1.03~$\mu$m region, with a cyan band marking the rest wavelengths of [\ion{S}{ii}] $\lambda\lambda1.029,1.034,1.037$. Right panel: full NIR wavelength range.}
         \label{fig:comparisonALLNIR}
   \end{figure}

For a broader comparison, we also include comparison of full spectra of SN~2024afyu to: the long-rising SN~1987A (given the original SN~II classification of SN~2024afyu); the well-studied SN~IIb 2011dh, which also has publicly available NIR spectroscopy; the luminous SN~IIb 2018gk; and the Type~Ic-BL SN~1998bw, motivated by the large inferred $^{56}$Ni mass (Sec.~\ref{sec:light-curve}) and by the fact that SN~1998bw has been shown to resemble SNe~Ib/IIb at nebular phases \citep{2001ApJ...555..900P}. We further include the NIR spectral templates for Type~II SNe from \cite{2019ApJ...887....4D} and for He-rich SESNe from \cite{2022ApJ...925..175S}. The optical and NIR spectroscopic comparisons are presented in Figs.~\ref{fig:comparisonALL} and \ref{fig:comparisonALLNIR}, respectively.
Comparison spectra were retrieved from WISeREP\footnote{\url{https://www.wiserep.org}.} \citep{2012PASP..124..668Y} and the Open Supernova Catalog \citep[OSC;][]{2017ApJ...835...64G}\footnote{The OSC contains all transients uploaded through 8 April 2022 and is available at \url{https://github.com/astrocatalogs/supernovae}.}. Specific references are included in the caption of the figures.

Figure~\ref{fig:comparisonALL} emphasizes the spectral peculiarities of SN~2024afyu relative to other CCSNe. At early epochs ($\sim$15--35 days), SN~2024afyu shows a bluer continuum and fewer spectral features than both SN~1987A and SN~2011dh at comparable phases. Overall, none of the comparison objects provides a particularly close match.
A broad emission is present in SN~2024afyu at a wavelength comparable to that of the narrow \ion{Mg}{i}]$\lambda$4571 line observed in SN~2011dh. Nebular spectral models of SN~2011dh presented by \citet{2015A&A...573A..12J} predict a broader feature at the position of \ion{Mg}{i}]$\lambda$4571 that is largely attributable to blends of \ion{Fe}{ii} lines. Thus, an identification of this feature as \ion{Fe}{ii} emission cannot be ruled out, although see Sec.~\ref{sec:sulfurANDSiliconID}.
The emission feature observed near 5000~\AA\ in SN~2024afyu could be associated with \ion{He}{i}$\lambda$5016, although none of the comparison SN exhibits a line of comparable strength at this wavelength. Likewise, the feature at $\sim$4070~\AA\ appears unique to SN~2024afyu and is absent from all comparison spectra considered here (see Sec.~\ref{sec:sulfurANDSiliconID}).
Figure~\ref{fig:comparisonALLNIR} shows that SN~2024afyu also exhibits notable differences from other CCSNe in the NIR, particularly with respect to Type~II events. In particular, several spectral features are weaker in SN~2024afyu, including the line at $\sim$1.5$\mu$m, typically associated to \ion{Mg}{i} $\lambda$1.5033. The strength of the emission \ion{He}{i} $\lambda$2.0581 emission feature is also weaker in SN~2024afyu. At similar wavelengths, \cite{2022ApJ...925..175S} also identify \ion{C}{i} $\lambda$2.1259 and/or \ion{Mg}{ii} $\lambda$2.1369 as main contributors. At $\sim$1.03~\AA\ we see a line in SN~2024afyu that is only marginally detected in some comparison spectra (see Sec.~\ref{sec:sulfurANDSiliconID}).

\begin{figure}
   \centering
   \includegraphics[width=\linewidth]{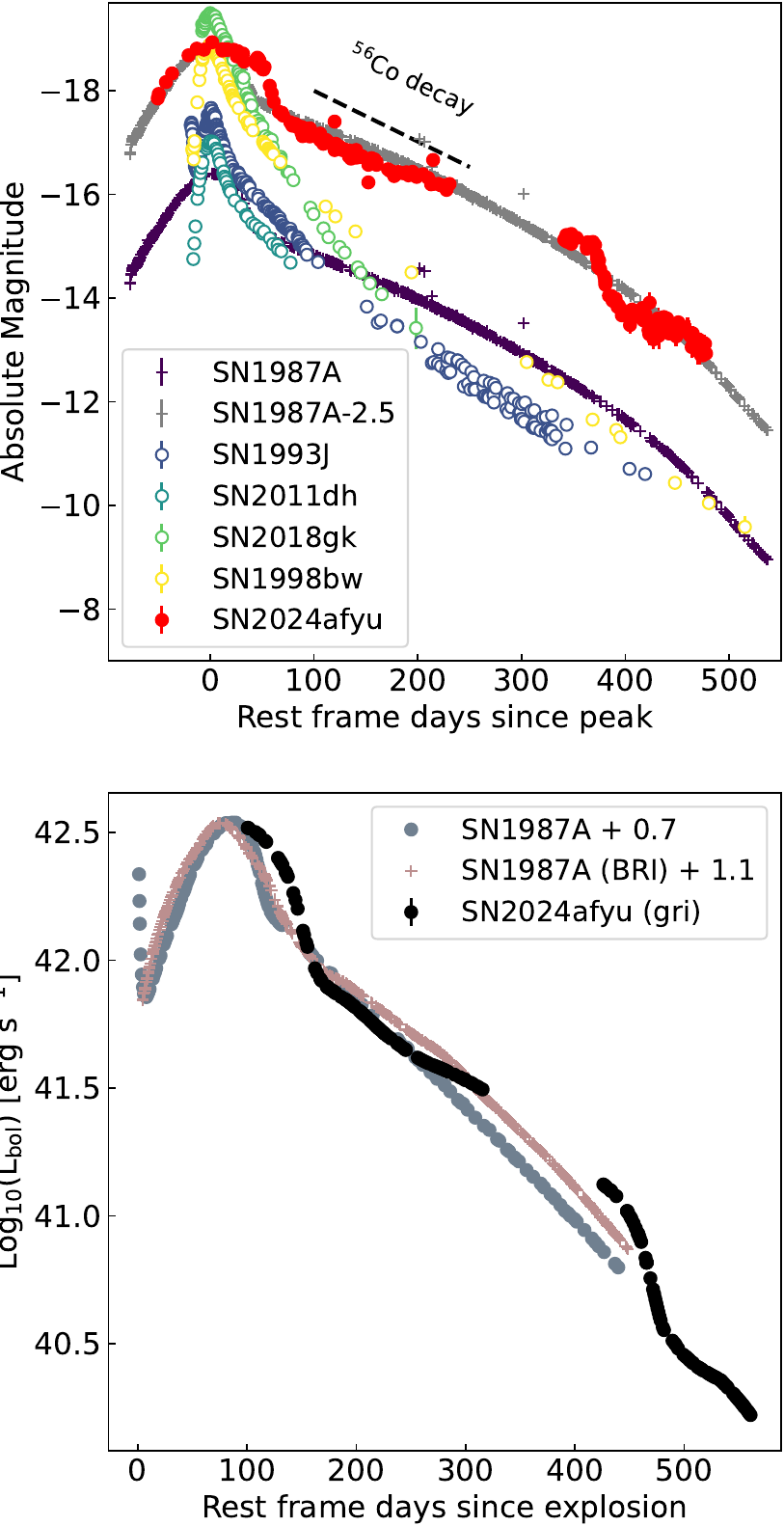}
      \caption{Top panel: $r$-band light curve of SN~2024afyu (red) compared to $r/R$-band light curves of Type~II SN~1987A \citep[retrieved from the OSC with the following data sources:][]{1987MNRAS.229P..15C,1987MNRAS.227P..39M,1988MNRAS.231P..75C,1988AJ.....96.1864S,1988MNRAS.234P...5W,1989MNRAS.237P..55C,1991PASP..103..958W,1992ApJ...384L..33S}, canonical Type~IIb SN~1993J \citep[retrieved from the OSC with the following data sources:][]{1993PASJ...45L..59V,1994AJ....107.1453B,1994AJ....107.1022R,1995A&AS..110..513B,1996AJ....112..732R}, Type~IIb SN~2011dh \citep[retrieved from][]{2014A&A...562A..17E}, luminous Type~IIb SN~2018gk \citep[retrieved from][]{2021MNRAS.503.3472B}, and Type~Ic-BL SN~1998bw \citep[retrieved from][]{2011AJ....141..163C}. Peak MJDs were obtained from GP interpolation. The light curve of SN~1987A is also presented in grey with a 2.5~mag shift for a more direct comparison to SN~2024afyu. A dashed black line marks the typical $^{56}$Co decay rate. Bottom panel: Comparison of the $gri$ pseudo-bolometric light curve of SN~2024afyu compared to that the $BRI$ and full bolometric light curves of SN~1987A (see Sec.~\ref{sec:light-curve}). These have been shifted 1.1 and 0.7 dex, respectively, in order for peaks to match.}
         \label{fig:compLC}
   \end{figure}

Figure~\ref{fig:compLC} compares the light curve of SN~2024afyu with those of the selected comparison sample. The rise time of SN~2024afyu ($\sim$85 days) is comparable to that of SN~1987A ($\sim$80 days; e.g., \citealt{1988Ap&SS.150..291M,1988PASA....7..401M,1989ARA&A..27..629A,2023ApJ...959..142S}). Yet, SN~2024afyu is more luminous, reaching a peak absolute magnitude comparable to that of SN~1998bw, although this has a narrower light curve. While SN~2024afyu is less luminous than the luminous Type~IIb SN~2018gk, its light curve is again substantially broader. At $\sim$370 days, SN~2024afyu shows a pronounced drop in brightness, deviating from the decline expected from the radioactive decay of $^{56}$Co that typically powers core-collapse supernovae. By $\sim$410 days, however, the overall decline rate once again resembles that of SN~1987A. Still, the late-time evolution is not smooth or linear; instead, the light curve displays several bumps (see Fig.~\ref{fig:obslc}). 

\subsubsection{Line strengths and velocities}
\label{sec:VelandpEW}

\ion{Fe}{ii}$\lambda$5169, H$\alpha$/\ion{Si}{ii}$\lambda$6355, and \ion{O}{i}$\lambda$7774 are among the commonly used diagnostic features in SESNe. In SN~2024afyu, however, none of these features can be confidently identified. In particular, the absorption feature immediately bluewards of 6563\AA\ may arise from a blend of H$\alpha$, \ion{Si}{ii}$\lambda$6355 and potentially \ion{Ca}{i}$\lambda$6572 (see Appendix~\ref{app:synapps}). To investigate whether these absorptions can plausibly be associated with the proposed ions, we measured the pseudo-equivalent width (pEW) and expansion velocity of the first absorption minimum seen bluewards of each corresponding rest wavelength, see Appendix~\ref{app:spec-measurements}. The resulting velocities and pEWs were compared with the SESN samples of \citet{2016ApJ...827...90L} and \citet{2023A&A...675A..83H}, as well as the SLSN~I sample of \citet{2017ApJ...845...85L}. Since these studies adopt rest frame phases relative to the $V$- \citep{2016ApJ...827...90L,2017ApJ...845...85L} or $B$-band maximum \citep{2023A&A...675A..83H}, we studied our $V$ band light curve and estimate a t$_{\mathrm{Peak}}^{V} = 60734 \pm 5.3$ days, this value was use to convert our measurements to the corresponding rest frame phases for a consistent comparison. 

Results are presented in Fig.~\ref{fig:pEWandVEL}. We see that the absorption feature of SN~2024afyu identified with \ion{Fe}{ii}$\lambda$5169, exhibits significantly lower expansion velocities than those observed in SLSNe~I throughout the available phases. These velocities occupy the lower end of the SESN distribution and are broadly comparable to those measured in some Type IIb and Ib supernovae. The feature bluewards of 6563\AA, associated with H$\alpha$/\ion{Si}{ii}$\lambda$6355, shows velocities intermediate between the Type Ib and Ic events, while the feature bluewards of \ion{O}{i}$\lambda$7774 displays velocities lying between the Type IIb and Ib distributions. These latter measurements should be interpreted with caution, as the spectral region is affected by telluric absorption, although spectra showing obvious telluric contamination were excluded from the analysis.
The pEW evolution of the lines of SN~2024afyu is more difficult to interpret. The measurements exhibit substantial scatter, reflecting the lower spectral resolution of the SEDM observations, showing no clear correspondence with any of the SNe subclasses considered here.

\section{Sulfur and Silicon in SN 2024afyu}
\label{sec:sulfurANDSiliconID}

As noted previously, the spectrum of SN~2024afyu exhibits a feature at $\sim$4070\AA\ that appears to be unique among the considered comparison sample (see Fig.~\ref{fig:comparisonALL}). Motivated by both its wavelength and the potentially large $^{56}$Ni mass ($\sim$0.4--1.0 \Msun{}, see Sec.~\ref{sec:light-curve}), we investigate the spectrum under the assumption that this line is caused by \SIIoptical. Such an identification is physically motivated, as explosions that synthesize large amounts of $^{56}$Ni are also expected to produce significant quantities of intermediate-mass elements, including sulfur, during explosive oxygen and silicon burning. Under this hypothesis, we found more lines from S and Si. At the luminosities seen in SN~2024afyu, the formation of these lines require multiple solar masses of S and Si in the ejecta. Such abundances cannot be produced in conventional CCSNe and are instead consistent only with the nucleosynthetic yields predicted for PISNe.

Here, we describe the line identification process,
inferring required elemental masses and temperatures to explain the observed luminosities. Key parameters and luminosity measurements are presented in Table \ref{tab:general_table}. 
To remain consistent with the SN nucleosynthesis literature, spectral epochs in this section are given relative to explosion, unless otherwise stated. 
We compare our results in detail to spectral models of PISNe in \citet{Jerkstrand_2016_PISN}. Specifically, 
to the 400d epochs of the \texttt{He80} and \texttt{He100} models, as their \el{56}{Ni} masses (0.13 and 5.8 \Msun{}, respectively) are closest to our estimated \el{56}{Ni} mass.
Still, we note that the \el{56}{Ni} mass of these models differs by at least a factor 4 with the one estimated for SN 2024afyu. 
Thus, we do not expect either model to provide a perfect match to the observed spectra, but rather use them as physical motivation to interpret the line identifications and to estimate the required elemental abundances.

\begin{table*}
\centering
\begin{tabular}{c|llll|l|llll}
Line & $\lambda_{\mathrm{centre}}$ & $E_{u}$ & $g_{u,\mathrm{eff}}$ & $A_{\mathrm{sum}}$ & $n_{e,\mathrm{crit,3500K}}$ & $L_{279d}$ &  $L_{312d}$ & $L_{479d}$ & $L_{554d}$   \\
\hline
& [Å/$\mu$m] & [eV] & & [s$^{-1}$] & [cm$^{-3}$] & [10$^{38}$ erg s$^{-1}$] & & &\\
\hline 
[S I] 4589 & 4589 & 2.750 & 1 & 2.98E-01 & 2.1 $\times$ 10$^{8}$ & -- & -- & --& --  \\ \xspace 
[S I] 7725 & 7725 & 2.750 & 1 & 1.38E+00 & 2.1 $\times$ 10$^{8}$& -- & -- & --& --  \\ \xspace 
[S I] 1.082 & 1.082 & 1.145 & 5 & 2.12E-02 & 4.0 $\times$ 10$^{6}$ & -- & -- & --& --  \\ \xspace 
[S I] 1.131 & 1.131 & 1.145 & 5 & 6.20E-03 & 4.0 $\times$ 10$^{6}$ & -- & -- & --& -- \\ 
\hline
[S II] 4069, 4076 & 4070  & 3.046 & 3.43 & 2.69E-01 & 1.3 $\times$ 10$^{6}$ & 92$\pm$9 & 95$\pm8$ & --& 3.3$\pm$1.8 \\ \xspace
[S II] 6716, 6731 & 6727 & 1.842 & 4.43 & 8.73E-04 & 9.5 $\times$ 10$^{3}$ &  12$\pm 2$ & 11$\pm2$ & --& $\geq$ 0.5 \\ \xspace 
[S II] NIR & 1.032 & 3.046 & 3.127 & 4.82E-01 & 1.3 $\times$ 10$^{6}$ & 65$\pm2$ & 66$\pm 2$ & -- & --  \\ 
\hline
[S III] 9069 & 9069 & 1.404 & 5 & 1.85E-02 & 3.3 $\times$ 10$^{5}$ & -- & -- & --& 0.87$\pm 0.08$ \\ \xspace 
[S III] 9531 & 9531 & 1.404 & 5 & 4.78E-02 & 3.3 $\times$ 10$^{5}$ & -- & -- & --& 2.1$\pm 0.2$ \\ 
\hline \hline
[Si I] 1.099 & 1.099 & 1.908 & 1 & 1.00E+00 & $^{*}$3.6 $\times$ 10$^{6}$ & -- & -- & -- & -- \\ \xspace
[Si I] 1.606 & 1.606 & 0.781 & 5 & 7.14E-04 & $^{*}$2.4 $\times$ 10$^{4}$ & 17.1 $\pm$ 0.2 & 22.0 $\pm$ 0.4 & 5.48 $\pm$ 0.03 & --\\ \xspace
[Si I] 1.645 & 1.645 & 0.781 & 5 & 2.01E-03 & $^{*}$2.4 $\times$ 10$^{4}$ & 18 $\pm$ 2 & 22 $\pm$ 4 & 4.9 $\pm$ 1.4 & --  \\
\hline 
\end{tabular}

\caption{Summary of prominent Si and S features. Columns 2, 3, 4 and 5 list the centroid wavelength, upper level energy, statistical weight and transition rate as obtained through NIST \citep{NIST_ASD}. For multiplets (e.g. \ion{S}{ii}), an effective $g_{u}$ (defined as $g_{u,\mathrm{eff}} \equiv \frac{\sum_{i}A_{i}g_{u,i}}{\sum_{i}A_{i}}$ ) and $A_{\mathrm{sum}} (= \sum_{i}A_{i})$ are given, so that these lines can be treated as a single line in equations like Equation \ref{eq:L_thin_LTE}. Column 6 notes the critical electron density $n_{e,\mathrm{crit}}$ for $T= 3500$ K, above which a transition may be considered to be in LTE, calculated following Barmentloo \& Jerkstrand (2026, submitted). For comparison, in the S- and Si-rich zones, $3 \times 10^{6}\lesssim n_{e} \lesssim 2\times 10^{8}$ cm $^{-3}$ for the \texttt{He80} and $10^{7}\lesssim n_{e} \lesssim 10^{8}$ cm $^{-3}$ for the \texttt{He100} models, and to first order $n_{e} \propto t^{-3}$. The remaining columns provide the measured line luminosities for each line in SN 2024afyu, at the given epoch post explosion. \\
$^{*}$: As no detailed collision strength calculations exist in the literature for Si I, these estimates were obtained using the same collision strengths as for S III (obtained from \citet{Grieve_2014_SIIIcollstrengths}), which has the same amount of electrons and thus a similar atomic structure. \\} 
\label{tab:general_table}
\end{table*}

\subsection{\SIIoptical and \SIINIR: The identification of Sulphur lines in SN 2024afyu}
\label{subsec:SII_mass}

The emission seen around 4070\AA\ in SN~2024afyu, that becomes much more prominent at nebular phases, is not typically seen in other SNe at similar phases. As is not the prominent emission line that shows up in the NIR spectra around 1.03 $\mu$m from $\sim$193 (281) days post peak (explosion)\footnote{Arguably already from 48.0 days post peak, albeit comparatively weaker.} (see Figs.~\ref{fig:comparisonALL} and \ref{fig:comparisonALLNIR}). We identify these two emission lines as \SIIoptical and [\ion{S}{ii}]\,$\lambda\lambda$1.029,1.032,1.034,1.037\,$\mu$m (\SIINIR from here on). 

Both \SIIoptical and \SIINIR originate from the same upper level doublet in \ion{S}{ii}, at 3.04 eV above the ground-state. This fact provides an excellent test for the identification of these lines; under LTE (see Table \ref{tab:general_table} and its caption) and optically thin conditions, the luminosity $L$ of an emission line is given by
\begin{equation}
L=M_{\text {ion }}\left(\mu m_p\right)^{-1} A h \nu \frac{g_u}{Z(T)} e^{-E_u / k T},
\label{eq:L_thin_LTE}
\end{equation}
where $M_{\text{ion}}$ is the mass of the emitting ion, $\mu$ is its mean atomic weight, $A$ is the Einstein coefficient for spontaneous emission (unit s$^{-1}$), $g_{u}$ is the statistical weight of the upper level, $E_u$ its energy, and $Z(T)$ the partition function of the ion, $Z(T)$ = $\sum_{i} g_{i} e^{-E_{i}/kT}$. For two optically thin lines from the same ion and upper level, their luminosity ratio is then simply given by
\begin{equation}
    \frac{L_1}{L_2} = \frac{A_1g_{u,1}\lambda_2}{A_2g_{u,2}\lambda_1}.
\label{eq:L_ratio}
\end{equation}
For the two S lines in question, this results in $L_{[\mathrm{S\,II}]\, \lambda\lambda\,4068,4076}$/$L_{\mathrm{[S\,II]\,NIR}}$ = 1.55. In case any of the lines are not completely optically thin, Equation \ref{eq:L_thin_LTE} is multiplied by the escape fraction $\beta \equiv \frac{1 - \exp(-\tau_{S})}{\tau_{S}}$, with $\tau$ the Sobolev optical depth, given by 
\begin{equation}
    \tau_S=\frac{1}{8 \pi} \frac{g_u}{g_l} A \lambda^3 n_l\left(1-\frac{g_l}{g_u} \frac{n_u}{n_l}\right) t.
    \label{eq:tau_sobolev}
\end{equation}
Here $t$ is the time since explosion in seconds, $g_{u}$ and $g_{l}$ are the statistical weights of the upper and lower levels of the transition, and $n_{l}$ is the number density of the lower level in cm$^{-3}$. The fact that the lower level of \SIINIR is an excited state while \SIIoptical is connected to the ground state means that we expect $\tau_{[\mathrm{S\,II}]\, \lambda\lambda\,4068,4076} > \tau_{\mathrm{[S\,II]\,NIR}}$ \citep{Jerkstrand_2017_handbook}, resulting in a lower ratio than 1.55. 

To compare the luminosities of the optical and NIR sulfur lines simultaneously, we combine pairs of absolute flux-calibrated ALFOSC and NIRES spectra obtained at closely matched epochs\footnote{At these late epochs, significant evolution of spectral lines is not expected during the few days separating the two spectra.}. The first pair, observed 193.7 and 192.8 days after peak, respectively, is hereafter referred to as the (ALFOSC+NIRES) 279 d (post-explosion) spectrum. The second pair, obtained 226.3 and 230.5 days after peak, is hereafter referred to as the (ALFOSC+NIRES) 312 d (post-explosion) spectrum.
We measure the ratio of the 279d and 312d spectra to be 1.41$_{-0.14}^{+0.16}$ and 1.44$_{-0.12}^{+0.13}$, respectively. This leads to an interpretation of these two lines as Sulphur lines, with $\tau_{[\mathrm{S\,II}]\, \lambda\lambda\,4068,4076} \sim 0.2$ (we check the consistency of this interpretation in Section \ref{subsec:SII_temperature}), assuming no contamination from other lines. As we will see in Sec.~\ref{subsec:SII_temperature} and \ref{subsec:SiI}, we find contamination from \SiIfouroneothree{} at later epochs \citep[another line predicted for PISNe, see e.g.][]{Dessart_2013_PISN, Jerkstrand_2016_PISN}.

\begin{figure*}
    \centering
    \includegraphics[scale=0.75]{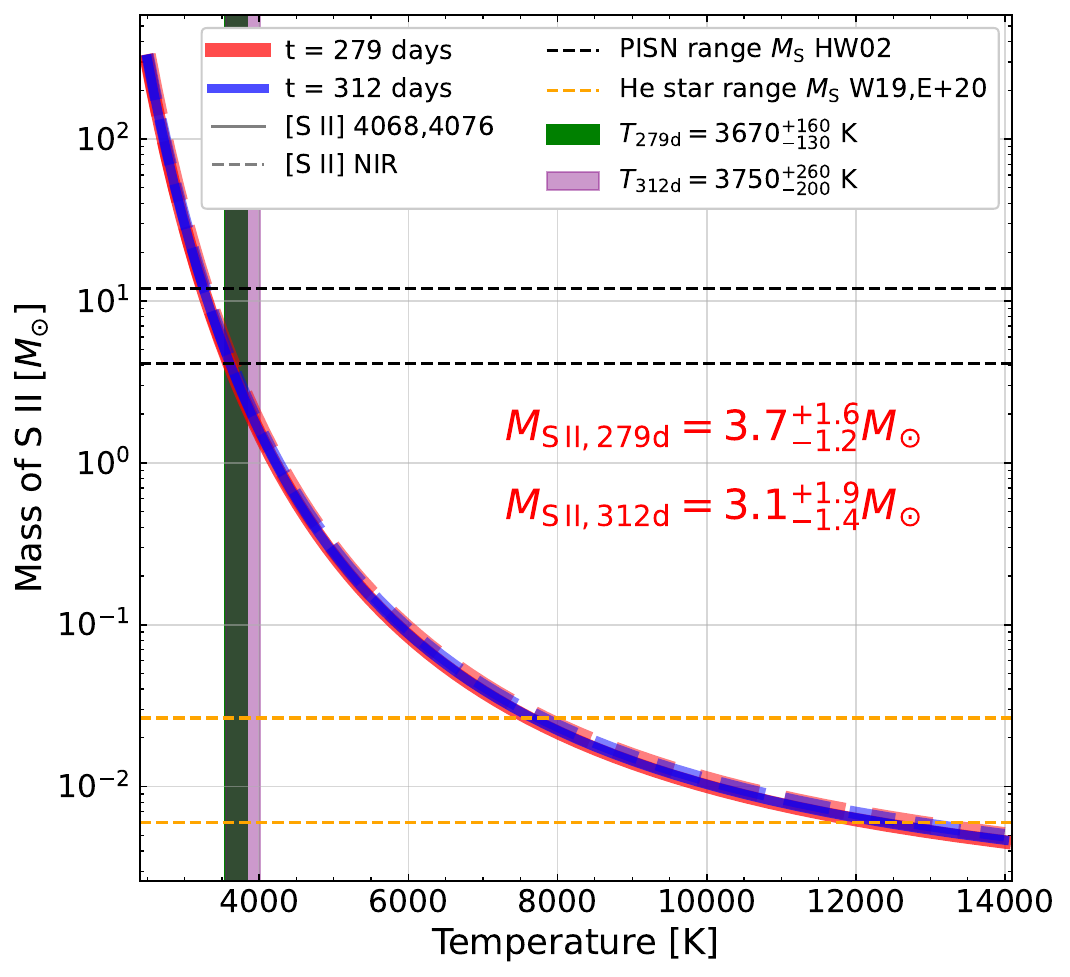}
    \caption{Allowed curves for $M_{\mathrm{S\,II}}$ and $T$ from luminosity estimates of \SIIoptical and \SIINIR at 279 and 312 days post explosion. Horizontal dashed lines indicate ranges for sulphur masses in the helium star grid of \citet[][W19]{Woosley_2019_Hestars} and \citet[][E+20]{Ertl_2020_Hestars}, and in the PISN grid of \citet[][HW02]{Heger_2002_PISNgrid}. The almost identical results for \SIIoptical and \SIINIR strengthen the interpretation of the $\sim$4070\AA and $\sim$1.03 $\mu$m emission in SN 2024afyu as sulphur lines. The figure shows that to explain such emission with helium star ejecta, unlikely high temperatures ($T > 7500$ K) are required, with more moderate values ($3200$ K $ \lesssim T \lesssim 4000 $K) in a PISN scenario. Indeed, our temperature estimates using \SIIoptical and \SIIsixseven (see Section \ref{subsec:SII_temperature}) are compatible with such a scenario.} 
    \label{fig:sulphur_curves}
\end{figure*}

With our identification strengthened, we further investigate these (otherwise atypical for nebular SNe) emission lines. 
Assuming LTE and optically thin conditions for the S lines, we use Equation \ref{eq:L_thin_LTE} to obtain permitted curves in $M_{\mathrm{S\,II}}$-T space from our line luminosity measurements (Table \ref{tab:general_table}) for the 279d and 312d spectra we show these curves in Fig.~\ref{fig:sulphur_curves}. In the figure, we also add ranges of sulphur masses for the grid of helium star progenitors (with initial helium core masses $M_{\mathrm{He,i}}$ between 3.3 and 12.0 \Msun) from \citet{Woosley_2019_Hestars} and \citet{Ertl_2020_Hestars} as well as for the grid of PISN progenitors (with initial helium core masses $M_{\mathrm{He,i}}$ between 70 and 130 \Msun) from \citet{Heger_2002_PISNgrid}. 

From Fig.~\ref{fig:sulphur_curves}, we see that to explain the luminosity of the S lines in SN 2024afyu from sulphur masses in typical SESN progenitors, a temperature of $T > 7500 $K is required. This is assuming that the sulphur lines are optically thin and that all sulphur is in the \ion{S}{ii} state ($x_{\mathrm{S\,II}} = 1$), with an even higher temperature required for $x_{\mathrm{S\,II}} < 1$ or non-negligible optical depths. For the \sumocodename models in Barmentloo \& Jerkstrand (2026, submitted)\footnote{Considering their $\chi = 1$ set for he3p3, he4p0 and he6p0 and $\chi = 3$ for he8p0.} using the aforementioned helium star progenitors as input models, the ionisation fraction spans a broad range, $x_{\mathrm{S\,II}} = 0.02 - 0.87$ at 300 days post explosion. For these same models, temperatures in the zone hosting the majority of the sulphur ejecta span a narrow range of only 2600--3000 K. The \texttt{CMFGEN} models in \citet{Dessart_2023_Hestars} have slightly higher temperatures of 3500--4500~K, but these are still much lower than the $T > 7500 $K required to explain the observed sulphur emission in SN 2024afyu. On the other hand, for sulphur masses in the PISN models in \citet{Heger_2002_PISNgrid}, temperatures of $T \sim$3200--4000 K (again, higher for $\tau > 0$ or $x_{\mathrm{S\,II}} < 1$) would suffice to explain the observed luminosities. These temperatures are compatible with what is found for \sumocodename \citep{Jerkstrand_2016_PISN} and \texttt{CMFGEN} \citep{Dessart_2013_PISN} NLTE models of the \citet{Heger_2002_PISNgrid} progenitors. Furthermore, they are consistent with the temperature estimates of $T =3670^{+160}_{-130}$ K and $T =3750^{+260}_{-200}$ K we obtain from the luminosity ratio of \SIINIR with \SIIsixseven in Sec.~\ref{subsec:SII_temperature}. 

In short, we have identified emission lines from \ion{S}{ii} at luminosities that require \ion{S}{ii}-masses on the order of a few solar masses, which are solely compatible with a PISN scenario. For such large \ion{S}{ii} masses, one should expect SN 2024afyu to show spectral signatures from other sulphur ions as well, providing further tests of a PISN scenario. We go through identifications of \ion{S}{i} and \ion{S}{iii} lines in Sec.~\ref{subsec:SI_highlying} to \ref{subsec:SIII}, but first we showcase how we obtained our estimates of the temperature in the S-rich zones for the 279d and 312d spectra.

\subsection{\SIIoptical, \SIINIR and \SIIsixseven: Temperature of S-rich Zones}
\label{subsec:SII_temperature}

While initially going unnoticed, we can see that the last and latest (554d) LRIS spectrum shows a clear broad emission component at the location of \SIIsixseven, which in this spectrum also has a narrow component from the host galaxy. 
After this clear detection, we re-examine the earlier spectra and see that  
\SIIsixseven is indeed present already at 232 days post explosion, and can be confused with the redder wing of the broad emission component identified as H$\alpha$ in the original Type~IIb classification (see Fig.~\ref{fig:obsspec}). 
This line originates from a different upper level than \SIIoptical and \SIINIR, the line ratio with either of these two provides the local temperature; assuming once more optically thin LTE, from Equation \ref{eq:L_thin_LTE} one can show that for lines from the same ion:
\begin{equation}
    T = \ln \left( \frac{A_{1}\nu_{1}g_{u,1}}{A_{2}\nu_{2}g_{u,2}} \right)/\ln \left( \frac{L_{1}}{L_{2}} \right) \times \frac{(E_{u,2} - E_{u,1})}{k}.
\label{eq:T_estimate}
\end{equation}
With the subscripts 1 and 2 indicating two distinct emission lines. As we found in Sec.~\ref{subsec:SII_mass} that \SIIoptical is likely not completely optically thin, we use \SIINIR for the temperature measurement. Performing this calculation for our 279d and 312d spectra, we obtain $T_{\mathrm{S,279d}} = 3670^{+160}_{-130}$ K and $T_{\mathrm{S,312d}} = 3740^{+270}_{-180}$ K. Both consistent with PISN model predictions from \citet{Dessart_2013_PISN} and \citet{Jerkstrand_2016_PISN}.  These result in $M_{\mathrm{S \,II,279d}} = 3.7^{+1.6}_{-1.2}$ \Msun{} and $M_{\mathrm{S \,II,312d}} = 3.1^{+1.9}_{-1.4}$ \Msun{} (see Fig.~\ref{fig:sulphur_curves}). This may be compared to the range of S masses in the PISN models of \citet{Heger_2002_PISNgrid} of 4.1\footnote{We exclude the \texttt{He65} and \texttt{He70} models when comparing, as they produced only $\leq$ 10$^{-2}$ \Msun{} of \el{56}{Ni}.} -- 12 \Msun{}, so that in any case a non-negligible fraction of the sulphur at this epoch is singly ionised. In the \citet{Jerkstrand_2016_PISN} \texttt{He80} and \texttt{He100} models at 400d post explosion, $M_{\mathrm{S\,II}} = 2.1 \times 10^{-2}$ \Msun{} and $M_{\mathrm{S\,II}} = 1.5$ \Msun{} respectively. When compared to their total sulphur masses of 4.7 and 10 \Msun{}, it indicates the sensitivity of the ionisation of S for different PISN models. We note that the free electron fraction in these models drops continuously with time, so that we would expect $x_{\mathrm{S\,II}}$ to have been higher for these models at 279 and 312 days than at the evaluated 400 days epoch. 

With the \ion{S}{II} mass estimates, we can now check our earlier interpretation of $\tau_{[\mathrm{S\,II}]\, \lambda\lambda\,4068,4076} \sim 0.2$ in Section \ref{subsec:SII_mass}. Using the values at 279d in Equation \ref{eq:tau_sobolev} (see also Table \ref{tab:general_table}), and assuming LTE populations as
\begin{equation}
    n_{i} = \frac{M_{\mathrm{ion}}}{\mu m_{p}} \times \frac{1}{V_{\mathrm{emit}}} \times \exp(-E_{i}/kT), 
\end{equation}
With $V_{\mathrm{emit}} \equiv \frac{4\pi}{3}\times (v_{exp}t)^{3} \times f_{S}$ the emitting volume of \ion{S}{II}, $v_{exp}$ the expansion velocity and $f_{S}$ the filling factor, we obtain:
\begin{equation}
    \tau_{[\mathrm{S\,II}]\, \lambda\lambda\,4068,4076} = 0.55 \times \frac{M_{\mathrm{S\,II}}}{3.7 M_{\odot}} \times \left(\frac{v_{exp}}{4000\, \mathrm{km/s}}\right)^{-3} \times \left(\frac{t}{279d}\right)^{-2} \times f_{S}^{-1}
\label{eq:tau_SII}
\end{equation}
\indent For assumed values of $f_{S} \sim1$ and $v_{exp} =4000 $ km/s, the theoretical $\tau$ values of 0.55 and 0.37 are then consistent with our considered observations\footnote{For e.g. $\tau$ = 0.55, we get $\beta = 0.77$, resulting in an expected ratio of 1.19, consistent within $\sim$1.5$\sigma$ of our observed 1.41$^{+0.16}_{-0.14}$.}. Still, obtaining accurate estimates for $v_{exp}$ (via the half width zero maximum of \SIIoptical{}) is non-trivial, as defining the 'edge' of the spectral line to obtain a width is made difficult by severe line blending. For a parabolic line profile fit to \SIIoptical \citep[representative of emission from a uniform sphere;][]{Jerkstrand_2017_handbook}, we obtain $v_{exp} =3600 \pm 150$ km/s. This is likely an underestimate, as real spectral lines typically have broader wings than a parabolic profile. Gaussian fits do indeed result in larger velocities, but putting an exact number on these is hindered by the fact that a Gaussian never reaches zero intensity. In any case, we can say that the line width and luminosity of \SIIoptical are not inconsistent with optically thin conditions.

\begin{figure}
    \centering
    \includegraphics[width=\linewidth]{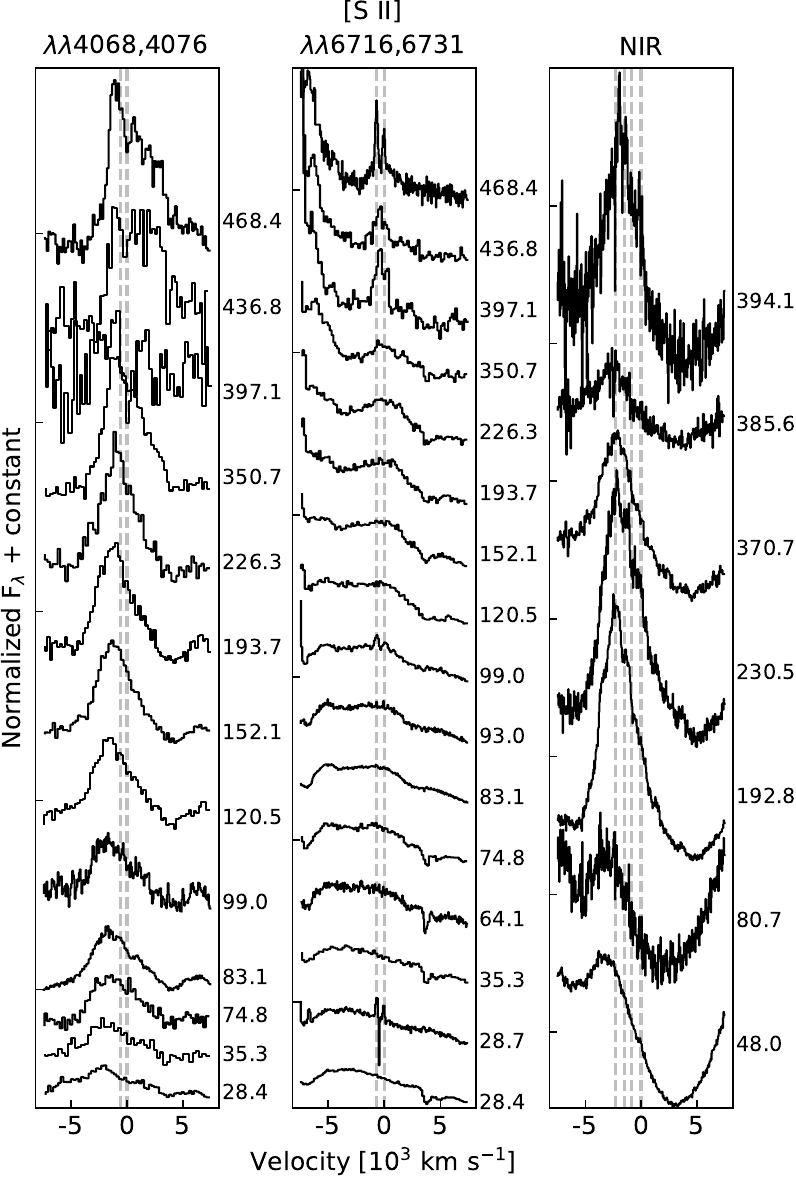}
    \caption{Evolution of the [\ion{S}{ii}] lines identified in SN 2024afyu. Zero velocity coincides with the velocity of the bluer multiplet component. Vertical dashed lines indicate the centroid wavelengths. The epochs are given with respect to peak; for explosion epochs, one should add 85d.}
    \label{fig:SII_evolution}
\end{figure}

At our last LRIS epoch, two complications arise to the methods used above. First, we have no NIR data at this epoch, so we can only rely on the luminosity ratio of \SIIoptical with \SIIsixseven to obtain a temperature estimate. Second, we see in Fig. \ref{fig:SII_evolution} that the line profile of \SIIoptical at this epoch is shifted towards the red, with the blue wing diminishing and the red one broadening. We do not observe the same behaviour in \SIIsixseven, and interpret this as follows: with time, the temperatures in the (mostly unmixed) PISN ejecta drop steadily, constraining the line forming region for the relatively high upper level energy line \SIIoptical to more inward regions where the temperature is higher, reducing its line width\footnote{The same happens to \SIIsixseven but to a lesser extent, due to its lower upper level energy (see Table \ref{tab:general_table}).}. At the same time, the line strength of \SiIfouroneothree{} (whose upper level is likely fed by recombinations from \ion{Si}{ii}) increases compared to \SIIoptical, leading to what looks like a rising red wing in the profile. Obtaining the \SIIoptical luminosity at this epoch is thus more complex. 

Performing then a double component fit to the \SIIoptical$+$ \SiIfouroneothree{} profile, forcing the centroids to their theoretical values and the width of \SIIoptical to equal that of \SIIsixseven, we obtain $L_{[\mathrm{S\,II}]\, \lambda\lambda\,4068,4076} = 3.3 \pm 1.8 \times 10^{38}$ erg s$^{-1}$. Due to potential contamination, we obtain a lower limit $L_{[\mathrm{S\,II}]\, \lambda\lambda\,6716,6731} = 0.5 \times 10^{38}$ erg s$^{-1}$. Combining the two using Equation \ref{eq:T_estimate}, we obtain an upper limit to the temperature in the S-rich zones of $T \lesssim $ 3390 K. For this temperature, we obtain a lower limit of $M_{\mathrm{S \,II}} \geq 0.27$\Msun{} using \SIIoptical and Equation \ref{eq:L_thin_LTE}. It seems then that the \ion{S}{II} mass has decreased from the previous epochs, which would be consistent with decreasing ionisation in the ejecta.

\subsection{\SIfourfiveeightnine{} and \SIsevenseventwofive{}}
\label{subsec:SI_highlying}

The latest LRIS spectrum of SN~2024afyu also shows \SIfourfiveeightnine{} and \SIsevenseventwofive{}. 
While their high \necrit{} of $2.1 \times 10^{8}$ cm$^{-3}$ does not allow \SIfourfiveeightnine{} and \SIsevenseventwofive{} to be treated as LTE lines, the fact that both of these lines arise from the same upper level means that their luminosity ratio is simply given by Equation \ref{eq:L_ratio}. This means that we expect $L_{[\mathrm{S\,I}]\, \lambda7725}$/$L_{\mathrm{[S\,I]\ \lambda4589}}$  = 2.75 in the optically thin case, which is likely the case at these late epochs. Measuring the luminosities in SN~2024afyu is complicated by potential blending with \ion{Mg}{i]}~$\lambda$4571 and clear blending with \ion{O}{i}~$\lambda$7774. The best we can say is that the observed line strengths in SN~2024afyu (assuming \SIfourfiveeightnine{} to dominate over \ion{Mg}{i]}$\lambda$4571) are not inconsistent with this ratio.

\subsection{\SIoneoeighttwo and \SIoneonethreeone}
\label{subsec:SI_lowlying}

\SIoneoeighttwo{} and \SIoneonethreeone{} are lines from the first excited state to the ground multiplet in \ion{S}{i}.
In all nebular NIRES spectra, lines are present around the centroids of these wavelengths (see Fig.~\ref{fig:speclines}). However, contamination is likely; for the feature around 1.09 $\mu$m, contributions in SESNe and PISNe may be expected from \ion{C}{i}$\lambda$1.07, \ion{He}{i}$\lambda$1.08 and \SiIoneoninenine{} \citep{Jerkstrand_2016_PISN, vanBaal_2024_grid}. Around 1.13$\mu$m, there instead exist multiple recombination lines of \ion{O}{i}, which may be expected to have luminosities on the order of \ion{O}{i}$\lambda$7774 \citep{2010MNRAS.408...87M, 2015A&A...573A..12J}, which is clearly present in SN~2024afyu. Indeed, fits to the line complex around 1.08$\mu$m require at least three gaussians for a satisfactory fit, and considering that some of the contaminating lines may be optically thick, we refrain from luminosity measurements for these epochs.

The line profile around 1.08--1.09 $\mu$m changes with time, such that it appears to have a separate red wing in the NIRES spectrum observed at 479 days post explosion. We interpret this as the presence of \SiIoneoninenine{} (see Sec.~\ref{subsec:SiI}).

\subsection{\SIIInineosixnine and \SIIIninefivethreeone}
\label{subsec:SIII}

\begin{figure}
    \centering
    \includegraphics[width=\linewidth]{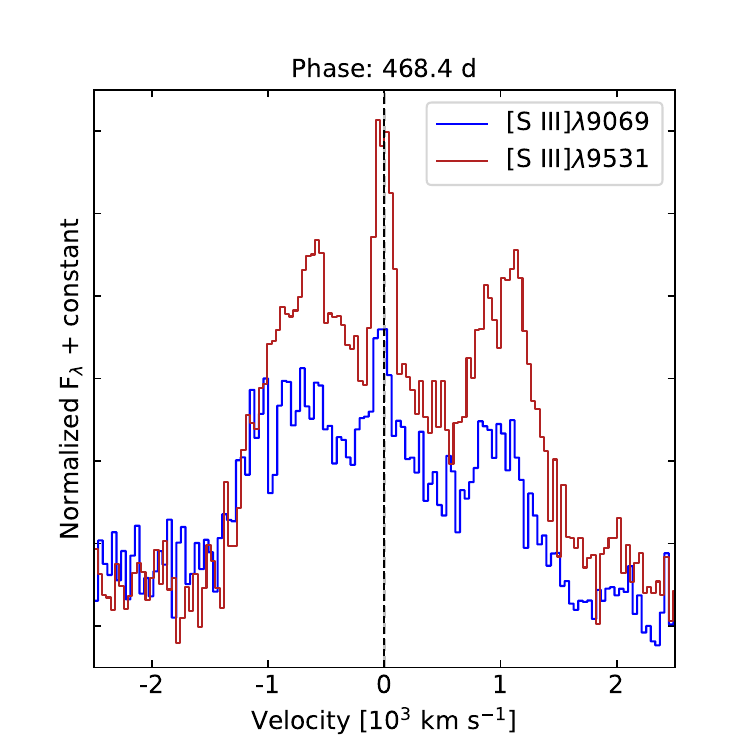}
    \caption{Features around 9070 and 9530 $\AA$ in the 468(554)d LRIS spectrum of SN 2024afyu. The similarity of the line profiles strengthens the hypothesis that both of these lines originate from \ion{S}{III}. The luminosities of these lines are consistent with $M_{\mathrm{S\,III}} = 1.0 \times 10^{-2}$ \Msun{} (\ref{subsec:SIII}).}
    \label{fig:SIII}
\end{figure}

While PISN spectra are predicted to consist mostly of neutral and singly ionised species \citep{Dessart_2013_PISN, Jerkstrand_2016_PISN}, we identify two emission lines with centroids consistent with the doubly ionised sulphur lines \SIIInineosixnine and \SIIIninefivethreeone (see Figure \ref{fig:speclines}) in the last LRIS spectrum. As these lines originate from the same upper level, we may again use equation \ref{eq:L_ratio} (assuming optical thinness) to obtain an expected luminosity ratio of $L_{[\mathrm{S\,III}]\, \lambda9531}$/$L_{\mathrm{[S\,III]\,\lambda9069}}$ = 2.46. Under this hypothesis, we consider the LRIS spectrum and measure a value of 2.45$^{+0.37}_{-0.33}$, thus perfectly consistent with the prediction. A further hint that these lines originate from the same zone is the similarity of their line profile shapes, as shown in Fig. \ref{fig:SIII}). Using the upper limit on the temperature of the S-rich zones 
(see Sec.~\ref{subsec:SII_temperature}), we obtain a lower limit $M_{\mathrm{S\, III}} \geq 1.0\times 10^{-2}$ \Msun{}. Considering the S masses presented by \citet{Heger_2002_PISNgrid} for their PISN models, this would correspond to an ionisation fraction $x_{\mathrm{S\,III}} \gtrsim$ 0.8--2.4 $\times 10^{-3}$. This may be compared to $M_{\mathrm{S\, III}} = 2\times 10^{-5}$ and $2\times 10^{-2}$ \Msun{} for the 400d \texttt{He80} and \texttt{He100} models respectively. \ion{S}{iii} mass compatible with the \texttt{He100} model, despite roughly a factor 10 less \el{56}{Ni} in SN~2024afyu, may hint at efficient \el{56}{Ni} mixing, different than typically found in hydrodynamical simulations of PISNe \citep[e.g.][]{Joggerst_2011_PISNmixing, Chatzopoulos_2013_PISNmixing, Chen_2014_PISNmixing, 2014ApJ...797....9W}. 

\subsection{\ion{Si}{i} lines}
\label{subsec:SiI}

Besides sulphur, the PISN models from \citet{Heger_2002_PISNgrid} also have some 10--20 \Msun{} of silicon in their ejecta, being mostly co-located with sulphur. Therefore, we would also expect to see Si signatures in spectra of PISNe. Due to unfavorable atomic structure for other ions, only lines of neutral Si are expected \citep[e.g.][]{Dessart_2013_PISN, Jerkstrand_2016_PISN}).

\SiIonesixosix and \SiIonesixfourfive are transitions from the first excited state to the ground state, with sufficiently low $n_{e,\mathrm{crit}}$ to be securely in LTE. There are features present in all our NIRES epochs around these centroid wavelengths, but contamination seems likely. 
If the lines are in the optically thin, LTE regime, the theoretical ratio $L_{\mathrm{[Si \,I]\,\lambda 1.645}}/L_{\mathrm{[Si \,I]\,\lambda 1.607}} = 2.75$ from Equation \ref{eq:L_ratio}, inconsistent with our luminosity estimates if we assume no contamination. If contamination is present, we can obtain an upper limit mass estimate from $L_{\mathrm{[Si \,I]\,\lambda 1.645}}$ using Equation \ref{eq:L_thin_LTE}. As Si and S are present in the same compositional zones, we can assume $T_{\mathrm{S}} \sim T_{\mathrm{Si}}$. For our luminosity estimate of \SiIonesixfourfive at 279d and $T = 3670$ K, we obtain $M_{\mathrm{Si\, I}} \lesssim 0.37$ \Msun{}. 

If the lines are instead in the optically thick, LTE regime, their luminosity is given by \citep[equation 45 in][]{Jerkstrand_2017_handbook}
\begin{equation}
L = 4 \pi V(c t)^{-1} \lambda \frac{2 h c^2}{\lambda^5} \frac{1}{e^{hc / \lambda k T}-1},
\label{eq:L_thick_LTE}
\end{equation}
with $V$ the emitting volume. From this formula, one would expect a luminosity ratio of \SiIonesixosix over \SiIonesixfourfive of $\sim$1, as their wavelengths are roughly equal, compatible with our observations (see Table \ref{tab:general_table}). For our luminosity estimate of \SiIonesixosix at 279d and $T = 3670$ K, we obtain an emitting volume for \ion{Si}{I} of $5.8_{-0.7}^{+0.5} \times 10^{46}$ cm$^{-3}$. Requiring that the sobolev optical depth (see Equation \ref{eq:tau_sobolev}) $\tau_{\mathrm{[Si \, I]\,\lambda 1.607}} \geq 3$ (i.e. when the optically thick regime is valid), we find a required $M_{\mathrm{Si\,I}} \gtrsim 3$ \Msun{}. 

Both obtained constraints have their difficulties. For the optically thin case, the $M_{\mathrm{Si\,I}}$ estimate would correspond to a low ionisation fraction of $x_{\mathrm{Si\,I}} \sim$0.014--0.03 for the PISN grid of \citet{Heger_2002_PISNgrid}. This can be compared to $x_{\mathrm{Si\,I}}$ = 0.96 and 0.44 for the \texttt{He80} and \texttt{He100} models at 400 days. Although, at such an earlier epoch, radioactive decays are more plentiful, likely ionising more of the \ion{Si}{I} into \ion{Si}{II}. Additionally, as also discussed in Section \ref{subsec:SIII}, larger mixing than assumed in models could lead to more efficient ionisation. For the optically thick case, the obtained emitting volume would correspond to a filling factor of only 0.015 for a sphere of $v_{\exp} = 4000$ km/s (see Section \ref{subsec:SII_temperature}). At such a low filling factor, the earlier discussed \SIIoptical would have an optical depth of $\tau_{S} \sim 36$ (assuming $f_{\mathrm{S}} \sim f_{\mathrm{Si}}$), which would result in a line ratio with \SIINIR inconsistent with our observations. In our opinion, an optically thin case is more likely, due to the firm identification of the sulphur lines presented in this work.

Another \ion{Si}{I} line from a low-lying level is \SiIoneoninenine. As mentioned before, we see progressively stronger signatures of this line in our nebular NIR spectra. Although always strongly blended with a neighboring complex (likely \SIoneoeighttwo), so that we refrain from attempting to extract its luminosity. 
We identify two more potential \ion{Si}{i} allowed transitions in our 554d LRIS spectrum. \SiIfouroneothree{} is found as a blend with \SIIoptical (see Sec.~\ref{subsec:SII_temperature}). A collection of \ion{Si}{i} transitions with $A \sim 10^{7}$ s$^{-1}$ clustered around 1.20$\mu$m (also identified in all nebular NIR spectra). Both of these lines are present at appreciable strength in the PISN models in \citet{Jerkstrand_2016_PISN}, but obtaining mass or temperature constraints from them is non-trivial as their high $A$-values mean these lines are most likely in the NLTE regime.

\section{Discussion}
\label{sec:discussion}

SN~2024afyu is a long-lived event whose photometric and spectroscopic properties differ substantially from those of normal Type II and IIb supernovae, the classifications initially assigned to the event. The light curve and spectra of SN~2024afyu is also different from other events in these classes (Figs.~\ref{fig:comparisonALL}, \ref{fig:comparisonALLNIR}, \ref{fig:compLC}), with expansion velocities inferred at the wavelengths corresponding to some absorption features being comparable to those measured in SESNe, but pseudo-equivalent widths not following the trends observed for any of the canonical SESN subclasses (Fig.~\ref{fig:pEWandVEL}). Furthermore, the spectra exhibit prominent sulfur emission, motivating us to investigate a pair-instability supernova (PISN) interpretation. Under this assumption, we identify additional sulfur and silicon spectral features that are naturally explained within a PISN framework (Sec.~\ref{sec:sulfurANDSiliconID}).

\subsection{SN~2024afyu as a PISN}
\label{sec:PISNinterpretation}

In this section, we compare the observational properties of SN~2024afyu with other PISN candidates and with publicly available PISN models.

\subsubsection{Other PISN candidates}

Several SNe have been proposed as PISN candidates, although many have subsequently been reinterpreted as being powered by alternative mechanisms. A notable example is OGLE-SN-2014-073 (OGLE14-073), which was initially suggested to originate from a PISN explosion by \cite{2017NatAs...1..713T}, with further support from \cite{2018MNRAS.479.3106K}. However, later studies argued instead for a magnetar-powered explosion \citep{2018A&A...613A...5D,2018A&A...619A.145O} or a fallback accretion scenario \citep{2018MNRAS.475L..11M}. We highlight this event because it illustrates how alternative powering mechanisms can reproduce the broad light curves, high luminosities, and large inferred $^{56}$Ni masses often considered indicative of a PISN.

A comprehensive review of proposed PISN candidates is provided by \citet{2024A&A...683A.223S}, who identify SN~2018ibb as the strongest PISN candidate to date. We compare the light curve and late time optical spectra of SN~2024afyu with the sample presented by \citet{2024A&A...683A.223S}, which consists predominantly of hydrogen-poor events (including SN~2007bi, \citealt{2009Natur.462..624G,2010A&A...512A..70Y,2017ApJ...835...13J}; PS1-11ap, \citealt{2014MNRAS.437..656M}; PTF12dam, \citealt{2013Natur.502..346N,2015MNRAS.452.1567C,2018ApJ...855....2Q}; PS1-14bj, \citealt{2016ApJ...831..144L}; LSQ14an, \citealt{2017MNRAS.468.4642I,2017ApJ...835...13J}; and SN~2015bn, \citealt{2016ApJ...826...39N,2016ApJ...828L..18N,2018ApJ...866L..24N}). We additionally include OGLE14-073, SN~2018lzi \citep[proposed as a PISN by][the light curve was obtained from \citealt{2025A&A...695A.142P}]{2022RNAAS...6..122P}, and SN~2023vbw \citep[proposed as a PISN by][the light curve was obtained from Skoglund et al. in prep.]{2026arXiv260516487H}, as they show hydrogen in their spectra, as does SN~2024afyu (see Appendix~\ref{app:synapps}). Although spectra are not publicly available for the latter two objects, we include them in the photometric comparison because their PISN interpretations were based primarily on their light curve properties. 

\begin{figure}
   \includegraphics[width=\linewidth]{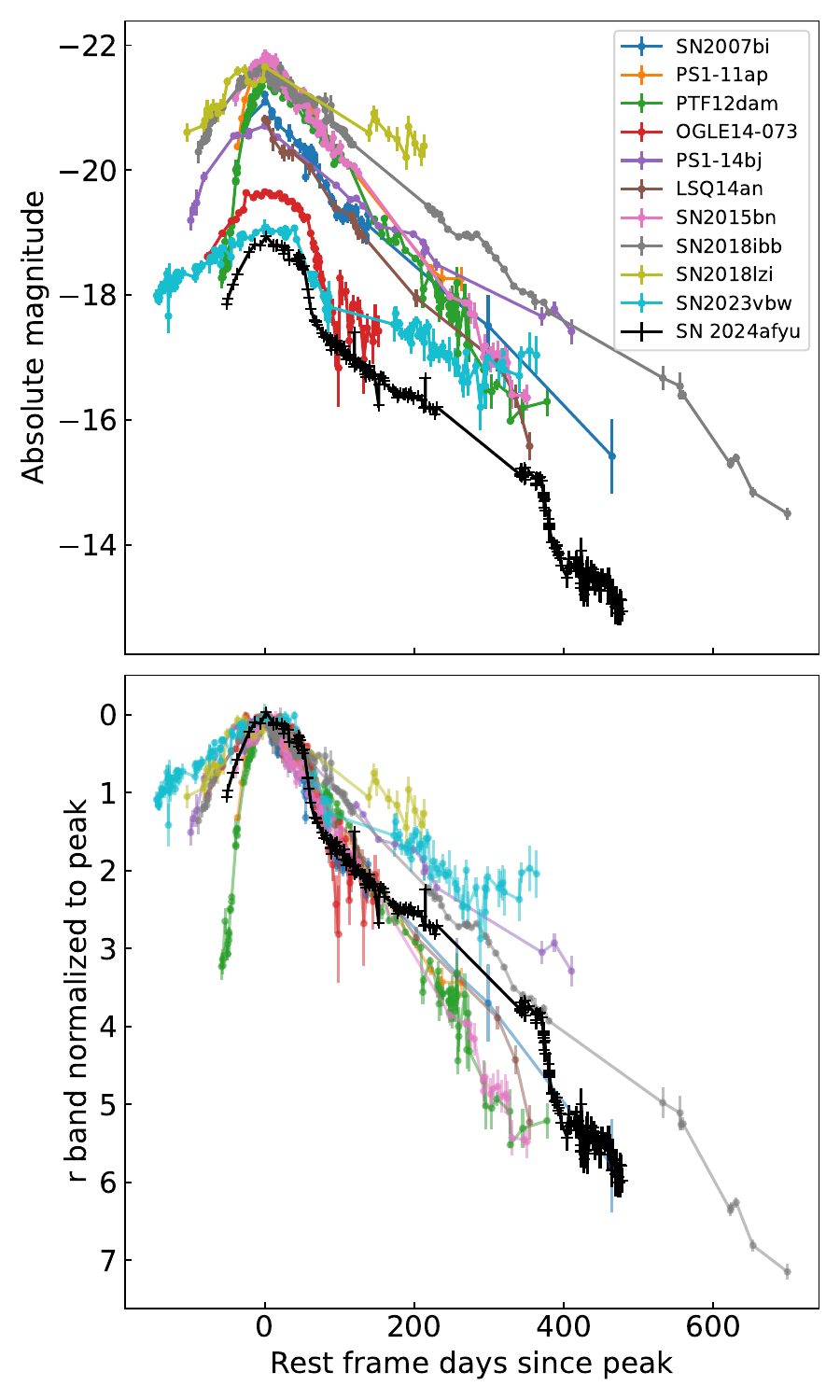}
      \caption{Light curves of PISN candidates compared to SN~2024afyu (black). Top: $r$ band absolute magnitudes, except for OGLE14-073 which is shown in the publicly available $i$ band. Bottom: Light curves normalized to their peak absolute magnitudes. If the peak is not observed, the first available photometric point is considered to be the peak.}
         \label{fig:comparisonPISN_LC}
   \end{figure}

\begin{figure}
   \includegraphics[width=\linewidth]{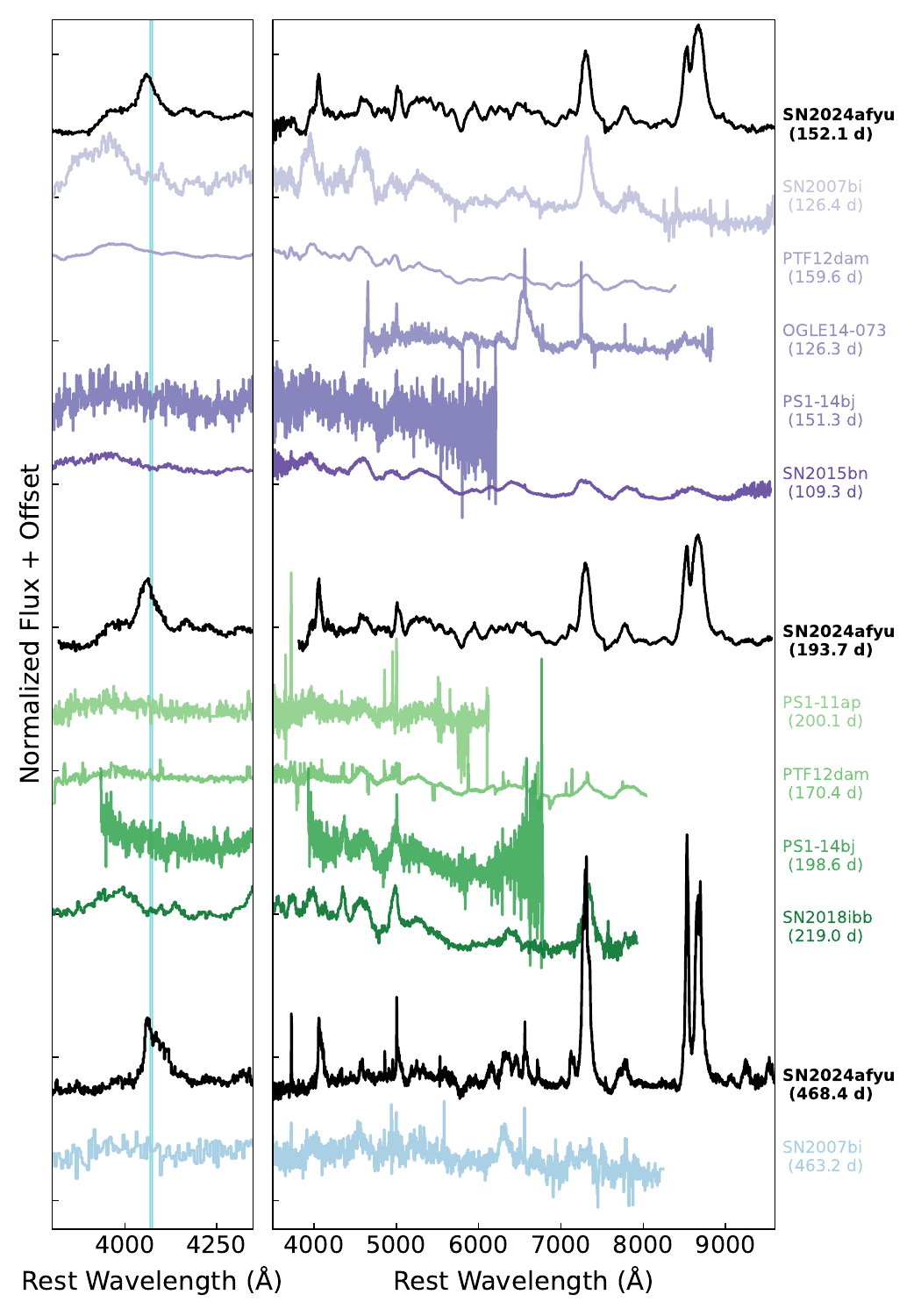}
      \caption{Optical spectral comparison of SN~2024afyu (black) with PISN candidates, colour-coded by spectral phase. Left panel: zoom into the $\sim$4070~\AA\ region, with a cyan band marking the rest wavelengths of \SIIoptical. Right panel: full optical wavelength range.}
         \label{fig:comparisonPISN}
   \end{figure}

Figure~\ref{fig:comparisonPISN_LC} shows the light curve comparison. SN~2024afyu is the faintest object in the sample. Despite this, the bottom panel of the figure shows that its overall light curve evolution is broadly similar to those of the other candidates. 
Figure~\ref{fig:comparisonPISN} shows the late time spectral comparison. Because spectra are not available at identical epochs for all objects, comparison spectra obtained within approximately 50 days of the corresponding SN~2024afyu epoch are grouped together in the same phase bin. The comparison reveals a remarkable spectroscopic diversity among different events. In particular, most of the comparison spectra show bluer continua at similar epochs and none of them exhibit the prominent emission feature around 4070~\AA\ that we identify with [\ion{S}{ii}] in SN~2024afyu. The spectrum most closely resembling SN~2024afyu is the one of SN~2007bi. However, SN~2007bi lacks the prominent \ion{Ca}{ii} NIR triplet seen in SN~2024afyu, and its feature near 4070~\AA\ is more naturally interpreted as Ca H\&K than as [\ion{S}{ii}], as it is centered at a significantly bluer wavelength. Thus, SN~2024afyu appears to be spectroscopically unique among PISN candidates.

Furthermore, Figure~\ref{fig:host:mass_vs_sfr} shows the positions of SN~2024afyu, ordinary Type~II and IIb SNe, and PISN candidates in the $M_\star$--SFR plane. The host of SN~2024afyu lies on the star-forming main sequence, within its intrinsic scatter, and within the 66 per cent contour of the Type~II/IIb host population, making it unremarkable among CCSN hosts. Among PISN candidates, it is the third most massive host, with a low SFRs relative to the main sequence. Nevertheless, its properties are consistent with the low-redshift PISN predictions of \citet{Briel2024a}, who find median values of $\log(M_\star/M_\odot) = 8.86^{+0.73}_{-0.80}$ and ${\rm SFR} = 0.15^{+0.12}_{-0.12}\,M_\odot\,{\rm yr^{-1}}$ at $z \simeq 0$.
The metallicity provides a complementary test of this picture.
We infer a gas-phase metallicity of $0.55\pm0.02$ solar (see Appendix~\ref{app:host}), which is typical for a galaxy of the inferred stellar mass \citep{Andrews2013a}, and it is also fully consistent with the the metallicity range predicted for low-redshift PISN hosts \citep{Briel2024a}. 

\subsubsection{PISN models}

\begin{figure*}
   \centering
   \includegraphics[scale=0.75]{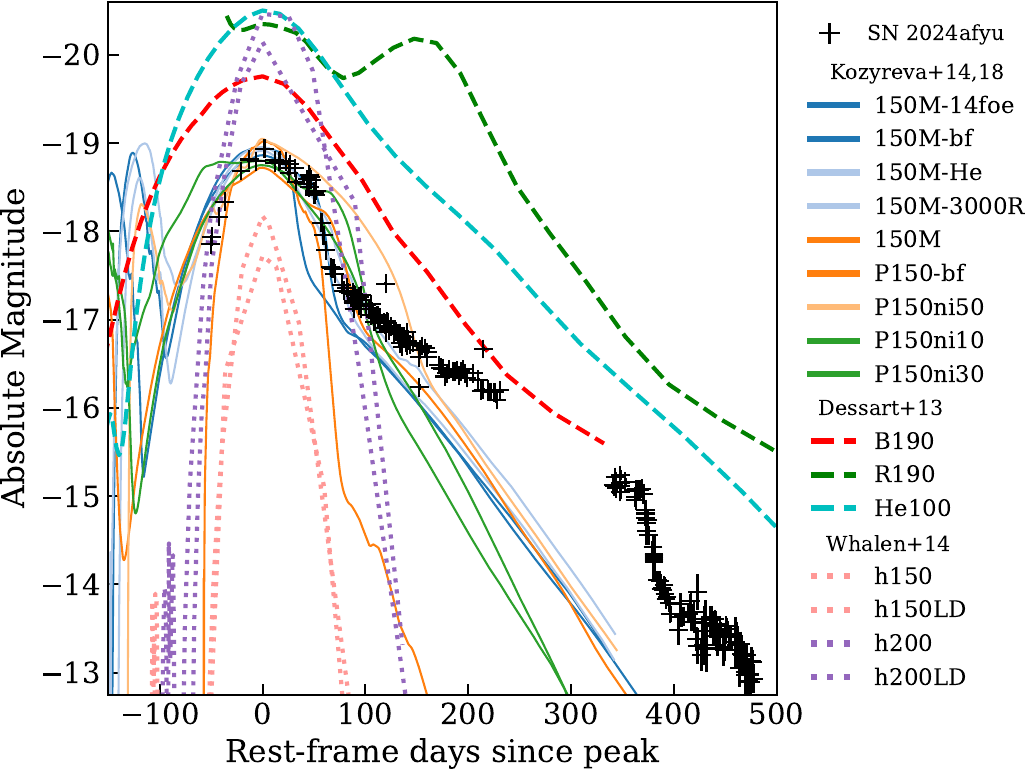}
      \caption{PISN $r$/$R$ light curve models compared to SN~2024afyu (black). Solid lines correspond to models by \cite{2014A&A...565A..70K,2018MNRAS.479.3106K}, dashed lines to models by \cite{Dessart_2013_PISN} and dotted lines to models by \cite{2014ApJ...797....9W}.}
         \label{fig:comparisonPISNmodels_LC}
   \end{figure*}

Figure~\ref{fig:comparisonPISNmodels_LC} compares the $r$-band light curve of SN~2024afyu with available PISN light curve models. The models of \cite{Dessart_2013_PISN}\footnote{\url{https://doi.org/10.5281/zenodo.5524918}.} explore three different progenitors: 190~\Msun{} red (R190) and blue (B190) supergiants, together with a bare helium core of 100~\Msun{} (He100), representing progenitors with progressively larger mass loss. None of these models provides a good match to SN~2024afyu. All predict broader and more luminous light curves that decline more rapidly than observed, while the R190 model additionally exhibits a secondary maximum that is absent in SN~2024afyu.
The models of \cite{2014ApJ...797....9W} specifically investigate low-redshift PISNe originating from relatively metal-rich\footnote{The metallicities considered are sub-solar but larger than one would expect for a Pop III star, typically consider as the progenitor of PISN.} progenitors that lose most of their hydrogen envelopes through line-driven winds before exploding as bare helium cores. These models produce light curves with peak widths comparable to that of SN~2024afyu and similarly exhibit an after peak ``flatness''. However, neither of the considered progenitor masses reproduces the observed peak luminosity: the 150~\Msun{} model is systematically too faint, whereas the 200~\Msun{} model is too bright. Moreover, both models decline significantly faster than SN~2024afyu after maximum.
The models of \cite{2014A&A...565A..70K,2018MNRAS.479.3106K}\footnote{\url{https://wwwmpa.mpa-garching.mpg.de/ccsnarchive/data/Kozyreva/PISN/}.}, which represent low-redshift PISN scenarios for progenitors with initial masses between 150 and 250~\Msun{}, provide the best overall agreement with the observations. In particular, the 150~\Msun{} models reproduce the observed peak luminosity without applying any magnitude offset, the peak width, and the epoch at which the post-maximum change in decline slope. Nevertheless, from approximately 100 days onwards, SN~2024afyu fades slower than any of the published models.

\begin{figure*}
   \centering
   \includegraphics[scale=0.5]{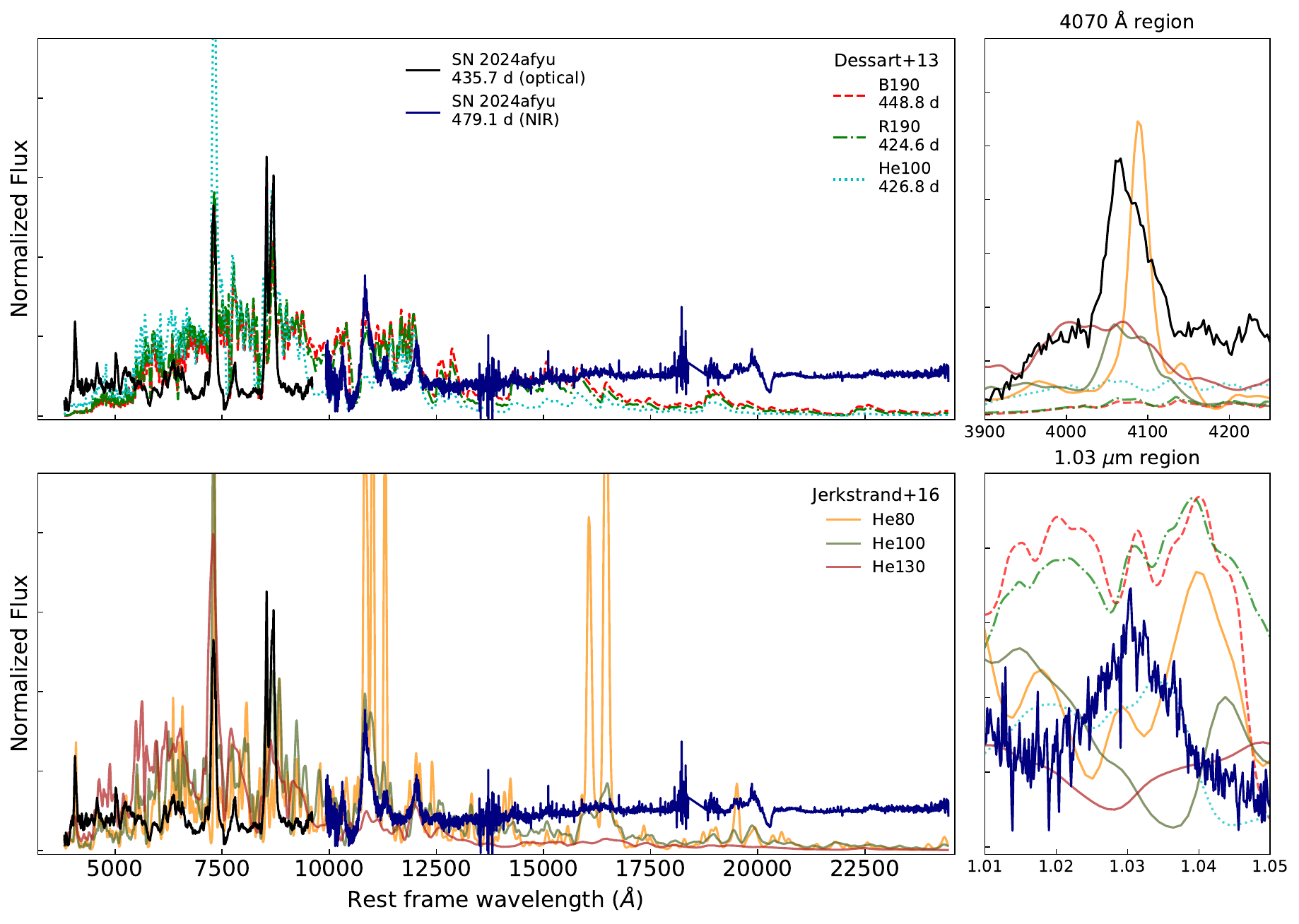}
      \caption{PISN spectral models compared to SN~2024afyu (black and navy). PISN spectral models compared to SN~2024afyu (black and navy). All spectral phases are indicated as rest frame days past explosion. Top left panel: comparison to models by \cite{Dessart_2013_PISN} at the closest day from observations. Bottom left panel: comparison to models by \cite{Jerkstrand_2016_PISN} at 400 days. Top right panel: Region around 4070\AA\ including all models. Bottom right panel: Region around 1.03$\mu$m including all models.}
         \label{fig:comparisonPISNmodels_spec}
   \end{figure*}

Figure~\ref{fig:comparisonPISNmodels_spec} compares the late-time spectra of SN~2024afyu with available PISN spectral models. Consistent with the photometric comparison, none of the models presented by \cite{Dessart_2013_PISN} reproduces the overall appearance of the spectrum. Nevertheless, both the R190 and B190 models provide a reasonable match to the [\ion{Ca}{ii}] emission.
The spectral models of \cite{Jerkstrand_2016_PISN}\footnote{\url{https://doi.org/10.5281/zenodo.5578624}.} provide a substantially better match, particularly across the optical wavelength range. Among the available models, He80 and He100 at 400 days show the closest resemblance to the observed spectrum. 
None of the published models successfully reproduces the near-infrared spectrum.

Although none of the available models reproduces all observed characteristics, the overall agreement is noteworthy given that the available models were not constructed specifically to match SN~2024afyu. We note that, compared to the models, the light curve of SN~2024afyu declines systematically more slowly after peak. However, \citet{2018MNRAS.479.3106K} show that increasing the degree of $^{56}$Ni mixing produces flatter maxima and slower post-peak declines. Interestingly, SN~2024afyu exhibits an extended, relatively flat maximum in all observed bands (Fig.~\ref{fig:obslc}), suggesting that enhanced nickel mixing may contribute to the observed light curve evolution.

\subsection{Alternative interpretations}

\subsubsection{CCSNe experiencing CSM interaction}

The light curve of SN~2024afyu shows a noticeable break in its decline at late times, followed by subtle ``wiggles''. A pure $^{56}$Ni-powered light curve cannot explain these wiggles. Similar light curve behaviour is commonly interpreted as signatures of interaction with CSM. 
In addition, [\ion{S}{ii}] lines have been predicted at nebular epochs for SESN \citep{1989ApJ...343..323F, 2010MNRAS.408...87M} and also for SNe~II undergoing CSM interaction \citep{1994ApJ...420..268C}, with iPTF14hls being an example of a SN~II that shows plenty of evidence of CSM interaction and also sulfur lines it its spectra appearing at $\sim$1092 days past discovery \citep{2019A&A...621A..30S}. So, the sulfur lines could be spectroscopic indication of ongoing interaction. However, we find two problems with this interpretation. First, evidence of sulfur lines appear in the spectra of SN~2024afyu as early as  $\sim$30 days past peak, which is very much within the photospheric phase, and none of the accompanying oxygen lines appear at similar epochs. Second, we estimate masses of $M_{\mathrm{S \,II,279d}} = 3.7^{+1.6}_{-1.2}$ \Msun{} and $M_{\mathrm{S \,II,312d}} = 3.1^{+1.9}_{-1.4}$ \Msun{} (see Sec.~\ref{subsec:SII_temperature}), requiring a sulphur mass so large that, independently of where it resides, can only result from a very massive progenitor star. 

The three latest optical spectra show a broad feature underlying a narrow H$\alpha$ emission line. While the narrow component is likely associated with the host environment, the broad component may result from CSM interaction. Similar late-time behavior was observed in SN~1993J and attributed to CSM interaction \citep[e.g.][]{1994AJ....108.2220F,2004ChPhL..21.1398W}. We find no clear evidence for interaction at earlier epochs in SN~2024afyu, although relatively low-density CSM could still affect the light-curve rise time and luminosity without producing prominent spectral signatures \citep{2017ApJ...838...28M,2019A&A...631A...8H}. The lack of sufficiently early spectra prevents us from investigating the presence of flash-ionization features expected in some interaction scenarios \citep[e.g.][]{2023ApJ...952..119B,2026arXiv260203638E}. At late times, the \SIIoptical lines develop a somewhat boxy profile with an enhanced red wing (Fig.~\ref{fig:SII_evolution}), a morphology previously associated with CSM interaction in other SN spectral features \citep[e.g., for oxygen lines][]{2024A&A...683A.223S}. No comparable evolution is seen in the \SIINIR lines, although our NIR observations do not cover the latest optical epochs.

\citet{2022A&A...660L...9D} found that, in H-rich interacting SNe, the strongest spectroscopic signatures of interaction are concentrated in the UV. This may be relevant given our hydrogen identification in SN~2024afyu (Appendix~\ref{app:synapps}). Although we lack UV coverage to test this possibility, we note that many of the previously proposed PISNe show a bluer continuum at shorter wavelengths compared to SN~2024afyu, that shows relatively flat spectra at all late phases (see Fig.~\ref{fig:comparisonPISN}). Also the latest available spectrum of iPTF14hls show such bluer continuum at shorter wavelengths (see Fig. 6 of \citealt{2019A&A...621A..30S}). If such behaviour is evidence of interaction, it is absent in SN~2024afyu.

It is worth mentioning that the appearance of the late-time light curve wiggles follow a steep decline that coincides with a sudden reddening of the observed colors (see bottom panel of Fig.~\ref{fig:obslc}). Such behaviour has been associated to dust formation, since dust absorbs bluer wavelenghts \citep[e.g.][]{2008MNRAS.389..141M}. Yet, we don't see any evidence of CO in our NIR spectra (see Fig.~\ref{fig:obsspecNIR}). However, \cite{2013ApJ...776....5M} argues that carbon dust grains can form even when there is no CO formation.

If the late-time light-curve wiggles are indeed caused by CSM interaction, and our $^{56}$Ni mass estimate is correct, a possible alternative to a PISN is a pulsational pair-instability supernova \citep[PPISN; e.g.][]{2023NatAs...7..779L}. In this scenario, the CSM originates from one or more episodes of mass ejection preceding the terminal explosion, which may occur over a wide range of timescales before core-collapse \citep[e.g.][]{2020A&A...640A..56R}. To search for evidence of such pre-explosion activity, we inspected the complete history of ZTF data releases at the position of SN~2024afyu using the SNAD viewer\footnote{\url{https://snad.space/}.} \citep{2023PASP..135b4503M}, covering more than six years prior to discovery. Although a few apparent brightenings are visible in the historical light curve, the corresponding ZTF forced photometry is consistent with non-detections. Therefore, if the progenitor experienced eruptive mass loss before explosion, it either occurred more than six years prior to discovery or remained below the ZTF detection limit. The PPISN interpretation is also difficult to reconcile with the spectroscopic properties of SN~2024afyu. PPI is expected for He-core masses below $\sim$80\Msun{} \citep{2020A&A...640A..56R}, whereas complete PI, capable of producing substantial amounts of intermediate-mass elements such as sulfur, requires He-core masses of approximately 90\Msun{} or greater \citep{2017ApJ...836..244W}. If the identification of sulfure features presented in Sect.~\ref{sec:sulfurANDSiliconID} is correct, their strengths favors a PISN over a PPISN origin.

Alternative core-collapse scenarios also face significant challenges. The nebular models of \citet{2010MNRAS.408...87M}, developed for the Type~Ic SN~2007gr, predict a $\sim$15\Msun{} progenitor and substantial \SIIoptical emission (stronger than that observed in SN~2007gr itself), but they also predict strong [\ion{O}{i}]~$\lambda\lambda6300,6363$ emission, which is notably absent in SN~2024afyu. Moreover, the models predict only a small amount of intermediate mass elements, with e.g. $\sim$0.005\Msun{} of silicon and 0.02\Msun{} of sulphur, some two to three orders of magnitude lower than our estimates. Still, the main difference in this case would be the light curve, SN~2007gr exhibits the rapidly evolving light curve characteristic of ordinary SESNe, which contrasts with the broad and slowly evolving light curve of SN~2024afyu.

Sulfur-rich spectra has also been seen in the so-called ultra-stripped SNe. \cite{2025Natur.644..634S} presents SN~2021yfj as an extreme case, where the sulfur and silicon rich material is interpreted as CSM surrounding an extremely stripped progenitor. However, such a scenario appears difficult to reconcile with the possible presence of H in SN~2024afyu (Appendix~\ref{app:synapps}), although it could be that there is no H in the spectra and the evidence of it is due to the shortcomings of the \texttt{SYNAPPS} spectral modeling. Even if the sulfur and silicon emission were instead produced in the CSM, the observed luminosity ratios of the \SIIoptical + \SIINIR and \SIIInineosixnine + \SIIIninefivethreeone line pairs (Sections \ref{subsec:SII_mass} and \ref{subsec:SIII}) are fully consistent with optically thin LTE, which allow us to infer a sulfur mass of approximately 3\Msun{}, implying an exceptionally large reservoir of intermediate-mass elements. Such a sulfur mass is difficult to produce in conventional core-collapse explosions and instead points to the explosion of a very massive progenitor, consistent with the expectations for a PISN.

\subsubsection{Central engine}

The two principal alternatives invoked to explain long-lived, luminous SNe are energy injection from the spin-down of a magnetar and fallback accretion onto a newly formed black hole. \citet{Dessart_2024_magnetar} simulated model spectra for helium star models (i.e. progenitors for standard CCSNe) under the influence of a magnetar. For lower mass models ($M_{\mathrm{He,i}} \lesssim 6.0$ \Msun{}, it was found that \SIIoptical and \SIIInineosixnine + \SIIIninefivethreeone become prominent in the spectra as early as $\sim 150 $ days after the magnetar started injecting energy, caused not by a high sulfur mass, but an increase in the temperature in the sulfur-rich zones (Fig. \ref{fig:sulphur_curves}). Crucially, just as most prototypical CCSNe, optical spectra of these models were dominated by \OIdoublet{} during the nebular phase (250d $\lesssim$ t $\lesssim$ 600d), in contrast with what is seen in SN~2024afyu. Furthermore, these models lacked any signatures of \CaIIdoublet and \CaNIR emission, which are the two most prominent lines in SN~2024afyu. The combination of these differences in key lines leads us to disfavour a magnetar scenario for SN~2024afyu. 

A fallback-powered explosion faces the difficulty that the larger the amount of fallback, the more efficiently the innermost nucleosynthetic products (such as S and Si) are accreted onto the compact remnant rather than ejected \citep{2002RvMP...74.1015W,2006NuPhA.777..424N}. Although hypernovae have been suggested to have some degree of fallback and simultaneously produce relatively larger yields of intermediate-mass elements than ordinary core-collapse supernovae \citep{Nomoto_2001_Hypernovae, Nomoto_2003_Hypernovae}, a representative example of these is SN~1998bw, whose light curve and spectroscopic evolution (see Figures~\ref{fig:compLC}, \ref{fig:comparisonALL}, and \ref{fig:comparisonALLNIR}) differ significantly from those of SN~2024afyu.

\section{Conclusion}
\label{sec:conclusion}

SN~2024afyu is an outstanding event. The first indication that it represented an unusual event was the early appearance of the [\ion{Ca}{ii}]~$\lambda\lambda7291,7324$ doublet without the accompanying [\ion{O}{i}]~$\lambda\lambda6300,6364$ emission. A similar combination was reported for SN~2019bkc by \citet{2020A&A...635A.186P}, who also identified several intermediate-mass elements in the corresponding spectra. However, SN~2019bkc shows a light curve evolution that is very different than that of SN~2024afyu, reaching peak brightness much faster (in only $\sim$6 days compared to $\sim$85 days) and being approximately 1.5 mag fainter. Given the observed spectral features and the rapidly evolving light curve, \cite{2020A&A...635A.186P} concluded that the available observations could not distinguish between a core-collapse explosion and a helium-detonation. The light curves of SN~2024afyu shows some post-peak flattening, reminiscent of the ``kink'' observed in the $i$-band light curves of some thermonuclear SNe~Ia. The physical origin of this feature remains uncertain, although it has been suggested that it could arise from Ca-rich plumes produced by instabilities in the deflagration front \citep{2022MNRAS.510.4929P,2025A&A...697A.125K}. This interpretation remains speculative, and detailed modelling is required to investigate whether such effects can produce the observed light-curve morphology. Nevertheless, if the flattening is confirmed to be intrinsic and SN~2024afyu is indeed a PISN, a similar mechanism could potentially provide a natural explanation for the prominent [\ion{Ca}{ii}] features observed in SN~2024afyu.

In this work, we interpret the observed features of SN~2024afyu, i.e. $\sim$85 day rise,  $^{56}$Ni mass estimate of around 0.4--1.0\Msun{}, and masses of $M_{\mathrm{S \,II,279d}} = 3.7^{+1.6}_{-1.2}$ \Msun{} and $M_{\mathrm{S \,II,312d}} = 3.1^{+1.9}_{-1.4}$ \Msun{}, as evidence of SN~2024afyu resulting from the explosion of a very massive star through the pair-instability mechanism. While alternative progenitor and explosion channels may be explored, we emphasize that light-curve modelling is subject to significant degeneracies, with physically distinct models capable of reproducing similar photometric evolution. Our primary objective was therefore to identify models that are consistent not only with the observed light curves, but also with the key spectroscopic properties of SN~2024afyu. This is particularly important in the context of PISN candidates, as their predicted nucleosynthetic yields give rise to distinctive spectral signatures that can provide complementary, and in some cases more discriminating, constraints on the explosion mechanism than photometry alone. In this regard, SN~2024afyu is notable because the PISN interpretation is supported by the consistency between its observed spectroscopic features and the spectral signatures predicted by PISN models.

\section*{Data availability}

Light curves can be found in the tables included in the Appendix. All the spectra can be found on WISeREP at \url{https://www.wiserep.org}.

\begin{acknowledgements}

P.J.P. thanks D. Whalen and A. Heger for kindly sharing the models presented in Whalen et al. (2014). P.J.P. also thanks E. Bordier for useful discussion about the origin of massive stars. 
BHTOM.space is based on the open-source TOM Toolkit by LCO and has been developed with funding from the OPTICON-RadioNet Pilot (ORP) of the European Union's Horizon 2020 research and innovation programme under grant agreement No 101004719 (2021-2025). This project has received funding from the European Union's Horizon Europe Research and Innovation programme ACME under grant agreement No 101131928 (2024-2028). {\L}.W. acknowledges support from the Polish National Science Centre DAINA grant No 2024/52/L/ST9/00210 and Polish MNiSW grant DIR/WK/2018/12.
Based on observations obtained with the Samuel Oschin Telescope 48-inch and the 60-inch Telescope at the Palomar Observatory as part of the Zwicky Transient Facility project. ZTF is supported by the National Science Foundation under Grants No. AST-1440341, AST-2034437, and currently Award \#2407588. ZTF receives additional funding from the ZTF partnership. Current members include Caltech, USA; Caltech/IPAC, USA; University of Maryland, USA; University of California, Berkeley, USA; Cornell University, USA; Drexel University, USA; University of North Carolina at Chapel Hill, USA; Institute of Science and Technology, Austria; National Central University, Taiwan, German Center for Astrophysics, Germany, and OKC, University of Stockholm, Sweden. Operations are conducted by Caltech's Optical Observatory (COO), Caltech/IPAC, and the University of Washington at Seattle, USA.
The ZTF forced-photometry service was funded under the Heising-Simons Foundation grant No. 12540303 (PI: Graham).
The SED Machine is based upon work supported by the National Science Foundation under Grant No. 1106171
The Liverpool Telescope is operated on the island of La Palma by Liverpool John Moores University in the Spanish Observatorio del Roque de los Muchachos of the Instituto de Astrofisica de Canarias with financial support from the UK Science and Technology Facilities Council.
The LBT is an international collaboration among institutions in the United States and Europe. At the time data were acquired for this research, LBT Corporation Members were the University of Arizona on behalf of the Arizona Board of Regents; Instituto Nazionale di Astrofisica, Italy; LBT Beteiligungsgesellschaft, Germany, representing the Max-Planck Society, the Leibniz Institute for Astrophysics Potsdam, and Heidelberg University; and The Ohio State University, representing The Ohio State University, University of Notre Dame, University of Minnesota, and University of Virginia.
This work has made use of data from the Asteroid Terrestrial-impact Last Alert System (ATLAS) project. The Asteroid Terrestrial-impact Last Alert System (ATLAS) project is primarily funded to search for near earth asteroids through NASA grants NN12AR55G, 80NSSC18K0284, and 80NSSC18K1575; byproducts of the NEO search include images and catalogs from the survey area. This work was partially funded by Kepler/K2 grant J1944/80NSSC19K0112 and HST GO-15889, and STFC grants ST/T000198/1 and ST/S006109/1. The ATLAS science products have been made possible through the contributions of the University of Hawaii Institute for Astronomy, the Queen’s University Belfast, the Space Telescope Science Institute, the South African Astronomical Observatory, and The Millennium Institute of Astrophysics (MAS), Chile.
This research has made use of the NASA/IPAC Extragalactic Database, which is funded by the National Aeronautics and Space Administration and operated by the California Institute of Technology.
Observations reported here were obtained at the MMT Observatory, a joint facility of the Smithsonian Institution and the University of Arizona.
A.J. and S.B. acknowledge funding by the Swedish National Research Council, grant 2018-03799
C.L. is supported by DoE award \#\,DE-SC0025599.
MMT Observatory access was supported by Northwestern University and the Center for Interdisciplinary Exploration and Research in Astrophysics (CIERA).
J.~M. Carrasco was (partially) supported by the Spanish MICIN/AEI/10.13039/501100011033 and by "ERDF A way of making Europe" by the “European Union” through grant PID2021-122842OB-C21 and PID2024-157964OB-C21, and the Institute of Cosmos Sciences University of Barcelona (ICCUB, Unidad de Excelencia ’Mar\'{\i}a de Maeztu’) through grant CEX2024-001451-M and the project 2021-SGR-00679 GRC de l'Agència de Gestió d'Ajuts Universitaris i de Recerca (Generalitat de Catalunya). The Joan Oró Telescope (TJO) of the Montsec Observatory (OdM) is owned by the Catalan Government and operated by the Institute for Space Studies of Catalonia (IEEC).
A. Singh acknowledges support from the Knut and Alice Wallenberg Foundation through the ``Gravity Meets Light" project.
C. V acknowledges the INAF project ``Supporto Arizona \& Italia''.
E.P., J.Z., E.S., M.M. acknowledge funding from the Research Council of Lithuania (LMTLT, grant No. S-LL-24-1).
P. King acknowledges The staff at Lick Observatory.
M.B. acknowledges the financial support from the Slovenian Research Agency (research core funding P1-0031, infrastructure program I0-0033). GoChile is a project funded by the University of Nova Gorica and the Astronomical Magazine Spika. We acknowledge financial support of the Slovenian Research and Innovation Agency through Action plan for Open Science.
N.R. is supported by a Northwestern University Presidential Fellowship
K.M acknowledges the support from the BRICS grant DST/ICD/BRICS/Call-5/CoNMuTraMO/2023 (G) funded by the Department of Science and Technology (DST), India.
A.C. has been supported by the Polish National Science Center project UMO-2023/51/D/ST9/00147.
This work made use of Astropy (\url{https://www.astropy.org}): a community-developed core Python package and an ecosystem of tools and resources for astronomy \citep{astropy:2013, astropy:2018, astropy:2022}.
This work made use of the \texttt{legacystamps} package (\url{https://github.com/tikk3r/legacystamps}).
The Legacy Surveys consist of three individual and complementary projects: the Dark Energy Camera Legacy Survey (DECaLS; Proposal ID \#2014B-0404; PIs: David Schlegel and Arjun Dey), the Beijing-Arizona Sky Survey (BASS; NOAO Prop. ID \#2015A-0801; PIs: Zhou Xu and Xiaohui Fan), and the Mayall z-band Legacy Survey (MzLS; Prop. ID \#2016A-0453; PI: Arjun Dey). DECaLS, BASS and MzLS together include data obtained, respectively, at the Blanco telescope, Cerro Tololo Inter-American Observatory, NSF's NOIRLab; the Bok telescope, Steward Observatory, University of Arizona; and the Mayall telescope, Kitt Peak National Observatory, NOIRLab. Pipeline processing and analyses of the data were supported by NOIRLab and the Lawrence Berkeley National Laboratory (LBNL). The Legacy Surveys project is honored to be permitted to conduct astronomical research on Iolkam Du'ag (Kitt Peak), a mountain with particular significance to the Tohono O'odham Nation.
NOIRLab is operated by the Association of Universities for Research in Astronomy (AURA) under a cooperative agreement with the National Science Foundation. LBNL is managed by the Regents of the University of California under contract to the U.S. Department of Energy.
This project used data obtained with the Dark Energy Camera (DECam), which was constructed by the Dark Energy Survey (DES) collaboration. Funding for the DES Projects has been provided by the U.S. Department of Energy, the U.S. National Science Foundation, the Ministry of Science and Education of Spain, the Science and Technology Facilities Council of the United Kingdom, the Higher Education Funding Council for England, the National Center for Supercomputing Applications at the University of Illinois at Urbana-Champaign, the Kavli Institute of Cosmological Physics at the University of Chicago, Center for Cosmology and Astro-Particle Physics at the Ohio State University, the Mitchell Institute for Fundamental Physics and Astronomy at Texas A\&M University, Financiadora de Estudos e Projetos, Fundacao Carlos Chagas Filho de Amparo, Financiadora de Estudos e Projetos, Fundacao Carlos Chagas Filho de Amparo a Pesquisa do Estado do Rio de Janeiro, Conselho Nacional de Desenvolvimento Cientifico e Tecnologico and the Ministerio da Ciencia, Tecnologia e Inovacao, the Deutsche Forschungsgemeinschaft and the Collaborating Institutions in the Dark Energy Survey. The Collaborating Institutions are Argonne National Laboratory, the University of California at Santa Cruz, the University of Cambridge, Centro de Investigaciones Energeticas, Medioambientales y Tecnologicas-Madrid, the University of Chicago, University College London, the DES-Brazil Consortium, the University of Edinburgh, the Eidgenossische Technische Hochschule (ETH) Zurich, Fermi National Accelerator Laboratory, the University of Illinois at Urbana-Champaign, the Institut de Ciencies de l'Espai (IEEC/CSIC), the Institut de Fisica d'Altes Energies, Lawrence Berkeley National Laboratory, the Ludwig Maximilians Universitat Munchen and the associated Excellence Cluster Universe, the University of Michigan, NSF's NOIRLab, the University of Nottingham, the Ohio State University, the University of Pennsylvania, the University of Portsmouth, SLAC National Accelerator Laboratory, Stanford University, the University of Sussex, and Texas A\&M University.
BASS is a key project of the Telescope Access Program (TAP), which has been funded by the National Astronomical Observatories of China, the Chinese Academy of Sciences (the Strategic Priority Research Program ``The Emergence of Cosmological Structures'' Grant \# XDB09000000), and the Special Fund for Astronomy from the Ministry of Finance. The BASS is also supported by the External Cooperation Program of Chinese Academy of Sciences (Grant \# 114A11KYSB20160057), and Chinese National Natural Science Foundation (Grant \# 12120101003, \# 11433005).
The Legacy Survey team makes use of data products from the Near-Earth Object Wide-field Infrared Survey Explorer (NEOWISE), which is a project of the Jet Propulsion Laboratory/California Institute of Technology. NEOWISE is funded by the National Aeronautics and Space Administration.
The Legacy Surveys imaging of the DESI footprint is supported by the Director, Office of Science, Office of High Energy Physics of the U.S. Department of Energy under Contract No. DE-AC02-05CH1123, by the National Energy Research Scientific Computing Center, a DOE Office of Science User Facility under the same contract; and by the U.S. National Science Foundation, Division of Astronomical Sciences under Contract No. AST-0950945 to NOAO.

\end{acknowledgements}

\bibliography{24afyu} 

@ARTICLE{Lotz2004,
       author = {{Lotz}, Jennifer M. and {Primack}, Joel and {Madau}, Piero},
        title = "{A New Nonparametric Approach to Galaxy Morphological Classification}",
      journal = {\aj},
         year = 2004,
        month = jul,
       volume = {128},
       number = {1},
        pages = {163-182},
          doi = {10.1086/421849},
archivePrefix = {arXiv},
       eprint = {astro-ph/0311352},
 primaryClass = {astro-ph},
       adsurl = {https://ui.adsabs.harvard.edu/abs/2004AJ....128..163L}
}

@ARTICLE{Lotz2008,
       author = {{Lotz}, Jennifer M. and {Jonsson}, Patrik and {Cox}, T.~J. and {Primack}, Joel R.},
        title = "{Galaxy merger morphologies and time-scales from simulations of equal-mass gas-rich disc mergers}",
      journal = {\mnras},
         year = 2008,
        month = dec,
       volume = {391},
       number = {3},
        pages = {1137-1162},
          doi = {10.1111/j.1365-2966.2008.14004.x},
archivePrefix = {arXiv},
       eprint = {0805.1246},
 primaryClass = {astro-ph},
       adsurl = {https://ui.adsabs.harvard.edu/abs/2008MNRAS.391.1137L}
}

@ARTICLE{Conselice2003,
       author = {{Conselice}, Christopher J.},
        title = "{The Relationship between Stellar Light Distributions of Galaxies and Their Formation Histories}",
      journal = {\apjs},
         year = 2003,
        month = jul,
       volume = {147},
       number = {1},
        pages = {1-28},
          doi = {10.1086/375001},
archivePrefix = {arXiv},
       eprint = {astro-ph/0303065},
 primaryClass = {astro-ph},
       adsurl = {https://ui.adsabs.harvard.edu/abs/2003ApJS..147....1C}
}

@ARTICLE{2024TNSTR5161....1S,
       author = {{Sollerman}, J. and {Fremling}, C. and {Perley}, D. and {Laz}, T.~D.},
        title = "{ZTF Transient Discovery Report for 2024-12-30}",
      journal = {Transient Name Server Discovery Report},
         year = 2024,
        month = dec,
       volume = {2024-5161},
        pages = {1},
       adsurl = {https://ui.adsabs.harvard.edu/abs/2024TNSTR5161....1S}
}

@ARTICLE{2025TNSCR..53....1B,
       author = {{Balcon}, C.},
        title = "{Transient Classification Report for 2025-01-04}",
      journal = {Transient Name Server Classification Report},
         year = 2025,
        month = jan,
       volume = {2025-53},
        pages = {1},
       adsurl = {https://ui.adsabs.harvard.edu/abs/2025TNSCR..53....1B}
}

@ARTICLE{2025TNSCR1202....1P,
       author = {{Palau}, C.~J. and {Pursiainen}, M. and {Godson}, B. and {Magee}, M. and {Lyman}, J. and {Galbany}, L. and {Killestein}, T.},
        title = "{GOTO Transient Classification Report for 2025-03-29}",
      journal = {Transient Name Server Classification Report},
         year = 2025,
        month = mar,
       volume = {2025-1202},
        pages = {1},
       adsurl = {https://ui.adsabs.harvard.edu/abs/2025TNSCR1202....1P}
}

@ARTICLE{2015MNRAS.450..317C,
       author = {{Carrick}, Jonathan and {Turnbull}, Stephen J. and {Lavaux}, Guilhem and {Hudson}, Michael J.},
        title = "{Cosmological parameters from the comparison of peculiar velocities with predictions from the 2M++ density field}",
      journal = {\mnras},
         year = 2015,
        month = jun,
       volume = {450},
       number = {1},
        pages = {317-332},
          doi = {10.1093/mnras/stv547},
archivePrefix = {arXiv},
       eprint = {1504.04627},
 primaryClass = {astro-ph.CO},
       adsurl = {https://ui.adsabs.harvard.edu/abs/2015MNRAS.450..317C}
}

@ARTICLE{2022PASA...39...46C,
       author = {{Carr}, Anthony and {Davis}, Tamara M. and {Scolnic}, Dan and {Said}, Khaled and {Brout}, Dillon and {Peterson}, Erik R. and {Kessler}, Richard},
        title = "{The Pantheon+ analysis: Improving the redshifts and peculiar velocities of Type Ia supernovae used in cosmological analyses}",
      journal = {\pasa},
         year = 2022,
        month = oct,
       volume = {39},
          eid = {e046},
        pages = {e046},
          doi = {10.1017/pasa.2022.41},
archivePrefix = {arXiv},
       eprint = {2112.01471},
 primaryClass = {astro-ph.CO},
       adsurl = {https://ui.adsabs.harvard.edu/abs/2022PASA...39...46C}
}

@ARTICLE{2022ApJ...938..110B,
       author = {{Brout}, Dillon and {Scolnic}, Dan and {Popovic}, Brodie and {Riess}, Adam G. and {Carr}, Anthony and {Zuntz}, Joe and {Kessler}, Rick and {Davis}, Tamara M. and {Hinton}, Samuel and {Jones}, David and {Kenworthy}, W. D'Arcy and {Peterson}, Erik R. and {Said}, Khaled and {Taylor}, Georgie and {Ali}, Noor and {Armstrong}, Patrick and {Charvu}, Pranav and {Dwomoh}, Arianna and {Meldorf}, Cole and {Palmese}, Antonella and {Qu}, Helen and {Rose}, Benjamin M. and {Sanchez}, Bruno and {Stubbs}, Christopher W. and {Vincenzi}, Maria and {Wood}, Charlotte M. and {Brown}, Peter J. and {Chen}, Rebecca and {Chambers}, Ken and {Coulter}, David A. and {Dai}, Mi and {Dimitriadis}, Georgios and {Filippenko}, Alexei V. and {Foley}, Ryan J. and {Jha}, Saurabh W. and {Kelsey}, Lisa and {Kirshner}, Robert P. and {M{\"o}ller}, Anais and {Muir}, Jessie and {Nadathur}, Seshadri and {Pan}, Yen-Chen and {Rest}, Armin and {Rojas-Bravo}, Cesar and {Sako}, Masao and {Siebert}, Matthew R. and {Smith}, Mat and {Stahl}, Benjamin E. and {Wiseman}, Phil},
        title = "{The Pantheon+ Analysis: Cosmological Constraints}",
      journal = {\apj},
         year = 2022,
        month = oct,
       volume = {938},
       number = {2},
          eid = {110},
        pages = {110},
          doi = {10.3847/1538-4357/ac8e04},
archivePrefix = {arXiv},
       eprint = {2202.04077},
 primaryClass = {astro-ph.CO},
       adsurl = {https://ui.adsabs.harvard.edu/abs/2022ApJ...938..110B}
}

@INPROCEEDINGS{2024lsstWyrzykowski,
       author = {{Wyrzykowski}, Lukasz},
        title = "{Power of many - BHTOM telescope network for time-domain astronomy}",
    booktitle = {What Was That? - Planning ESO Follow up for Transients, Variables, and Solar System Objects in the Era of LSST},
         year = 2024,
        month = jan,
          eid = {4},
        pages = {4},
          doi = {10.5281/zenodo.10571539},
       adsurl = {https://ui.adsabs.harvard.edu/abs/2024lsst.confE...4W}
}

@INPROCEEDINGS{2025RMxACMikolajczyk,
       author = {{Mikolajczyk}, P.~J. and {Zieli{\'n}ski}, P. and {Wyrzykowski}, L. and {Krawczyk}, A. and {Kotysz}, K.},
        title = "{Black Hole TOM - an Automatic Tool for Photometric Time-Domain Data}",
    booktitle = {Revista Mexicana de Astronomia y Astrofisica Conference Series},
         year = 2025,
       series = {Revista Mexicana de Astronomia y Astrofisica Conference Series},
       volume = {59},
        month = jul,
        pages = {167-172},
          doi = {10.22201/ia.14052059p.2025.59.26},
       adsurl = {https://ui.adsabs.harvard.edu/abs/2025RMxAC..59..167M}
}

@ARTICLE{2019PASP..131a8002B,
       author = {{Bellm}, Eric C. and {Kulkarni}, Shrinivas R. and {Graham}, Matthew J. and {Dekany}, Richard and {Smith}, Roger M. and {Riddle}, Reed and {Masci}, Frank J. and {Helou}, George and {Prince}, Thomas A. and {Adams}, Scott M. and {Barbarino}, C. and {Barlow}, Tom and {Bauer}, James and {Beck}, Ron and {Belicki}, Justin and {Biswas}, Rahul and {Blagorodnova}, Nadejda and {Bodewits}, Dennis and {Bolin}, Bryce and {Brinnel}, Valery and {Brooke}, Tim and {Bue}, Brian and {Bulla}, Mattia and {Burruss}, Rick and {Cenko}, S. Bradley and {Chang}, Chan-Kao and {Connolly}, Andrew and {Coughlin}, Michael and {Cromer}, John and {Cunningham}, Virginia and {De}, Kishalay and {Delacroix}, Alex and {Desai}, Vandana and {Duev}, Dmitry A. and {Eadie}, Gwendolyn and {Farnham}, Tony L. and {Feeney}, Michael and {Feindt}, Ulrich and {Flynn}, David and {Franckowiak}, Anna and {Frederick}, S. and {Fremling}, C. and {Gal-Yam}, Avishay and {Gezari}, Suvi and {Giomi}, Matteo and {Goldstein}, Daniel A. and {Golkhou}, V. Zach and {Goobar}, Ariel and {Groom}, Steven and {Hacopians}, Eugean and {Hale}, David and {Henning}, John and {Ho}, Anna Y.~Q. and {Hover}, David and {Howell}, Justin and {Hung}, Tiara and {Huppenkothen}, Daniela and {Imel}, David and {Ip}, Wing-Huen and {Ivezi{\'c}}, {\v{Z}}eljko and {Jackson}, Edward and {Jones}, Lynne and {Juric}, Mario and {Kasliwal}, Mansi M. and {Kaspi}, S. and {Kaye}, Stephen and {Kelley}, Michael S.~P. and {Kowalski}, Marek and {Kramer}, Emily and {Kupfer}, Thomas and {Landry}, Walter and {Laher}, Russ R. and {Lee}, Chien-De and {Lin}, Hsing Wen and {Lin}, Zhong-Yi and {Lunnan}, Ragnhild and {Giomi}, Matteo and {Mahabal}, Ashish and {Mao}, Peter and {Miller}, Adam A. and {Monkewitz}, Serge and {Murphy}, Patrick and {Ngeow}, Chow-Choong and {Nordin}, Jakob and {Nugent}, Peter and {Ofek}, Eran and {Patterson}, Maria T. and {Penprase}, Bryan and {Porter}, Michael and {Rauch}, Ludwig and {Rebbapragada}, Umaa and {Reiley}, Dan and {Rigault}, Mickael and {Rodriguez}, Hector and {van Roestel}, Jan and {Rusholme}, Ben and {van Santen}, Jakob and {Schulze}, S. and {Shupe}, David L. and {Singer}, Leo P. and {Soumagnac}, Maayane T. and {Stein}, Robert and {Surace}, Jason and {Sollerman}, Jesper and {Szkody}, Paula and {Taddia}, F. and {Terek}, Scott and {Van Sistine}, Angela and {van Velzen}, Sjoert and {Vestrand}, W. Thomas and {Walters}, Richard and {Ward}, Charlotte and {Ye}, Quan-Zhi and {Yu}, Po-Chieh and {Yan}, Lin and {Zolkower}, Jeffry},
        title = "{The Zwicky Transient Facility: System Overview, Performance, and First Results}",
      journal = {\pasp},
         year = 2019,
        month = jan,
       volume = {131},
       number = {995},
        pages = {018002},
          doi = {10.1088/1538-3873/aaecbe},
archivePrefix = {arXiv},
       eprint = {1902.01932},
 primaryClass = {astro-ph.IM},
       adsurl = {https://ui.adsabs.harvard.edu/abs/2019PASP..131a8002B}
}

@ARTICLE{2019PASP..131g8001G,
       author = {{Graham}, Matthew J. and {Kulkarni}, S.~R. and {Bellm}, Eric C. and {Adams}, Scott M. and {Barbarino}, Cristina and {Blagorodnova}, Nadejda and {Bodewits}, Dennis and {Bolin}, Bryce and {Brady}, Patrick R. and {Cenko}, S. Bradley and {Chang}, Chan-Kao and {Coughlin}, Michael W. and {De}, Kishalay and {Eadie}, Gwendolyn and {Farnham}, Tony L. and {Feindt}, Ulrich and {Franckowiak}, Anna and {Fremling}, Christoffer and {Gezari}, Suvi and {Ghosh}, Shaon and {Goldstein}, Daniel A. and {Golkhou}, V. Zach and {Goobar}, Ariel and {Ho}, Anna Y.~Q. and {Huppenkothen}, Daniela and {Ivezi{\'c}}, {\v{Z}}eljko and {Jones}, R. Lynne and {Juric}, Mario and {Kaplan}, David L. and {Kasliwal}, Mansi M. and {Kelley}, Michael S.~P. and {Kupfer}, Thomas and {Lee}, Chien-De and {Lin}, Hsing Wen and {Lunnan}, Ragnhild and {Mahabal}, Ashish A. and {Miller}, Adam A. and {Ngeow}, Chow-Choong and {Nugent}, Peter and {Ofek}, Eran O. and {Prince}, Thomas A. and {Rauch}, Ludwig and {van Roestel}, Jan and {Schulze}, Steve and {Singer}, Leo P. and {Sollerman}, Jesper and {Taddia}, Francesco and {Yan}, Lin and {Ye}, Quan-Zhi and {Yu}, Po-Chieh and {Barlow}, Tom and {Bauer}, James and {Beck}, Ron and {Belicki}, Justin and {Biswas}, Rahul and {Brinnel}, Valery and {Brooke}, Tim and {Bue}, Brian and {Bulla}, Mattia and {Burruss}, Rick and {Connolly}, Andrew and {Cromer}, John and {Cunningham}, Virginia and {Dekany}, Richard and {Delacroix}, Alex and {Desai}, Vandana and {Duev}, Dmitry A. and {Feeney}, Michael and {Flynn}, David and {Frederick}, Sara and {Gal-Yam}, Avishay and {Giomi}, Matteo and {Groom}, Steven and {Hacopians}, Eugean and {Hale}, David and {Helou}, George and {Henning}, John and {Hover}, David and {Hillenbrand}, Lynne A. and {Howell}, Justin and {Hung}, Tiara and {Imel}, David and {Ip}, Wing-Huen and {Jackson}, Edward and {Kaspi}, Shai and {Kaye}, Stephen and {Kowalski}, Marek and {Kramer}, Emily and {Kuhn}, Michael and {Landry}, Walter and {Laher}, Russ R. and {Mao}, Peter and {Masci}, Frank J. and {Monkewitz}, Serge and {Murphy}, Patrick and {Nordin}, Jakob and {Patterson}, Maria T. and {Penprase}, Bryan and {Porter}, Michael and {Rebbapragada}, Umaa and {Reiley}, Dan and {Riddle}, Reed and {Rigault}, Mickael and {Rodriguez}, Hector and {Rusholme}, Ben and {van Santen}, Jakob and {Shupe}, David L. and {Smith}, Roger M. and {Soumagnac}, Maayane T. and {Stein}, Robert and {Surace}, Jason and {Szkody}, Paula and {Terek}, Scott and {Van Sistine}, Angela and {van Velzen}, Sjoert and {Vestrand}, W. Thomas and {Walters}, Richard and {Ward}, Charlotte and {Zhang}, Chaoran and {Zolkower}, Jeffry},
        title = "{The Zwicky Transient Facility: Science Objectives}",
      journal = {\pasp},
         year = 2019,
        month = jul,
       volume = {131},
       number = {1001},
        pages = {078001},
          doi = {10.1088/1538-3873/ab006c},
archivePrefix = {arXiv},
       eprint = {1902.01945},
 primaryClass = {astro-ph.IM},
       adsurl = {https://ui.adsabs.harvard.edu/abs/2019PASP..131g8001G}
}

@ARTICLE{2019PASP..131a8003M,
       author = {{Masci}, Frank J. and {Laher}, Russ R. and {Rusholme}, Ben and {Shupe}, David L. and {Groom}, Steven and {Surace}, Jason and {Jackson}, Edward and {Monkewitz}, Serge and {Beck}, Ron and {Flynn}, David and {Terek}, Scott and {Landry}, Walter and {Hacopians}, Eugean and {Desai}, Vandana and {Howell}, Justin and {Brooke}, Tim and {Imel}, David and {Wachter}, Stefanie and {Ye}, Quan-Zhi and {Lin}, Hsing-Wen and {Cenko}, S. Bradley and {Cunningham}, Virginia and {Rebbapragada}, Umaa and {Bue}, Brian and {Miller}, Adam A. and {Mahabal}, Ashish and {Bellm}, Eric C. and {Patterson}, Maria T. and {Juri{\'c}}, Mario and {Golkhou}, V. Zach and {Ofek}, Eran O. and {Walters}, Richard and {Graham}, Matthew and {Kasliwal}, Mansi M. and {Dekany}, Richard G. and {Kupfer}, Thomas and {Burdge}, Kevin and {Cannella}, Christopher B. and {Barlow}, Tom and {Van Sistine}, Angela and {Giomi}, Matteo and {Fremling}, Christoffer and {Blagorodnova}, Nadejda and {Levitan}, David and {Riddle}, Reed and {Smith}, Roger M. and {Helou}, George and {Prince}, Thomas A. and {Kulkarni}, Shrinivas R.},
        title = "{The Zwicky Transient Facility: Data Processing, Products, and Archive}",
      journal = {\pasp},
         year = 2019,
        month = jan,
       volume = {131},
       number = {995},
        pages = {018003},
          doi = {10.1088/1538-3873/aae8ac},
archivePrefix = {arXiv},
       eprint = {1902.01872},
 primaryClass = {astro-ph.IM},
       adsurl = {https://ui.adsabs.harvard.edu/abs/2019PASP..131a8003M}
}

@ARTICLE{2020PASP..132c8001D,
       author = {{Dekany}, Richard and {Smith}, Roger M. and {Riddle}, Reed and {Feeney}, Michael and {Porter}, Michael and {Hale}, David and {Zolkower}, Jeffry and {Belicki}, Justin and {Kaye}, Stephen and {Henning}, John and {Walters}, Richard and {Cromer}, John and {Delacroix}, Alex and {Rodriguez}, Hector and {Reiley}, Daniel J. and {Mao}, Peter and {Hover}, David and {Murphy}, Patrick and {Burruss}, Rick and {Baker}, John and {Kowalski}, Marek and {Reif}, Klaus and {Mueller}, Phillip and {Bellm}, Eric and {Graham}, Matthew and {Kulkarni}, Shrinivas R.},
        title = "{The Zwicky Transient Facility: Observing System}",
      journal = {\pasp},
         year = 2020,
        month = mar,
       volume = {132},
       number = {1009},
          eid = {038001},
        pages = {038001},
          doi = {10.1088/1538-3873/ab4ca2},
archivePrefix = {arXiv},
       eprint = {2008.04923},
 primaryClass = {astro-ph.IM},
       adsurl = {https://ui.adsabs.harvard.edu/abs/2020PASP..132c8001D}
}

@ARTICLE{2018PASP..130f4505T,
       author = {{Tonry}, J.~L. and {Denneau}, L. and {Heinze}, A.~N. and {Stalder}, B. and {Smith}, K.~W. and {Smartt}, S.~J. and {Stubbs}, C.~W. and {Weiland}, H.~J. and {Rest}, A.},
        title = "{ATLAS: A High-cadence All-sky Survey System}",
      journal = {\pasp},
         year = 2018,
        month = jun,
       volume = {130},
       number = {988},
        pages = {064505},
          doi = {10.1088/1538-3873/aabadf},
archivePrefix = {arXiv},
       eprint = {1802.00879},
 primaryClass = {astro-ph.IM},
       adsurl = {https://ui.adsabs.harvard.edu/abs/2018PASP..130f4505T}
}

@ARTICLE{2020PASP..132h5002S,
       author = {{Smith}, K.~W. and {Smartt}, S.~J. and {Young}, D.~R. and {Tonry}, J.~L. and {Denneau}, L. and {Flewelling}, H. and {Heinze}, A.~N. and {Weiland}, H.~J. and {Stalder}, B. and {Rest}, A. and {Stubbs}, C.~W. and {Anderson}, J.~P. and {Chen}, T.-W. and {Clark}, P. and {Do}, A. and {F{\"o}rster}, F. and {Fulton}, M. and {Gillanders}, J. and {McBrien}, O.~R. and {O'Neill}, D. and {Srivastav}, S. and {Wright}, D.~E.},
        title = "{Design and Operation of the ATLAS Transient Science Server}",
      journal = {\pasp},
         year = 2020,
        month = aug,
       volume = {132},
       number = {1014},
          eid = {085002},
        pages = {085002},
          doi = {10.1088/1538-3873/ab936e},
archivePrefix = {arXiv},
       eprint = {2003.09052},
 primaryClass = {astro-ph.IM},
       adsurl = {https://ui.adsabs.harvard.edu/abs/2020PASP..132h5002S}
}

@ARTICLE{2018PASP..130c5003B,
       author = {{Blagorodnova}, Nadejda and {Neill}, James D. and {Walters}, Richard and {Kulkarni}, Shrinivas R. and {Fremling}, Christoffer and {Ben-Ami}, Sagi and {Dekany}, Richard G. and {Fucik}, Jason R. and {Konidaris}, Nick and {Nash}, Reston and {Ngeow}, Chow-Choong and {Ofek}, Eran O. and {O' Sullivan}, Donal and {Quimby}, Robert and {Ritter}, Andreas and {Vyhmeister}, Karl E.},
        title = "{The SED Machine: A Robotic Spectrograph for Fast Transient Classification}",
      journal = {\pasp},
         year = 2018,
        month = mar,
       volume = {130},
       number = {985},
        pages = {035003},
          doi = {10.1088/1538-3873/aaa53f},
archivePrefix = {arXiv},
       eprint = {1710.02917},
 primaryClass = {astro-ph.IM},
       adsurl = {https://ui.adsabs.harvard.edu/abs/2018PASP..130c5003B}
}

@ARTICLE{2010AJ....140.1868W,
       author = {{Wright}, Edward L. and {Eisenhardt}, Peter R.~M. and {Mainzer}, Amy K. and {Ressler}, Michael E. and {Cutri}, Roc M. and {Jarrett}, Thomas and {Kirkpatrick}, J. Davy and {Padgett}, Deborah and {McMillan}, Robert S. and {Skrutskie}, Michael and {Stanford}, S.~A. and {Cohen}, Martin and {Walker}, Russell G. and {Mather}, John C. and {Leisawitz}, David and {Gautier}, III, Thomas N. and {McLean}, Ian and {Benford}, Dominic and {Lonsdale}, Carol J. and {Blain}, Andrew and {Mendez}, Bryan and {Irace}, William R. and {Duval}, Valerie and {Liu}, Fengchuan and {Royer}, Don and {Heinrichsen}, Ingolf and {Howard}, Joan and {Shannon}, Mark and {Kendall}, Martha and {Walsh}, Amy L. and {Larsen}, Mark and {Cardon}, Joel G. and {Schick}, Scott and {Schwalm}, Mark and {Abid}, Mohamed and {Fabinsky}, Beth and {Naes}, Larry and {Tsai}, Chao-Wei},
        title = "{The Wide-field Infrared Survey Explorer (WISE): Mission Description and Initial On-orbit Performance}",
      journal = {\aj},
         year = 2010,
        month = dec,
       volume = {140},
       number = {6},
        pages = {1868-1881},
          doi = {10.1088/0004-6256/140/6/1868},
archivePrefix = {arXiv},
       eprint = {1008.0031},
 primaryClass = {astro-ph.IM},
       adsurl = {https://ui.adsabs.harvard.edu/abs/2010AJ....140.1868W}
}

@ARTICLE{2019PASP..131f8003B,
       author = {{Bellm}, Eric C. and {Kulkarni}, Shrinivas R. and {Barlow}, Tom and {Feindt}, Ulrich and {Graham}, Matthew J. and {Goobar}, Ariel and {Kupfer}, Thomas and {Ngeow}, Chow-Choong and {Nugent}, Peter and {Ofek}, Eran and {Prince}, Thomas A. and {Riddle}, Reed and {Walters}, Richard and {Ye}, Quan-Zhi},
        title = "{The Zwicky Transient Facility: Surveys and Scheduler}",
      journal = {\pasp},
         year = 2019,
        month = jun,
       volume = {131},
       number = {1000},
        pages = {068003},
          doi = {10.1088/1538-3873/ab0c2a},
archivePrefix = {arXiv},
       eprint = {1905.02209},
 primaryClass = {astro-ph.IM},
       adsurl = {https://ui.adsabs.harvard.edu/abs/2019PASP..131f8003B}
}

@ARTICLE{2025A&A...695A.142P,
       author = {{Pessi}, P.~J. and {Lunnan}, R. and {Sollerman}, J. and {Schulze}, S. and {Gkini}, A. and {Gangopadhyay}, A. and {Yan}, L. and {Gal-Yam}, A. and {Perley}, D.~A. and {Chen}, T.-W. and {Hinds}, K.~R. and {Brennan}, S.~J. and {Hu}, Y. and {Singh}, A. and {Andreoni}, I. and {Cook}, D.~O. and {Fremling}, C. and {Ho}, A.~Y.~Q. and {Sharma}, Y. and {van Velzen}, S. and {Kangas}, T. and {Wold}, A. and {Bellm}, E.~C. and {Bloom}, J.~S. and {Graham}, M.~J. and {Kasliwal}, M.~M. and {Kulkarni}, S.~R. and {Riddle}, R. and {Rusholme}, B.},
        title = "{Sample of hydrogen-rich superluminous supernovae from the Zwicky Transient Facility}",
      journal = {\aap},
         year = 2025,
        month = mar,
       volume = {695},
          eid = {A142},
        pages = {A142},
          doi = {10.1051/0004-6361/202452014},
archivePrefix = {arXiv},
       eprint = {2408.15086},
 primaryClass = {astro-ph.HE},
       adsurl = {https://ui.adsabs.harvard.edu/abs/2025A&A...695A.142P}
}

@INPROCEEDINGS{1995AAS...186.4405G,
       author = {{Gunn}, James E.},
        title = "{The Sloan Digital Sky Survey}",
    booktitle = {American Astronomical Society Meeting Abstracts \#186},
         year = 1995,
       series = {American Astronomical Society Meeting Abstracts},
       volume = {186},
        month = may,
          eid = {44.05},
        pages = {44.05},
       adsurl = {https://ui.adsabs.harvard.edu/abs/1995AAS...186.4405G}
}

@ARTICLE{2016A&A...593A..68F,
       author = {{Fremling}, C. and {Sollerman}, J. and {Taddia}, F. and {Ergon}, M. and {Fraser}, M. and {Karamehmetoglu}, E. and {Valenti}, S. and {Jerkstrand}, A. and {Arcavi}, I. and {Bufano}, F. and {Elias Rosa}, N. and {Filippenko}, A.~V. and {Fox}, D. and {Gal-Yam}, A. and {Howell}, D.~A. and {Kotak}, R. and {Mazzali}, P. and {Milisavljevic}, D. and {Nugent}, P.~E. and {Nyholm}, A. and {Pian}, E. and {Smartt}, S.},
        title = "{PTF12os and iPTF13bvn. Two stripped-envelope supernovae from low-mass progenitors in NGC 5806}",
      journal = {\aap},
         year = 2016,
        month = sep,
       volume = {593},
          eid = {A68},
        pages = {A68},
          doi = {10.1051/0004-6361/201628275},
archivePrefix = {arXiv},
       eprint = {1606.03074},
 primaryClass = {astro-ph.HE},
       adsurl = {https://ui.adsabs.harvard.edu/abs/2016A&A...593A..68F}
}

@ARTICLE{2021TNSAN...7....1S,
       author = {{Shingles}, L. and {Smith}, K.~W. and {Young}, D.~R. and {Smartt}, S.~J. and {Tonry}, J. and {Denneau}, L. and {Heinze}, A. and {Weiland}, H. and {Flewelling}, H. and {Stalder}, B. and {Clocchiatti}, A. and {F{\"o}rster}, F. and {Pignata}, G. and {Rest}, A. and {Anderson}, J. and {Stubbs}, C. and {Erasmus}, N.},
        title = "{Release of the ATLAS Forced Photometry server for public use}",
      journal = {Transient Name Server AstroNote},
         year = 2021,
        month = jan,
       volume = {7},
        pages = {1-7},
       adsurl = {https://ui.adsabs.harvard.edu/abs/2021TNSAN...7....1S}
}

@software{Young_plot_atlas_fp,
    author = {Young, David R.},
    doi = {10.5281/zenodo.10978968},
    license = {GPL-3.0-only},
    title = {{plot\_atlas\_fp.py}},
    url = {https://zenodo.org/doi/10.5281/zenodo.10978968},
    year = {2020}
}

@INPROCEEDINGS{StreetTOM2024,
       author = {{Street}, R.~A. and {Lindstrom}, W. and {Heinrich-Josties}, E. and {Collom}, D. and {Riba}, A. and {Nation}, J. and {McCully}, C. and {Bowman}, M.},
        title = "{The TOM Toolkit: Power tools to enhance science and observations}",
    booktitle = {Astronomical Data Analysis Software and Systems XXXI},
         year = 2024,
       editor = {{Hugo}, B.~V. and {Van Rooyen}, R. and {Smirnov}, O.~M.},
       series = {Astronomical Society of the Pacific Conference Series},
       volume = {535},
        month = may,
        pages = {3},
       adsurl = {https://ui.adsabs.harvard.edu/abs/2024ASPC..535....3S}
}

@INPROCEEDINGS{2020past.conf..190Z,
       author = {{Zieli{\'n}ski}, Pawe{\l} and {Wyrzykowski}, {\l}ukasz and {Miko{\l}ajczyk}, Przemys{\l}aw and {Rybicki}, Krzysztof and {Ko{\l}aczkowski}, Zbigniew},
        title = "{Towards an automatic processing of CCD images with CPCS 2.0}",
    booktitle = {XXXIX Polish Astronomical Society Meeting},
         year = 2020,
       editor = {{Ma{\l}ek}, Katarzyna and {Poli{\'n}ska}, Magdalena and {Majczyna}, Agnieszka and {Stachowski}, Grzegorz and {Poleski}, Rados{\l}aw and {Wyrzykowski}, {\L}ukasz and {R{\'o}{\.z}a{\'n}ska}, Agata},
       volume = {10},
        month = oct,
        pages = {190-193},
          doi = {10.48550/arXiv.2006.05160},
archivePrefix = {arXiv},
       eprint = {2006.05160},
 primaryClass = {astro-ph.IM},
       adsurl = {https://ui.adsabs.harvard.edu/abs/2020past.conf..190Z}
}

@ARTICLE{2019CoSka..49..125Z,
       author = {{Zieli{\'n}ski}, P. and {Wyrzykowski}, {\L}. and {Rybicki}, K. and {Ko{\l}aczkowski}, Z. and {Bru{\'s}}, P. and {Miko{\l}ajczyk}, P.},
        title = "{CPCS 2.0 {\textemdash} new automatic tool for time-domain astronomy}",
      journal = {Contributions of the Astronomical Observatory Skalnate Pleso},
         year = 2019,
        month = may,
       volume = {49},
       number = {2},
        pages = {125-131},
       adsurl = {https://ui.adsabs.harvard.edu/abs/2019CoSka..49..125Z}
}

@ARTICLE{1996A&ASBertin,
       author = {{Bertin}, E. and {Arnouts}, S.},
        title = "{SExtractor: Software for source extraction.}",
      journal = {\aaps},
         year = 1996,
        month = jun,
       volume = {117},
        pages = {393-404},
          doi = {10.1051/aas:1996164},
       adsurl = {https://ui.adsabs.harvard.edu/abs/1996A&AS..117..393B}
}

@INPROCEEDINGS{2006ASPCBertin,
       author = {{Bertin}, E.},
        title = "{Automatic Astrometric and Photometric Calibration with SCAMP}",
    booktitle = {Astronomical Data Analysis Software and Systems XV},
         year = 2006,
       editor = {{Gabriel}, C. and {Arviset}, C. and {Ponz}, D. and {Enrique}, S.},
       series = {Astronomical Society of the Pacific Conference Series},
       volume = {351},
        month = jul,
        pages = {112},
       adsurl = {https://ui.adsabs.harvard.edu/abs/2006ASPC..351..112B}
}

@ARTICLE{1987PASPStetson,
       author = {{Stetson}, Peter B.},
        title = "{DAOPHOT: A Computer Program for Crowded-Field Stellar Photometry}",
      journal = {\pasp},
         year = 1987,
        month = mar,
       volume = {99},
        pages = {191},
          doi = {10.1086/131977},
       adsurl = {https://ui.adsabs.harvard.edu/abs/1987PASP...99..191S}
}

@ARTICLE{2023A&AGaiaCollaboration,
       author = {{Gaia Collaboration} and {Montegriffo}, P. and {Bellazzini}, M. and {De Angeli}, F. and {Andrae}, R. and {Barstow}, M.~A. and {Bossini}, D. and {Bragaglia}, A. and {Burgess}, P.~W. and {Cacciari}, C. and {Carrasco}, J.~M. and {Chornay}, N. and {Delchambre}, L. and {Evans}, D.~W. and {Fouesneau}, M. and {Fr{\'e}mat}, Y. and {Garabato}, D. and {Jordi}, C. and {Manteiga}, M. and {Massari}, D. and {Palaversa}, L. and {Pancino}, E. and {Riello}, M. and {Ruz Mieres}, D. and {Sanna}, N. and {Santove{\~n}a}, R. and {Sordo}, R. and {Vallenari}, A. and {Walton}, N.~A. and {Brown}, A.~G.~A. and {Prusti}, T. and {de Bruijne}, J.~H.~J. and {Arenou}, F. and {Babusiaux}, C. and {Biermann}, M. and {Creevey}, O.~L. and {Ducourant}, C. and {Eyer}, L. and {Guerra}, R. and {Hutton}, A. and {Klioner}, S.~A. and {Lammers}, U.~L. and {Lindegren}, L. and {Luri}, X. and {Mignard}, F. and {Panem}, C. and {Pourbaix}, D. and {Randich}, S. and {Sartoretti}, P. and {Soubiran}, C. and {Tanga}, P. and {Bailer-Jones}, C.~A.~L. and {Bastian}, U. and {Drimmel}, R. and {Jansen}, F. and {Katz}, D. and {Lattanzi}, M.~G. and {van Leeuwen}, F. and {Bakker}, J. and {Casta{\~n}eda}, J. and {Fabricius}, C. and {Galluccio}, L. and {Guerrier}, A. and {Heiter}, U. and {Masana}, E. and {Messineo}, R. and {Mowlavi}, N. and {Nicolas}, C. and {Nienartowicz}, K. and {Pailler}, F. and {Panuzzo}, P. and {Riclet}, F. and {Roux}, W. and {Seabroke}, G.~M. and {Th{\'e}venin}, F. and {Gracia-Abril}, G. and {Portell}, J. and {Teyssier}, D. and {Altmann}, M. and {Audard}, M. and {Bellas-Velidis}, I. and {Benson}, K. and {Berthier}, J. and {Blomme}, R. and {Busonero}, D. and {Busso}, G. and {C{\'a}novas}, H. and {Carry}, B. and {Cellino}, A. and {Cheek}, N. and {Clementini}, G. and {Damerdji}, Y. and {Davidson}, M. and {de Teodoro}, P. and {Nu{\~n}ez Campos}, M. and {Dell'Oro}, A. and {Esquej}, P. and {Fern{\'a}ndez-Hern{\'a}ndez}, J. and {Fraile}, E. and {Garc{\'\i}a-Lario}, P. and {Gosset}, E. and {Haigron}, R. and {Halbwachs}, J.-L. and {Hambly}, N.~C. and {Harrison}, D.~L. and {Hern{\'a}ndez}, J. and {Hestroffer}, D. and {Hodgkin}, S.~T. and {Holl}, B. and {Jan{\ss}en}, K. and {Jevardat de Fombelle}, G. and {Jordan}, S. and {Krone-Martins}, A. and {Lanzafame}, A.~C. and {L{\"o}ffler}, W. and {Marchal}, O. and {Marrese}, P.~M. and {Moitinho}, A. and {Muinonen}, K. and {Osborne}, P. and {Pauwels}, T. and {Recio-Blanco}, A. and {Reyl{\'e}}, C. and {Rimoldini}, L. and {Roegiers}, T. and {Rybizki}, J. and {Sarro}, L.~M. and {Siopis}, C. and {Smith}, M. and {Sozzetti}, A. and {Utrilla}, E. and {van Leeuwen}, M. and {Abbas}, U. and {{\'A}brah{\'a}m}, P. and {Abreu Aramburu}, A. and {Aerts}, C. and {Aguado}, J.~J. and {Ajaj}, M. and {Aldea-Montero}, F. and {Altavilla}, G. and {{\'A}lvarez}, M.~A. and {Alves}, J. and {Anderson}, R.~I. and {Anglada Varela}, E. and {Antoja}, T. and {Baines}, D. and {Baker}, S.~G. and {Balaguer-N{\'u}{\~n}ez}, L. and {Balbinot}, E. and {Balog}, Z. and {Barache}, C. and {Barbato}, D. and {Barros}, M. and {Bartolom{\'e}}, S. and {Bassilana}, J.-L. and {Bauchet}, N. and {Becciani}, U. and {Berihuete}, A. and {Bernet}, M. and {Bertone}, S. and {Bianchi}, L. and {Binnenfeld}, A. and {Blanco-Cuaresma}, S. and {Boch}, T. and {Bombrun}, A. and {Bouquillon}, S. and {Bramante}, L. and {Breedt}, E. and {Bressan}, A. and {Brouillet}, N. and {Brugaletta}, E. and {Bucciarelli}, B. and {Burlacu}, A. and {Butkevich}, A.~G. and {Buzzi}, R. and {Caffau}, E. and {Cancelliere}, R. and {Cantat-Gaudin}, T. and {Carballo}, R. and {Carlucci}, T. and {Carnerero}, M.~I. and {Casamiquela}, L. and {Castellani}, M. and {Castro-Ginard}, A. and {Chaoul}, L. and {Charlot}, P. and {Chemin}, L. and {Chiaramida}, V. and {Chiavassa}, A. and {Comoretto}, G. and {Contursi}, G. and {Cooper}, W.~J. and {Cornez}, T. and {Cowell}, S. and {Crifo}, F. and {Cropper}, M. and {Crosta}, M. and {Crowley}, C. and {Dafonte}, C. and {Dapergolas}, A.},
        title = "{Gaia Data Release 3. The Galaxy in your preferred colours: Synthetic photometry from Gaia low-resolution spectra}",
      journal = {\aap},
         year = 2023,
        month = jun,
       volume = {674},
          eid = {A33},
        pages = {A33},
          doi = {10.1051/0004-6361/202243709},
archivePrefix = {arXiv},
       eprint = {2206.06215},
 primaryClass = {astro-ph.SR},
       adsurl = {https://ui.adsabs.harvard.edu/abs/2023A&A...674A..33G}
}

@ARTICLE{2015AJZacharias,
       author = {{Zacharias}, N. and {Finch}, C. and {Subasavage}, J. and {Bredthauer}, G. and {Crockett}, C. and {Divittorio}, M. and {Ferguson}, E. and {Harris}, F. and {Harris}, H. and {Henden}, A. and {Kilian}, C. and {Munn}, J. and {Rafferty}, T. and {Rhodes}, A. and {Schultheiss}, M. and {Tilleman}, T. and {Wieder}, G.},
        title = "{The First U.S. Naval Observatory Robotic Astrometric Telescope Catalog}",
      journal = {\aj},
         year = 2015,
        month = oct,
       volume = {150},
       number = {4},
          eid = {101},
        pages = {101},
          doi = {10.1088/0004-6256/150/4/101},
archivePrefix = {arXiv},
       eprint = {1508.04637},
 primaryClass = {astro-ph.IM},
       adsurl = {https://ui.adsabs.harvard.edu/abs/2015AJ....150..101Z}
}

@ARTICLE{2013AJZacharias,
       author = {{Zacharias}, N. and {Finch}, C.~T. and {Girard}, T.~M. and {Henden}, A. and {Bartlett}, J.~L. and {Monet}, D.~G. and {Zacharias}, M.~I.},
        title = "{The Fourth US Naval Observatory CCD Astrograph Catalog (UCAC4)}",
      journal = {\aj},
         year = 2013,
        month = feb,
       volume = {145},
       number = {2},
          eid = {44},
        pages = {44},
          doi = {10.1088/0004-6256/145/2/44},
archivePrefix = {arXiv},
       eprint = {1212.6182},
 primaryClass = {astro-ph.IM},
       adsurl = {https://ui.adsabs.harvard.edu/abs/2013AJ....145...44Z}
}

@ARTICLE{2003AJMonet,
       author = {{Monet}, David G. and {Levine}, Stephen E. and {Canzian}, Blaise and {Ables}, Harold D. and {Bird}, Alan R. and {Dahn}, Conard C. and {Guetter}, Harry H. and {Harris}, Hugh C. and {Henden}, Arne A. and {Leggett}, Sandy K. and {Levison}, Harold F. and {Luginbuhl}, Christian B. and {Martini}, Joan and {Monet}, Alice K.~B. and {Munn}, Jeffrey A. and {Pier}, Jeffrey R. and {Rhodes}, Albert R. and {Riepe}, Betty and {Sell}, Stephen and {Stone}, Ronald C. and {Vrba}, Frederick J. and {Walker}, Richard L. and {Westerhout}, Gart and {Brucato}, Robert J. and {Reid}, I. Neill and {Schoening}, William and {Hartley}, M. and {Read}, M.~A. and {Tritton}, S.~B.},
        title = "{The USNO-B Catalog}",
      journal = {\aj},
         year = 2003,
        month = feb,
       volume = {125},
       number = {2},
        pages = {984-993},
          doi = {10.1086/345888},
archivePrefix = {arXiv},
       eprint = {astro-ph/0210694},
 primaryClass = {astro-ph},
       adsurl = {https://ui.adsabs.harvard.edu/abs/2003AJ....125..984M}
}

@ARTICLE{2023A&AGaiaDR3,
       author = {{Gaia Collaboration} and {Vallenari}, A. and {Brown}, A.~G.~A. and {Prusti}, T. and {de Bruijne}, J.~H.~J. and {Arenou}, F. and {Babusiaux}, C. and {Biermann}, M. and {Creevey}, O.~L. and {Ducourant}, C. and {Evans}, D.~W. and {Eyer}, L. and {Guerra}, R. and {Hutton}, A. and {Jordi}, C. and {Klioner}, S.~A. and {Lammers}, U.~L. and {Lindegren}, L. and {Luri}, X. and {Mignard}, F. and {Panem}, C. and {Pourbaix}, D. and {Randich}, S. and {Sartoretti}, P. and {Soubiran}, C. and {Tanga}, P. and {Walton}, N.~A. and {Bailer-Jones}, C.~A.~L. and {Bastian}, U. and {Drimmel}, R. and {Jansen}, F. and {Katz}, D. and {Lattanzi}, M.~G. and {van Leeuwen}, F. and {Bakker}, J. and {Cacciari}, C. and {Casta{\~n}eda}, J. and {De Angeli}, F. and {Fabricius}, C. and {Fouesneau}, M. and {Fr{\'e}mat}, Y. and {Galluccio}, L. and {Guerrier}, A. and {Heiter}, U. and {Masana}, E. and {Messineo}, R. and {Mowlavi}, N. and {Nicolas}, C. and {Nienartowicz}, K. and {Pailler}, F. and {Panuzzo}, P. and {Riclet}, F. and {Roux}, W. and {Seabroke}, G.~M. and {Sordo}, R. and {Th{\'e}venin}, F. and {Gracia-Abril}, G. and {Portell}, J. and {Teyssier}, D. and {Altmann}, M. and {Andrae}, R. and {Audard}, M. and {Bellas-Velidis}, I. and {Benson}, K. and {Berthier}, J. and {Blomme}, R. and {Burgess}, P.~W. and {Busonero}, D. and {Busso}, G. and {C{\'a}novas}, H. and {Carry}, B. and {Cellino}, A. and {Cheek}, N. and {Clementini}, G. and {Damerdji}, Y. and {Davidson}, M. and {de Teodoro}, P. and {Nu{\~n}ez Campos}, M. and {Delchambre}, L. and {Dell'Oro}, A. and {Esquej}, P. and {Fern{\'a}ndez-Hern{\'a}ndez}, J. and {Fraile}, E. and {Garabato}, D. and {Garc{\'\i}a-Lario}, P. and {Gosset}, E. and {Haigron}, R. and {Halbwachs}, J.-L. and {Hambly}, N.~C. and {Harrison}, D.~L. and {Hern{\'a}ndez}, J. and {Hestroffer}, D. and {Hodgkin}, S.~T. and {Holl}, B. and {Jan{\ss}en}, K. and {Jevardat de Fombelle}, G. and {Jordan}, S. and {Krone-Martins}, A. and {Lanzafame}, A.~C. and {L{\"o}ffler}, W. and {Marchal}, O. and {Marrese}, P.~M. and {Moitinho}, A. and {Muinonen}, K. and {Osborne}, P. and {Pancino}, E. and {Pauwels}, T. and {Recio-Blanco}, A. and {Reyl{\'e}}, C. and {Riello}, M. and {Rimoldini}, L. and {Roegiers}, T. and {Rybizki}, J. and {Sarro}, L.~M. and {Siopis}, C. and {Smith}, M. and {Sozzetti}, A. and {Utrilla}, E. and {van Leeuwen}, M. and {Abbas}, U. and {{\'A}brah{\'a}m}, P. and {Abreu Aramburu}, A. and {Aerts}, C. and {Aguado}, J.~J. and {Ajaj}, M. and {Aldea-Montero}, F. and {Altavilla}, G. and {{\'A}lvarez}, M.~A. and {Alves}, J. and {Anders}, F. and {Anderson}, R.~I. and {Anglada Varela}, E. and {Antoja}, T. and {Baines}, D. and {Baker}, S.~G. and {Balaguer-N{\'u}{\~n}ez}, L. and {Balbinot}, E. and {Balog}, Z. and {Barache}, C. and {Barbato}, D. and {Barros}, M. and {Barstow}, M.~A. and {Bartolom{\'e}}, S. and {Bassilana}, J.-L. and {Bauchet}, N. and {Becciani}, U. and {Bellazzini}, M. and {Berihuete}, A. and {Bernet}, M. and {Bertone}, S. and {Bianchi}, L. and {Binnenfeld}, A. and {Blanco-Cuaresma}, S. and {Blazere}, A. and {Boch}, T. and {Bombrun}, A. and {Bossini}, D. and {Bouquillon}, S. and {Bragaglia}, A. and {Bramante}, L. and {Breedt}, E. and {Bressan}, A. and {Brouillet}, N. and {Brugaletta}, E. and {Bucciarelli}, B. and {Burlacu}, A. and {Butkevich}, A.~G. and {Buzzi}, R. and {Caffau}, E. and {Cancelliere}, R. and {Cantat-Gaudin}, T. and {Carballo}, R. and {Carlucci}, T. and {Carnerero}, M.~I. and {Carrasco}, J.~M. and {Casamiquela}, L. and {Castellani}, M. and {Castro-Ginard}, A. and {Chaoul}, L. and {Charlot}, P. and {Chemin}, L. and {Chiaramida}, V. and {Chiavassa}, A. and {Chornay}, N. and {Comoretto}, G. and {Contursi}, G. and {Cooper}, W.~J. and {Cornez}, T. and {Cowell}, S. and {Crifo}, F. and {Cropper}, M. and {Crosta}, M. and {Crowley}, C. and {Dafonte}, C. and {Dapergolas}, A. and {David}, M. and {David}, P. and {de Laverny}, P. and {De Luise}, F. and {De March}, R.},
        title = "{Gaia Data Release 3. Summary of the content and survey properties}",
      journal = {\aap},
         year = 2023,
        month = jun,
       volume = {674},
          eid = {A1},
        pages = {A1},
          doi = {10.1051/0004-6361/202243940},
archivePrefix = {arXiv},
       eprint = {2208.00211},
 primaryClass = {astro-ph.GA},
       adsurl = {https://ui.adsabs.harvard.edu/abs/2023A&A...674A...1G}
}

@ARTICLE{2006AJSkrutskie,
       author = {{Skrutskie}, M.~F. and {Cutri}, R.~M. and {Stiening}, R. and {Weinberg}, M.~D. and {Schneider}, S. and {Carpenter}, J.~M. and {Beichman}, C. and {Capps}, R. and {Chester}, T. and {Elias}, J. and {Huchra}, J. and {Liebert}, J. and {Lonsdale}, C. and {Monet}, D.~G. and {Price}, S. and {Seitzer}, P. and {Jarrett}, T. and {Kirkpatrick}, J.~D. and {Gizis}, J.~E. and {Howard}, E. and {Evans}, T. and {Fowler}, J. and {Fullmer}, L. and {Hurt}, R. and {Light}, R. and {Kopan}, E.~L. and {Marsh}, K.~A. and {McCallon}, H.~L. and {Tam}, R. and {Van Dyk}, S. and {Wheelock}, S.},
        title = "{The Two Micron All Sky Survey (2MASS)}",
      journal = {\aj},
         year = 2006,
        month = feb,
       volume = {131},
       number = {2},
        pages = {1163-1183},
          doi = {10.1086/498708},
       adsurl = {https://ui.adsabs.harvard.edu/abs/2006AJ....131.1163S}
}

@ARTICLE{2012PASP..124..140B,
       author = {{Bessell}, Michael and {Murphy}, Simon},
        title = "{Spectrophotometric Libraries, Revised Photonic Passbands, and Zero Points for UBVRI, Hipparcos, and Tycho Photometry}",
      journal = {\pasp},
         year = 2012,
        month = feb,
       volume = {124},
       number = {912},
        pages = {140},
          doi = {10.1086/664083},
archivePrefix = {arXiv},
       eprint = {1112.2698},
 primaryClass = {astro-ph.SR},
       adsurl = {https://ui.adsabs.harvard.edu/abs/2012PASP..124..140B}
}

@ARTICLE{2004ApJ...611.1005G,
       author = {{Gehrels}, N. and {Chincarini}, G. and {Giommi}, P. and {Mason}, K.~O. and {Nousek}, J.~A. and {Wells}, A.~A. and {White}, N.~E. and {Barthelmy}, S.~D. and {Burrows}, D.~N. and {Cominsky}, L.~R. and {Hurley}, K.~C. and {Marshall}, F.~E. and {M{\'e}sz{\'a}ros}, P. and {Roming}, P.~W.~A. and {Angelini}, L. and {Barbier}, L.~M. and {Belloni}, T. and {Campana}, S. and {Caraveo}, P.~A. and {Chester}, M.~M. and {Citterio}, O. and {Cline}, T.~L. and {Cropper}, M.~S. and {Cummings}, J.~R. and {Dean}, A.~J. and {Feigelson}, E.~D. and {Fenimore}, E.~E. and {Frail}, D.~A. and {Fruchter}, A.~S. and {Garmire}, G.~P. and {Gendreau}, K. and {Ghisellini}, G. and {Greiner}, J. and {Hill}, J.~E. and {Hunsberger}, S.~D. and {Krimm}, H.~A. and {Kulkarni}, S.~R. and {Kumar}, P. and {Lebrun}, F. and {Lloyd-Ronning}, N.~M. and {Markwardt}, C.~B. and {Mattson}, B.~J. and {Mushotzky}, R.~F. and {Norris}, J.~P. and {Osborne}, J. and {Paczynski}, B. and {Palmer}, D.~M. and {Park}, H.-S. and {Parsons}, A.~M. and {Paul}, J. and {Rees}, M.~J. and {Reynolds}, C.~S. and {Rhoads}, J.~E. and {Sasseen}, T.~P. and {Schaefer}, B.~E. and {Short}, A.~T. and {Smale}, A.~P. and {Smith}, I.~A. and {Stella}, L. and {Tagliaferri}, G. and {Takahashi}, T. and {Tashiro}, M. and {Townsley}, L.~K. and {Tueller}, J. and {Turner}, M.~J.~L. and {Vietri}, M. and {Voges}, W. and {Ward}, M.~J. and {Willingale}, R. and {Zerbi}, F.~M. and {Zhang}, W.~W.},
        title = "{The Swift Gamma-Ray Burst Mission}",
      journal = {\apj},
         year = 2004,
        month = aug,
       volume = {611},
       number = {2},
        pages = {1005-1020},
          doi = {10.1086/422091},
archivePrefix = {arXiv},
       eprint = {astro-ph/0405233},
 primaryClass = {astro-ph},
       adsurl = {https://ui.adsabs.harvard.edu/abs/2004ApJ...611.1005G}
}

@ARTICLE{2005SSRv..120...95R,
       author = {{Roming}, Peter W.~A. and {Kennedy}, Thomas E. and {Mason}, Keith O. and {Nousek}, John A. and {Ahr}, Lindy and {Bingham}, Richard E. and {Broos}, Patrick S. and {Carter}, Mary J. and {Hancock}, Barry K. and {Huckle}, Howard E. and {Hunsberger}, S.~D. and {Kawakami}, Hajime and {Killough}, Ronnie and {Koch}, T. Scott and {McLelland}, Michael K. and {Smith}, Kelly and {Smith}, Philip J. and {Soto}, Juan Carlos and {Boyd}, Patricia T. and {Breeveld}, Alice A. and {Holland}, Stephen T. and {Ivanushkina}, Mariya and {Pryzby}, Michael S. and {Still}, Martin D. and {Stock}, Joseph},
        title = "{The Swift Ultra-Violet/Optical Telescope}",
      journal = {\ssr},
         year = 2005,
        month = oct,
       volume = {120},
       number = {3-4},
        pages = {95-142},
          doi = {10.1007/s11214-005-5095-4},
archivePrefix = {arXiv},
       eprint = {astro-ph/0507413},
 primaryClass = {astro-ph},
       adsurl = {https://ui.adsabs.harvard.edu/abs/2005SSRv..120...95R}
}

@ARTICLE{2019PASP..131g5004F,
       author = {{Fabricant}, Daniel and {Fata}, Robert and {Epps}, Harland and {Gauron}, Thomas and {Mueller}, Mark and {Zajac}, Joseph and {Amato}, Stephen and {Barberis}, Jack and {Bergner}, Henry and {Brennan}, Patricia and {Brown}, Warren and {Chilingarian}, Igor and {Geary}, John and {Kradinov}, Vladimir and {McLeod}, Brian and {Smith}, Matthew and {Woods}, Deborah},
        title = "{Binospec: A Wide-field Imaging Spectrograph for the MMT}",
      journal = {\pasp},
         year = 2019,
        month = jul,
       volume = {131},
       number = {1001},
        pages = {075004},
          doi = {10.1088/1538-3873/ab1d78},
archivePrefix = {arXiv},
       eprint = {1905.03320},
 primaryClass = {astro-ph.IM},
       adsurl = {https://ui.adsabs.harvard.edu/abs/2019PASP..131g5004F}
}

@MISC{2012ivoa.rept.1015R,
       author = {{Rodrigo}, Carlos and {Solano}, Enrique and {Bayo}, Amelia},
        title = "{SVO Filter Profile Service Version 1.0}",
 howpublished = {IVOA Working Draft 15 October 2012},
         year = 2012,
        month = oct,
        pages = {1015},
          doi = {10.5479/ADS/bib/2012ivoa.rept.1015R},
       adsurl = {https://ui.adsabs.harvard.edu/abs/2012ivoa.rept.1015R}
}

@INPROCEEDINGS{2020sea..confE.182R,
       author = {{Rodrigo}, C. and {Solano}, E.},
        title = "{The SVO Filter Profile Service}",
    booktitle = {XIV.0 Scientific Meeting (virtual) of the Spanish Astronomical Society},
         year = 2020,
        month = jul,
          eid = {182},
        pages = {182},
       adsurl = {https://ui.adsabs.harvard.edu/abs/2020sea..confE.182R}
}

@ARTICLE{2019JOSS....4.1247V,
       author = {{van der Walt}, St{\'e}fan and {Crellin-Quick}, Arien and {Bloom}, Joshua},
        title = "{SkyPortal: An Astronomical Data Platform}",
      journal = {The Journal of Open Source Software},
         year = 2019,
        month = may,
       volume = {4},
       number = {37},
          eid = {1247},
        pages = {1247},
          doi = {10.21105/joss.01247},
       adsurl = {https://ui.adsabs.harvard.edu/abs/2019JOSS....4.1247V}
}

@ARTICLE{2023ApJS..267...31C,
       author = {{Coughlin}, Michael W. and {Bloom}, Joshua S. and {Nir}, Guy and {Antier}, Sarah and {du Laz}, Theophile Jegou and {van der Walt}, St{\'e}fan and {Crellin-Quick}, Arien and {Culino}, Thomas and {Duev}, Dmitry A. and {Goldstein}, Daniel A. and {Healy}, Brian F. and {Karambelkar}, Viraj and {Lilleboe}, Jada and {Shin}, Kyung Min and {Singer}, Leo P. and {Ahumada}, Tom{\'a}s and {Anand}, Shreya and {Bellm}, Eric C. and {Dekany}, Richard and {Graham}, Matthew J. and {Kasliwal}, Mansi M. and {Kostadinova}, Ivona and {Kiendrebeogo}, R. Weizmann and {Kulkarni}, Shrinivas R. and {Jenkins}, Sydney and {LeBaron}, Natalie and {Mahabal}, Ashish A. and {Neill}, James D. and {Parazin}, B. and {Peloton}, Julien and {Perley}, Daniel A. and {Riddle}, Reed and {Rusholme}, Ben and {van Santen}, Jakob and {Sollerman}, Jesper and {Stein}, Robert and {Turpin}, D. and {Wold}, Avery and {Amat}, Carla and {Bonnefon}, Adrien and {Bonnefoy}, Adrien and {Flament}, Manon and {Kerkow}, Frank and {Kishore}, Sulekha and {Jani}, Shloke and {Mahanty}, Stephen K. and {Liu}, C{\'e}line and {Llinares}, Laura and {Makarison}, Jolyane and {Olli{\'e}ric}, Alix and {Perez}, In{\`e}s and {Pont}, Lydie and {Sharma}, Vyom},
        title = "{A Data Science Platform to Enable Time-domain Astronomy}",
      journal = {\apjs},
         year = 2023,
        month = aug,
       volume = {267},
       number = {2},
          eid = {31},
        pages = {31},
          doi = {10.3847/1538-4365/acdee1},
archivePrefix = {arXiv},
       eprint = {2305.00108},
 primaryClass = {astro-ph.IM},
       adsurl = {https://ui.adsabs.harvard.edu/abs/2023ApJS..267...31C}
}

@ARTICLE{2019A&A...627A.115R,
       author = {{Rigault}, M. and {Neill}, J.~D. and {Blagorodnova}, N. and {Dugas}, A. and {Feeney}, M. and {Walters}, R. and {Brinnel}, V. and {Copin}, Y. and {Fremling}, C. and {Nordin}, J. and {Sollerman}, J.},
        title = "{Fully automated integral field spectrograph pipeline for the SEDMachine: pysedm}",
      journal = {\aap},
         year = 2019,
        month = jul,
       volume = {627},
          eid = {A115},
        pages = {A115},
          doi = {10.1051/0004-6361/201935344},
archivePrefix = {arXiv},
       eprint = {1902.08526},
 primaryClass = {astro-ph.IM},
       adsurl = {https://ui.adsabs.harvard.edu/abs/2019A&A...627A.115R}
}

@ARTICLE{2022PASP..134b4505K,
       author = {{Kim}, Y.-L. and {Rigault}, M. and {Neill}, J.~D. and {Briday}, M. and {Copin}, Y. and {Lezmy}, J. and {Nicolas}, N. and {Riddle}, R. and {Sharma}, Y. and {Smith}, M. and {Sollerman}, J. and {Walters}, R.},
        title = "{New Modules for the SEDMachine to Remove Contaminations from Cosmic Rays and Non-target Light: BYECR and CONTSEP}",
      journal = {\pasp},
         year = 2022,
        month = feb,
       volume = {134},
       number = {1032},
          eid = {024505},
        pages = {024505},
          doi = {10.1088/1538-3873/ac50a0},
archivePrefix = {arXiv},
       eprint = {2203.01346},
 primaryClass = {astro-ph.IM},
       adsurl = {https://ui.adsabs.harvard.edu/abs/2022PASP..134b4505K}
}

@ARTICLE{2020JOSS....5.2308P,
       author = {{Prochaska}, J. and {Hennawi}, Joseph and {Westfall}, Kyle and {Cooke}, Ryan and {Wang}, Feige and {Hsyu}, Tiffany and {Davies}, Frederick and {Farina}, Emanuele and {Pelliccia}, Debora},
        title = "{PypeIt: The Python Spectroscopic Data Reduction Pipeline}",
      journal = {The Journal of Open Source Software},
         year = 2020,
        month = dec,
       volume = {5},
       number = {56},
          eid = {2308},
        pages = {2308},
          doi = {10.21105/joss.02308},
archivePrefix = {arXiv},
       eprint = {2005.06505},
 primaryClass = {astro-ph.IM},
       adsurl = {https://ui.adsabs.harvard.edu/abs/2020JOSS....5.2308P}
}

@BOOK{2006gpml.book.....R,
       author = {{Rasmussen}, Carl Edward and {Williams}, Christopher K.~I.},
        title = "{Gaussian Processes for Machine Learning}",
         year = 2006,
       adsurl = {https://ui.adsabs.harvard.edu/abs/2006gpml.book.....R}
}

@ARTICLE{2023ApJ...943...41C,
       author = {{Chen}, Z.~H. and {Yan}, Lin and {Kangas}, T. and {Lunnan}, R. and {Schulze}, S. and {Sollerman}, J. and {Perley}, D.~A. and {Chen}, T. -W. and {Taggart}, K. and {Hinds}, K.~R. and {Gal-Yam}, A. and {Wang}, X.~F. and {Andreoni}, I. and {Bellm}, E. and {Bloom}, J.~S. and {Burdge}, K. and {Burgos}, A. and {Cook}, D. and {Dahiwale}, A. and {De}, K. and {Dekany}, R. and {Dugas}, A. and {Frederik}, S. and {Fremling}, C. and {Graham}, M. and {Hankins}, M. and {Ho}, A. and {Jencson}, J. and {Karambelkar}, V. and {Kasliwal}, M. and {Kulkarni}, S. and {Laher}, R. and {Rusholme}, B. and {Sharma}, Y. and {Taddia}, F. and {Tartaglia}, L. and {Thomas}, B.~P. and {Tzanidakis}, A. and {Van Roestel}, J. and {Walter}, R. and {Yang}, Y. and {Yao}, Y.~H. and {Yaron}, O.},
        title = "{The Hydrogen-poor Superluminous Supernovae from the Zwicky Transient Facility Phase I Survey. I. Light Curves and Measurements}",
      journal = {\apj},
         year = 2023,
        month = jan,
       volume = {943},
       number = {1},
          eid = {41},
        pages = {41},
          doi = {10.3847/1538-4357/aca161},
archivePrefix = {arXiv},
       eprint = {2202.02059},
 primaryClass = {astro-ph.HE},
       adsurl = {https://ui.adsabs.harvard.edu/abs/2023ApJ...943...41C}
}

@ARTICLE{2019MNRAS.483.5459R,
       author = {{Rodr{\'\i}guez}, {\'O}. and {Pignata}, G. and {Hamuy}, M. and {Clocchiatti}, A. and {Phillips}, M.~M. and {Krisciunas}, K. and {Morrell}, N.~I. and {Folatelli}, G. and {Roth}, M. and {Castell{\'o}n}, S. and {Jang}, I.~S. and {Apostolovski}, Y. and {L{\'o}pez}, P. and {Marchi}, S. and {Ram{\'\i}rez}, R. and {S{\'a}nchez}, P.},
        title = "{Type II supernovae as distance indicators at near-IR wavelengths}",
      journal = {\mnras},
         year = 2019,
        month = mar,
       volume = {483},
       number = {4},
        pages = {5459-5479},
          doi = {10.1093/mnras/sty3396},
archivePrefix = {arXiv},
       eprint = {1812.04982},
 primaryClass = {astro-ph.CO},
       adsurl = {https://ui.adsabs.harvard.edu/abs/2019MNRAS.483.5459R}
}

@article{astropy:2013,
        Adsurl = {https://adsabs.harvard.edu/abs/2013A%26A...558A..33A},
        Archiveprefix = {arXiv},
        Author = {{Astropy Collaboration} and {Robitaille}, T.~P. and {Tollerud}, E.~J. and {Greenfield}, P. and {Droettboom}, M. and {Bray}, E. and {Aldcroft}, T. and {Davis}, M. and {Ginsburg}, A. and {Price-Whelan}, A.~M. and {Kerzendorf}, W.~E. and {Conley}, A. and {Crighton}, N. and {Barbary}, K. and {Muna}, D. and {Ferguson}, H. and {Grollier}, F. and {Parikh}, M.~M. and {Nair}, P.~H. and {Unther}, H.~M. and {Deil}, C. and {Woillez}, J. and {Conseil}, S. and {Kramer}, R. and {Turner}, J.~E.~H. and {Singer}, L. and {Fox}, R. and {Weaver}, B.~A. and {Zabalza}, V. and {Edwards}, Z.~I. and {Azalee Bostroem}, K. and {Burke}, D.~J. and {Casey}, A.~R. and {Crawford}, S.~M. and {Dencheva}, N. and {Ely}, J. and {Jenness}, T. and {Labrie}, K. and {Lim}, P.~L. and {Pierfederici}, F. and {Pontzen}, A. and {Ptak}, A. and {Refsdal}, B. and {Servillat}, M. and {Streicher}, O.},
        Doi = {10.1051/0004-6361/201322068},
        Eid = {A33},
        Eprint = {1307.6212},
        Journal = {\aap},
        Month = oct,
        Pages = {A33},
        Primaryclass = {astro-ph.IM},
        Title = {{Astropy: A community Python package for astronomy}},
        Volume = 558,
        Year = 2013}

@ARTICLE{astropy:2018,
               author = {{Astropy Collaboration} and {Price-Whelan}, A.~M. and
                 {Sip{\H{o}}cz}, B.~M. and {G{\"u}nther}, H.~M. and {Lim}, P.~L. and
                 {Crawford}, S.~M. and {Conseil}, S. and {Shupe}, D.~L. and
                 {Craig}, M.~W. and {Dencheva}, N. and {Ginsburg}, A. and {Vand
                erPlas}, J.~T. and {Bradley}, L.~D. and {P{\'e}rez-Su{\'a}rez}, D. and
                 {de Val-Borro}, M. and {Aldcroft}, T.~L. and {Cruz}, K.~L. and
                 {Robitaille}, T.~P. and {Tollerud}, E.~J. and {Ardelean}, C. and
                 {Babej}, T. and {Bach}, Y.~P. and {Bachetti}, M. and {Bakanov}, A.~V. and
                 {Bamford}, S.~P. and {Barentsen}, G. and {Barmby}, P. and
                 {Baumbach}, A. and {Berry}, K.~L. and {Biscani}, F. and {Boquien}, M. and
                 {Bostroem}, K.~A. and {Bouma}, L.~G. and {Brammer}, G.~B. and
                 {Bray}, E.~M. and {Breytenbach}, H. and {Buddelmeijer}, H. and
                 {Burke}, D.~J. and {Calderone}, G. and {Cano Rodr{\'\i}guez}, J.~L. and
                 {Cara}, M. and {Cardoso}, J.~V.~M. and {Cheedella}, S. and {Copin}, Y. and
                 {Corrales}, L. and {Crichton}, D. and {D'Avella}, D. and {Deil}, C. and
                 {Depagne}, {\'E}. and {Dietrich}, J.~P. and {Donath}, A. and
                 {Droettboom}, M. and {Earl}, N. and {Erben}, T. and {Fabbro}, S. and
                 {Ferreira}, L.~A. and {Finethy}, T. and {Fox}, R.~T. and
                 {Garrison}, L.~H. and {Gibbons}, S.~L.~J. and {Goldstein}, D.~A. and
                 {Gommers}, R. and {Greco}, J.~P. and {Greenfield}, P. and
                 {Groener}, A.~M. and {Grollier}, F. and {Hagen}, A. and {Hirst}, P. and
                 {Homeier}, D. and {Horton}, A.~J. and {Hosseinzadeh}, G. and {Hu}, L. and
                 {Hunkeler}, J.~S. and {Ivezi{\'c}}, {\v{Z}}. and {Jain}, A. and
                 {Jenness}, T. and {Kanarek}, G. and {Kendrew}, S. and {Kern}, N.~S. and
                 {Kerzendorf}, W.~E. and {Khvalko}, A. and {King}, J. and {Kirkby}, D. and
                 {Kulkarni}, A.~M. and {Kumar}, A. and {Lee}, A. and {Lenz}, D. and
                 {Littlefair}, S.~P. and {Ma}, Z. and {Macleod}, D.~M. and
                 {Mastropietro}, M. and {McCully}, C. and {Montagnac}, S. and
                 {Morris}, B.~M. and {Mueller}, M. and {Mumford}, S.~J. and {Muna}, D. and
                 {Murphy}, N.~A. and {Nelson}, S. and {Nguyen}, G.~H. and
                 {Ninan}, J.~P. and {N{\"o}the}, M. and {Ogaz}, S. and {Oh}, S. and
                 {Parejko}, J.~K. and {Parley}, N. and {Pascual}, S. and {Patil}, R. and
                 {Patil}, A.~A. and {Plunkett}, A.~L. and {Prochaska}, J.~X. and
                 {Rastogi}, T. and {Reddy Janga}, V. and {Sabater}, J. and
                 {Sakurikar}, P. and {Seifert}, M. and {Sherbert}, L.~E. and
                 {Sherwood-Taylor}, H. and {Shih}, A.~Y. and {Sick}, J. and
                 {Silbiger}, M.~T. and {Singanamalla}, S. and {Singer}, L.~P. and
                 {Sladen}, P.~H. and {Sooley}, K.~A. and {Sornarajah}, S. and
                 {Streicher}, O. and {Teuben}, P. and {Thomas}, S.~W. and
                 {Tremblay}, G.~R. and {Turner}, J.~E.~H. and {Terr{\'o}n}, V. and
                 {van Kerkwijk}, M.~H. and {de la Vega}, A. and {Watkins}, L.~L. and
                 {Weaver}, B.~A. and {Whitmore}, J.~B. and {Woillez}, J. and
                 {Zabalza}, V. and {Astropy Contributors}},
                title = "{The Astropy Project: Building an Open-science Project and Status of the v2.0 Core Package}",
              journal = {\aj},
                 year = 2018,
                month = sep,
               volume = {156},
               number = {3},
                  eid = {123},
                pages = {123},
                  doi = {10.3847/1538-3881/aabc4f},
        archivePrefix = {arXiv},
               eprint = {1801.02634},
         primaryClass = {astro-ph.IM},
               adsurl = {https://ui.adsabs.harvard.edu/abs/2018AJ....156..123A}
        }

@ARTICLE{astropy:2022,
               author = {{Astropy Collaboration} and {Price-Whelan}, Adrian M. and {Lim}, Pey Lian and {Earl}, Nicholas and {Starkman}, Nathaniel and {Bradley}, Larry and {Shupe}, David L. and {Patil}, Aarya A. and {Corrales}, Lia and {Brasseur}, C.~E. and {N{"o}the}, Maximilian and {Donath}, Axel and {Tollerud}, Erik and {Morris}, Brett M. and {Ginsburg}, Adam and {Vaher}, Eero and {Weaver}, Benjamin A. and {Tocknell}, James and {Jamieson}, William and {van Kerkwijk}, Marten H. and {Robitaille}, Thomas P. and {Merry}, Bruce and {Bachetti}, Matteo and {G{"u}nther}, H. Moritz and {Aldcroft}, Thomas L. and {Alvarado-Montes}, Jaime A. and {Archibald}, Anne M. and {B{'o}di}, Attila and {Bapat}, Shreyas and {Barentsen}, Geert and {Baz{'a}n}, Juanjo and {Biswas}, Manish and {Boquien}, M{'e}d{'e}ric and {Burke}, D.~J. and {Cara}, Daria and {Cara}, Mihai and {Conroy}, Kyle E. and {Conseil}, Simon and {Craig}, Matthew W. and {Cross}, Robert M. and {Cruz}, Kelle L. and {D'Eugenio}, Francesco and {Dencheva}, Nadia and {Devillepoix}, Hadrien A.~R. and {Dietrich}, J{"o}rg P. and {Eigenbrot}, Arthur Davis and {Erben}, Thomas and {Ferreira}, Leonardo and {Foreman-Mackey}, Daniel and {Fox}, Ryan and {Freij}, Nabil and {Garg}, Suyog and {Geda}, Robel and {Glattly}, Lauren and {Gondhalekar}, Yash and {Gordon}, Karl D. and {Grant}, David and {Greenfield}, Perry and {Groener}, Austen M. and {Guest}, Steve and {Gurovich}, Sebastian and {Handberg}, Rasmus and {Hart}, Akeem and {Hatfield-Dodds}, Zac and {Homeier}, Derek and {Hosseinzadeh}, Griffin and {Jenness}, Tim and {Jones}, Craig K. and {Joseph}, Prajwel and {Kalmbach}, J. Bryce and {Karamehmetoglu}, Emir and {Ka{l}uszy{'n}ski}, Miko{l}aj and {Kelley}, Michael S.~P. and {Kern}, Nicholas and {Kerzendorf}, Wolfgang E. and {Koch}, Eric W. and {Kulumani}, Shankar and {Lee}, Antony and {Ly}, Chun and {Ma}, Zhiyuan and {MacBride}, Conor and {Maljaars}, Jakob M. and {Muna}, Demitri and {Murphy}, N.~A. and {Norman}, Henrik and {O'Steen}, Richard and {Oman}, Kyle A. and {Pacifici}, Camilla and {Pascual}, Sergio and {Pascual-Granado}, J. and {Patil}, Rohit R. and {Perren}, Gabriel I. and {Pickering}, Timothy E. and {Rastogi}, Tanuj and {Roulston}, Benjamin R. and {Ryan}, Daniel F. and {Rykoff}, Eli S. and {Sabater}, Jose and {Sakurikar}, Parikshit and {Salgado}, Jes{'u}s and {Sanghi}, Aniket and {Saunders}, Nicholas and {Savchenko}, Volodymyr and {Schwardt}, Ludwig and {Seifert-Eckert}, Michael and {Shih}, Albert Y. and {Jain}, Anany Shrey and {Shukla}, Gyanendra and {Sick}, Jonathan and {Simpson}, Chris and {Singanamalla}, Sudheesh and {Singer}, Leo P. and {Singhal}, Jaladh and {Sinha}, Manodeep and {Sip{H{o}}cz}, Brigitta M. and {Spitler}, Lee R. and {Stansby}, David and {Streicher}, Ole and {{{S}}umak}, Jani and {Swinbank}, John D. and {Taranu}, Dan S. and {Tewary}, Nikita and {Tremblay}, Grant R. and {Val-Borro}, Miguel de and {Van Kooten}, Samuel J. and {Vasovi{'c}}, Zlatan and {Verma}, Shresth and {de Miranda Cardoso}, Jos{'e} Vin{'i}cius and {Williams}, Peter K.~G. and {Wilson}, Tom J. and {Winkel}, Benjamin and {Wood-Vasey}, W.~M. and {Xue}, Rui and {Yoachim}, Peter and {Zhang}, Chen and {Zonca}, Andrea and {Astropy Project Contributors}},
                title = "{The Astropy Project: Sustaining and Growing a Community-oriented Open-source Project and the Latest Major Release (v5.0) of the Core Package}",
              journal = {\apj},
                 year = 2022,
                month = aug,
               volume = {935},
               number = {2},
                  eid = {167},
                pages = {167},
                  doi = {10.3847/1538-4357/ac7c74},
        archivePrefix = {arXiv},
               eprint = {2206.14220},
         primaryClass = {astro-ph.IM},
               adsurl = {https://ui.adsabs.harvard.edu/abs/2022ApJ...935..167A}
        }

@ARTICLE{2007ApJ...666.1024B,
       author = {{Blondin}, St{\'e}phane and {Tonry}, John L.},
        title = "{Determining the Type, Redshift, and Age of a Supernova Spectrum}",
      journal = {\apj},
         year = 2007,
        month = sep,
       volume = {666},
       number = {2},
        pages = {1024-1047},
          doi = {10.1086/520494},
archivePrefix = {arXiv},
       eprint = {0709.4488},
 primaryClass = {astro-ph},
       adsurl = {https://ui.adsabs.harvard.edu/abs/2007ApJ...666.1024B}
}

@ARTICLE{1993ApJ...415L.103F,
       author = {{Filippenko}, Alexei V. and {Matheson}, Thomas and {Ho}, Luis C.},
        title = "{The ``Type IIb'' Supernova 1993J in M81: A Close Relative of Type Ib Supernovae}",
      journal = {\apjl},
         year = 1993,
        month = oct,
       volume = {415},
        pages = {L103},
          doi = {10.1086/187043},
       adsurl = {https://ui.adsabs.harvard.edu/abs/1993ApJ...415L.103F}
}

@ARTICLE{2012ApJ...755..161K,
       author = {{Kasliwal}, Mansi M. and {Kulkarni}, S.~R. and {Gal-Yam}, Avishay and {Nugent}, Peter E. and {Sullivan}, Mark and {Bildsten}, Lars and {Yaron}, Ofer and {Perets}, Hagai B. and {Arcavi}, Iair and {Ben-Ami}, Sagi and {Bhalerao}, Varun B. and {Bloom}, Joshua S. and {Cenko}, S. Bradley and {Filippenko}, Alexei V. and {Frail}, Dale A. and {Ganeshalingam}, Mohan and {Horesh}, Assaf and {Howell}, D. Andrew and {Law}, Nicholas M. and {Leonard}, Douglas C. and {Li}, Weidong and {Ofek}, Eran O. and {Polishook}, David and {Poznanski}, Dovi and {Quimby}, Robert M. and {Silverman}, Jeffrey M. and {Sternberg}, Assaf and {Xu}, Dong},
        title = "{Calcium-rich Gap Transients in the Remote Outskirts of Galaxies}",
      journal = {\apj},
         year = 2012,
        month = aug,
       volume = {755},
       number = {2},
          eid = {161},
        pages = {161},
          doi = {10.1088/0004-637X/755/2/161},
archivePrefix = {arXiv},
       eprint = {1111.6109},
 primaryClass = {astro-ph.HE},
       adsurl = {https://ui.adsabs.harvard.edu/abs/2012ApJ...755..161K}
}

@ARTICLE{2022ApJ...928..151F,
       author = {{Fang}, Qiliang and {Maeda}, Keiichi and {Kuncarayakti}, Hanindyo and {Tanaka}, Masaomi and {Kawabata}, Koji S. and {Hattori}, Takashi and {Aoki}, Kentaro and {Moriya}, Takashi J. and {Yamanaka}, Masayuki},
        title = "{Statistical Properties of the Nebular Spectra of 103 Stripped-envelope Core-collapse Supernovae}",
      journal = {\apj},
         year = 2022,
        month = apr,
       volume = {928},
       number = {2},
          eid = {151},
        pages = {151},
          doi = {10.3847/1538-4357/ac4f60},
archivePrefix = {arXiv},
       eprint = {2201.11467},
 primaryClass = {astro-ph.HE},
       adsurl = {https://ui.adsabs.harvard.edu/abs/2022ApJ...928..151F}
}

@ARTICLE{2022MNRAS.514.5686P,
       author = {{Prentice}, S.~J. and {Maguire}, K. and {Siebenaler}, L. and {Jerkstrand}, A.},
        title = "{Oxygen and calcium nebular emission line relationships in core-collapse supernovae and Ca-rich transients}",
      journal = {\mnras},
         year = 2022,
        month = aug,
       volume = {514},
       number = {4},
        pages = {5686-5705},
          doi = {10.1093/mnras/stac1657},
archivePrefix = {arXiv},
       eprint = {2206.06062},
 primaryClass = {astro-ph.HE},
       adsurl = {https://ui.adsabs.harvard.edu/abs/2022MNRAS.514.5686P}
}

@ARTICLE{2022ApJ...925..175S,
       author = {{Shahbandeh}, M. and {Hsiao}, E.~Y. and {Ashall}, C. and {Teffs}, J. and {Hoeflich}, P. and {Morrell}, N. and {Phillips}, M.~M. and {Anderson}, J.~P. and {Baron}, E. and {Burns}, C.~R. and {Contreras}, C. and {Davis}, S. and {Diamond}, T.~R. and {Folatelli}, G. and {Galbany}, L. and {Gall}, C. and {Hachinger}, S. and {Holmbo}, S. and {Karamehmetoglu}, E. and {Kasliwal}, M.~M. and {Kirshner}, R.~P. and {Krisciunas}, K. and {Kumar}, S. and {Lu}, J. and {Marion}, G.~H. and {Mazzali}, P.~A. and {Piro}, A.~L. and {Sand}, D.~J. and {Stritzinger}, M.~D. and {Suntzeff}, N.~B. and {Taddia}, F. and {Uddin}, S.~A.},
        title = "{Carnegie Supernova Project-II: Near-infrared Spectroscopy of Stripped-envelope Core-collapse Supernovae}",
      journal = {\apj},
         year = 2022,
        month = feb,
       volume = {925},
       number = {2},
          eid = {175},
        pages = {175},
          doi = {10.3847/1538-4357/ac4030},
archivePrefix = {arXiv},
       eprint = {2110.12083},
 primaryClass = {astro-ph.HE},
       adsurl = {https://ui.adsabs.harvard.edu/abs/2022ApJ...925..175S}
}

@ARTICLE{2019ApJ...887....4D,
       author = {{Davis}, S. and {Hsiao}, E.~Y. and {Ashall}, C. and {Hoeflich}, P. and {Phillips}, M.~M. and {Marion}, G.~H. and {Kirshner}, R.~P. and {Morrell}, N. and {Sand}, D.~J. and {Burns}, C. and {Contreras}, C. and {Stritzinger}, M. and {Anderson}, J.~P. and {Baron}, E. and {Diamond}, T. and {Guti{\'e}rrez}, C.~P. and {Hamuy}, M. and {Holmbo}, S. and {Kasliwal}, M.~M. and {Krisciunas}, K. and {Kumar}, S. and {Lu}, J. and {Pessi}, P.~J. and {Piro}, A.~L. and {Prieto}, J.~L. and {Shahbandeh}, M. and {Suntzeff}, N.~B.},
        title = "{Carnegie Supernova Project-II: Near-infrared Spectroscopic Diversity of Type II Supernovae}",
      journal = {\apj},
         year = 2019,
        month = dec,
       volume = {887},
       number = {1},
          eid = {4},
        pages = {4},
          doi = {10.3847/1538-4357/ab4c40},
archivePrefix = {arXiv},
       eprint = {1910.03410},
 primaryClass = {astro-ph.HE},
       adsurl = {https://ui.adsabs.harvard.edu/abs/2019ApJ...887....4D}
}

@ARTICLE{2012PASP..124..668Y,
       author = {{Yaron}, Ofer and {Gal-Yam}, Avishay},
        title = "{WISeREP{\textemdash}An Interactive Supernova Data Repository}",
      journal = {\pasp},
         year = 2012,
        month = jul,
       volume = {124},
       number = {917},
        pages = {668},
          doi = {10.1086/666656},
archivePrefix = {arXiv},
       eprint = {1204.1891},
 primaryClass = {astro-ph.IM},
       adsurl = {https://ui.adsabs.harvard.edu/abs/2012PASP..124..668Y}
}

@ARTICLE{1995A&AS..110..513B,
       author = {{Barbon}, R. and {Benetti}, S. and {Cappellaro}, E. and {Patat}, F. and {Turatto}, M. and {Iijima}, T.},
        title = "{SN 1993J in M 81: One year of observations at Asiago.}",
      journal = {\aaps},
         year = 1995,
        month = may,
       volume = {110},
        pages = {513},
       adsurl = {https://ui.adsabs.harvard.edu/abs/1995A&AS..110..513B}
}

@ARTICLE{2017ApJ...835...64G,
       author = {{Guillochon}, James and {Parrent}, Jerod and {Kelley}, Luke Zoltan and {Margutti}, Raffaella},
        title = "{An Open Catalog for Supernova Data}",
      journal = {\apj},
         year = 2017,
        month = jan,
       volume = {835},
       number = {1},
          eid = {64},
        pages = {64},
          doi = {10.3847/1538-4357/835/1/64},
archivePrefix = {arXiv},
       eprint = {1605.01054},
 primaryClass = {astro-ph.SR},
       adsurl = {https://ui.adsabs.harvard.edu/abs/2017ApJ...835...64G}
}

@ARTICLE{2015A&A...573A..12J,
       author = {{Jerkstrand}, A. and {Ergon}, M. and {Smartt}, S.~J. and {Fransson}, C. and {Sollerman}, J. and {Taubenberger}, S. and {Bersten}, M. and {Spyromilio}, J.},
        title = "{Late-time spectral line formation in Type IIb supernovae, with application to SN 1993J, SN 2008ax, and SN 2011dh}",
      journal = {\aap},
         year = 2015,
        month = jan,
       volume = {573},
          eid = {A12},
        pages = {A12},
          doi = {10.1051/0004-6361/201423983},
archivePrefix = {arXiv},
       eprint = {1408.0732},
 primaryClass = {astro-ph.HE},
       adsurl = {https://ui.adsabs.harvard.edu/abs/2015A&A...573A..12J}
}

@ARTICLE{2000AJ....120.1499M,
       author = {{Matheson}, Thomas and {Filippenko}, Alexei V. and {Ho}, Luis C. and {Barth}, Aaron J. and {Leonard}, Douglas C.},
        title = "{Detailed Analysis of Early to Late-Time Spectra of Supernova 1993J}",
      journal = {\aj},
         year = 2000,
        month = sep,
       volume = {120},
       number = {3},
        pages = {1499-1515},
          doi = {10.1086/301519},
archivePrefix = {arXiv},
       eprint = {astro-ph/0006264},
 primaryClass = {astro-ph},
       adsurl = {https://ui.adsabs.harvard.edu/abs/2000AJ....120.1499M}
}

@ARTICLE{1995ApJS...99..223P,
       author = {{Pun}, Chun S.~J. and {Kirshner}, Robert P. and {Sonneborn}, George and {Challis}, Peter and {Nassiopoulos}, George and {Arquilla}, Richard and {Crenshaw}, D. Michael and {Shrader}, Chris and {Teays}, Terry and {Cassatella}, Angelo and {Gilmozzi}, Roberto and {Talavera}, Antonio and {Wamsteker}, Willem and {Fransson}, Claes and {Panagia}, Nino},
        title = "{Ultraviolet Observations of SN 1987A with the IUE Satellite}",
      journal = {\apjs},
         year = 1995,
        month = jul,
       volume = {99},
        pages = {223},
          doi = {10.1086/192185},
       adsurl = {https://ui.adsabs.harvard.edu/abs/1995ApJS...99..223P}
}

@ARTICLE{2014A&A...562A..17E,
       author = {{Ergon}, M. and {Sollerman}, J. and {Fraser}, M. and {Pastorello}, A. and {Taubenberger}, S. and {Elias-Rosa}, N. and {Bersten}, M. and {Jerkstrand}, A. and {Benetti}, S. and {Botticella}, M.~T. and {Fransson}, C. and {Harutyunyan}, A. and {Kotak}, R. and {Smartt}, S. and {Valenti}, S. and {Bufano}, F. and {Cappellaro}, E. and {Fiaschi}, M. and {Howell}, A. and {Kankare}, E. and {Magill}, L. and {Mattila}, S. and {Maund}, J. and {Naves}, R. and {Ochner}, P. and {Ruiz}, J. and {Smith}, K. and {Tomasella}, L. and {Turatto}, M.},
        title = "{Optical and near-infrared observations of SN 2011dh - The first 100 days}",
      journal = {\aap},
         year = 2014,
        month = feb,
       volume = {562},
          eid = {A17},
        pages = {A17},
          doi = {10.1051/0004-6361/201321850},
archivePrefix = {arXiv},
       eprint = {1305.1851},
 primaryClass = {astro-ph.SR},
       adsurl = {https://ui.adsabs.harvard.edu/abs/2014A&A...562A..17E}
}

@ARTICLE{2013MNRAS.436.3614S,
       author = {{Shivvers}, Isaac and {Mazzali}, Paolo and {Silverman}, Jeffrey M. and {Boty{\'a}nszki}, J{\'a}nos and {Cenko}, S. Bradley and {Filippenko}, Alexei V. and {Kasen}, Daniel and {Van Dyk}, Schuyler D. and {Clubb}, Kelsey I.},
        title = "{Nebular spectroscopy of the nearby Type IIb supernova 2011dh}",
      journal = {\mnras},
         year = 2013,
        month = dec,
       volume = {436},
       number = {4},
        pages = {3614-3625},
          doi = {10.1093/mnras/stt1839},
archivePrefix = {arXiv},
       eprint = {1307.2246},
 primaryClass = {astro-ph.HE},
       adsurl = {https://ui.adsabs.harvard.edu/abs/2013MNRAS.436.3614S}
}

@ARTICLE{2021MNRAS.503.3472B,
       author = {{Bose}, Subhash and {Dong}, Subo and {Kochanek}, C.~S. and {Stritzinger}, M.~D. and {Ashall}, Chris and {Benetti}, Stefano and {Falco}, E. and {Filippenko}, Alexei V. and {Pastorello}, Andrea and {Prieto}, Jose L. and {Somero}, Auni and {Sukhbold}, Tuguldur and {Zhang}, Junbo and {Auchettl}, Katie and {Brink}, Thomas G. and {Brown}, J.~S. and {Chen}, Ping and {Fiore}, A. and {Grupe}, Dirk and {Holoien}, T.~W.-S. and {Lundqvist}, Peter and {Mattila}, Seppo and {Mutel}, Robert and {Pooley}, David and {Post}, R.~S. and {Reddy}, Naveen and {Reynolds}, Thomas M. and {Shappee}, Benjamin J. and {Stanek}, K.~Z. and {Thompson}, Todd A. and {Villanueva}, Jr., S. and {Zheng}, WeiKang},
        title = "{ASASSN-18am/SN 2018gk: an overluminous Type IIb supernova from a massive progenitor}",
      journal = {\mnras},
         year = 2021,
        month = may,
       volume = {503},
       number = {3},
        pages = {3472-3491},
          doi = {10.1093/mnras/stab629},
archivePrefix = {arXiv},
       eprint = {2007.00008},
 primaryClass = {astro-ph.HE},
       adsurl = {https://ui.adsabs.harvard.edu/abs/2021MNRAS.503.3472B}
}

@ARTICLE{2001ApJ...555..900P,
       author = {{Patat}, Ferdinando and {Cappellaro}, Enrico and {Danziger}, John and {Mazzali}, Paolo A. and {Sollerman}, Jesper and {Augusteijn}, Thomas and {Brewer}, James and {Doublier}, Vanessa and {Gonzalez}, Jean Fran{\c{c}}ois and {Hainaut}, Olivier and {Lidman}, Chris and {Leibundgut}, Bruno and {Nomoto}, Ken'ichi and {Nakamura}, Takayoshi and {Spyromilio}, Jason and {Rizzi}, Luca and {Turatto}, Massimo and {Walsh}, Jeremy and {Galama}, Titus J. and {van Paradijs}, Jan and {Kouveliotou}, Chryssa and {Vreeswijk}, Paul M. and {Frontera}, Filippo and {Masetti}, Nicola and {Palazzi}, Eliana and {Pian}, Elena},
        title = "{The Metamorphosis of SN 1998bw}",
      journal = {\apj},
         year = 2001,
        month = jul,
       volume = {555},
       number = {2},
        pages = {900-917},
          doi = {10.1086/321526},
archivePrefix = {arXiv},
       eprint = {astro-ph/0103111},
 primaryClass = {astro-ph},
       adsurl = {https://ui.adsabs.harvard.edu/abs/2001ApJ...555..900P}
}

@ARTICLE{2015A&A...580A.142E,
       author = {{Ergon}, M. and {Jerkstrand}, A. and {Sollerman}, J. and {Elias-Rosa}, N. and {Fransson}, C. and {Fraser}, M. and {Pastorello}, A. and {Kotak}, R. and {Taubenberger}, S. and {Tomasella}, L. and {Valenti}, S. and {Benetti}, S. and {Helou}, G. and {Kasliwal}, M.~M. and {Maund}, J. and {Smartt}, S.~J. and {Spyromilio}, J.},
        title = "{The Type IIb SN 2011dh: Two years of observations and modelling of the lightcurves}",
      journal = {\aap},
         year = 2015,
        month = aug,
       volume = {580},
          eid = {A142},
        pages = {A142},
          doi = {10.1051/0004-6361/201424592},
archivePrefix = {arXiv},
       eprint = {1408.0731},
 primaryClass = {astro-ph.SR},
       adsurl = {https://ui.adsabs.harvard.edu/abs/2015A&A...580A.142E}
}

@ARTICLE{1992ApJ...384L..33S,
       author = {{Suntzeff}, Nicholas B. and {Phillips}, M.~M. and {Elias}, J.~H. and {Depoy}, D.~L. and {Walker}, A.~R.},
        title = "{The Energy Sources Powering the Late-Time Bolometric Evolution of SN 1987A}",
      journal = {\apjl},
         year = 1992,
        month = jan,
       volume = {384},
        pages = {L33},
          doi = {10.1086/186256},
       adsurl = {https://ui.adsabs.harvard.edu/abs/1992ApJ...384L..33S}
}

@ARTICLE{1991PASP..103..958W,
       author = {{Walker}, Alistair R. and {Suntzeff}, Nicholas B.},
        title = "{CCD Photometry of SN 1987A . I. Days 680 to 1469}",
      journal = {\pasp},
         year = 1991,
        month = sep,
       volume = {103},
        pages = {958},
          doi = {10.1086/132912},
       adsurl = {https://ui.adsabs.harvard.edu/abs/1991PASP..103..958W}
}

@ARTICLE{1989MNRAS.237P..55C,
       author = {{Catchpole}, R.~M. and {Whitelock}, P.~A. and {Menzies}, J.~W. and {Feast}, M.~W. and {Marang}, F. and {Sekiguchi}, K. and {van Wyk}, F. and {Roberts}, G. and {Balona}, L.~A. and {Egan}, J.~M. and {Carter}, B.~S. and {Laney}, C.~D. and {Laing}, J.~D. and {Spencer Jones}, J.~H. and {Glass}, I.~S. and {Winkler}, H. and {Fairall}, A.~P. and {Lloyd Evans}, T.~H.~H. and {Cropper}, M.~S. and {Shenton}, M. and {Hill}, P.~W. and {Payne}, P. and {Jones}, K.~N. and {Wargau}, W. and {Mason}, K.~O. and {Jeffery}, C.~S. and {Hellier}, C. and {Parker}, Q.~A. and {Chini}, R. and {James}, P.~A. and {Doyle}, J.~G. and {Butler}, C.~J. and {Bromage}, G.},
        title = "{Spectroscopic and photometric observations of SN 1987A- V. Days 386-616.}",
      journal = {\mnras},
         year = 1989,
        month = mar,
       volume = {237},
        pages = {55P-68},
          doi = {10.1093/mnras/237.1.55P},
       adsurl = {https://ui.adsabs.harvard.edu/abs/1989MNRAS.237P..55C}
}

@ARTICLE{1988MNRAS.231P..75C,
       author = {{Catchpole}, R.~M. and {Whitelock}, P.~A. and {Feast}, M.~W. and {Menzies}, J.~M. and {Glass}, I.~S. and {Marang}, F. and {Laing}, J.~D. and {Spencer Jones}, J.~H. and {Roberts}, G. and {Balona}, L.~A. and {Carter}, B.~S. and {Laney}, C.~D. and {Evans}, Lloyd T. and {Sekiguchi}, K. and {Hutchinson}, G.~G. and {Maddison}, R. and {Albinson}, J. and {Evans}, A. and {Allen}, F.~A. and {Winkler}, H. and {Fairall}, A. and {Corbally}, C. and {Davies}, J.~K. and {Parker}, Q.~A.},
        title = "{Spectroscopic and photometric observations of SN 1987A - III. Days 135 to 260.}",
      journal = {\mnras},
         year = 1988,
        month = apr,
       volume = {231},
        pages = {75P-89},
          doi = {10.1093/mnras/231.1.75P},
       adsurl = {https://ui.adsabs.harvard.edu/abs/1988MNRAS.231P..75C}
}

@ARTICLE{1988AJ.....96.1864S,
       author = {{Suntzeff}, Nicholas B. and {Hamuy}, Mario and {Martin}, Gabriel and {Gomez}, Arturo and {Gonzalez}, Ricardo},
        title = "{SN 1987A in the LMC. II. Optical Photometry at Cerro Tololo}",
      journal = {\aj},
         year = 1988,
        month = dec,
       volume = {96},
        pages = {1864},
          doi = {10.1086/114933},
       adsurl = {https://ui.adsabs.harvard.edu/abs/1988AJ.....96.1864S}
}

@ARTICLE{1988MNRAS.234P...5W,
       author = {{Whitelock}, P.~A. and {Catchpole}, R.~M. and {Menzies}, J.~W. and {Feast}, M.~W. and {Winkler}, H. and {Marang}, F. and {Glass}, I.~S. and {Balona}, L.~A. and {Egan}, J. and {Carter}, B.~S. and {Roberts}, G. and {Sekiguchi}, K. and {Laney}, C.~D. and {Lloyd Evans}, T. and {Laing}, J.~D. and {Spencer Jones}, J. and {Fernley}, J. and {James}, P. and {Fairall}, A.~P. and {Monk}, A.~S. and {van Wyk}, F.},
        title = "{Spectroscopic and photometric observations of SN 1987A - IV. Days 260-385.}",
      journal = {\mnras},
         year = 1988,
        month = sep,
       volume = {234},
        pages = {5P-18},
          doi = {10.1093/mnras/234.1.5P},
       adsurl = {https://ui.adsabs.harvard.edu/abs/1988MNRAS.234P...5W}
}

@ARTICLE{1987MNRAS.229P..15C,
       author = {{Catchpole}, R.~M. and {Menzies}, J.~W. and {Monk}, A.~S. and {Wargau}, W.~F. and {Pollaco}, D. and {Carter}, B.~S. and {Whitelock}, P.~A. and {Marang}, F. and {Laney}, C.~D. and {Balona}, L.~A. and {Feast}, M.~W. and {Lloyd Evans}, T.~H.~H. and {Sekiguchi}, K. and {Laing}, J.~D. and {Kilkenny}, D.~M. and {Spencer Jones}, J. and {Roberts}, G. and {Cousins}, A.~W.~J. and {van Vuuren}, G. and {Winkler}, H.},
        title = "{Spectroscopic and photometric observations of SN 1987A- II. Days 51 to134.}",
      journal = {\mnras},
         year = 1987,
        month = nov,
       volume = {229},
        pages = {15P-25},
          doi = {10.1093/mnras/229.1.15P},
       adsurl = {https://ui.adsabs.harvard.edu/abs/1987MNRAS.229P..15C}
}

@ARTICLE{1987MNRAS.227P..39M,
       author = {{Menzies}, J.~W. and {Catchpole}, R.~M. and {van Vuuren}, G. and {Winkler}, H. and {Laney}, C.~D. and {Whitelock}, P.~A. and {Cousins}, A.~W.~J. and {Carter}, B.~S. and {Marang}, F. and {Lloyd Evans}, T.~H.~H. and {Roberts}, G. and {Kilkenny}, D. and {Spencer Jones}, J. and {Sekiguchi}, K. and {Fairall}, A.~P. and {Wolstencroft}, R.~D.},
        title = "{Spectroscopic and photometric observations of SN 1987A : the first 50days.}",
      journal = {\mnras},
         year = 1987,
        month = aug,
       volume = {227},
        pages = {39P-49},
          doi = {10.1093/mnras/227.1.39P},
       adsurl = {https://ui.adsabs.harvard.edu/abs/1987MNRAS.227P..39M}
}

@ARTICLE{1996AJ....112..732R,
       author = {{Richmond}, Michael W. and {Treffers}, Richard R. and {Filippenko}, Alexei V. and {Paik}, Young},
        title = "{UBVRI Photometry of SN 1993J in M81: Days 3 to 365}",
      journal = {\aj},
         year = 1996,
        month = aug,
       volume = {112},
        pages = {732},
          doi = {10.1086/118048},
       adsurl = {https://ui.adsabs.harvard.edu/abs/1996AJ....112..732R}
}

@ARTICLE{1994AJ....107.1022R,
       author = {{Richmond}, Michael W. and {Treffers}, Richard R. and {Filippenko}, Alexei V. and {Paik}, Young and {Leibundgut}, Bruno and {Schulman}, Eric and {Cox}, Caroline V.},
        title = "{UBVRI Photometry of SN 1993J in M81: The First 120 Days}",
      journal = {\aj},
         year = 1994,
        month = mar,
       volume = {107},
        pages = {1022},
          doi = {10.1086/116915},
       adsurl = {https://ui.adsabs.harvard.edu/abs/1994AJ....107.1022R}
}

@ARTICLE{1994AJ....107.1453B,
       author = {{Benson}, P.~J. and {Herbst}, W. and {Salzer}, J.~J. and {Vinton}, G. and {Hanson}, G.~J. and {Ratcliff}, S.~J. and {Winkler}, P.~F. and {Elmegreen}, D.~M. and {Chromey}, F. and {Strom}, C. and {Balonek}, T.~J. and {Elmegreen}, B.~G.},
        title = "{Light Curves of SN 1993J From The Keck Northeast Astronomy Consortium}",
      journal = {\aj},
         year = 1994,
        month = apr,
       volume = {107},
        pages = {1453},
          doi = {10.1086/116958},
       adsurl = {https://ui.adsabs.harvard.edu/abs/1994AJ....107.1453B}
}

@ARTICLE{1993PASJ...45L..59V,
       author = {{van Driel}, Wim and {Yoshida}, Shigeomi and {Nakada}, Yoshikazu and {Aoki}, Tsutomu and {Soyano}, Takao and {Tarusawa}, Ken'ichi and {Ichikawa}, Takashi and {Kakehashi}, Takuya and {Nomoto}, Ken'ichi and {Wakamatsu}, Ken'ichi},
        title = "{BVRI Photometry of SN 1993J in M81. I. Period April 01─30 1993}",
      journal = {\pasj},
         year = 1993,
        month = oct,
       volume = {45},
       number = {5},
        pages = {L59-L62},
          doi = {10.1093/pasj/45.5.L59},
       adsurl = {https://ui.adsabs.harvard.edu/abs/1993PASJ...45L..59V}
}

@ARTICLE{2011AJ....141..163C,
       author = {{Clocchiatti}, Alejandro and {Suntzeff}, Nicholas B. and {Covarrubias}, Ricardo and {Candia}, Pablo},
        title = "{The Ultimate Light Curve of SN 1998bw/GRB 980425}",
      journal = {\aj},
         year = 2011,
        month = may,
       volume = {141},
       number = {5},
          eid = {163},
        pages = {163},
          doi = {10.1088/0004-6256/141/5/163},
archivePrefix = {arXiv},
       eprint = {1106.1695},
 primaryClass = {astro-ph.HE},
       adsurl = {https://ui.adsabs.harvard.edu/abs/2011AJ....141..163C}
}

@ARTICLE{Thomas2011,
       author = {{Thomas}, R.~C. and {Nugent}, P.~E. and {Meza}, J.~C.},
        title = "{SYNAPPS: Data-Driven Analysis for Supernova Spectroscopy}",
      journal = {\pasp},
         year = 2011,
        month = feb,
       volume = {123},
       number = {900},
        pages = {237},
          doi = {10.1086/658673},
       adsurl = {https://ui.adsabs.harvard.edu/abs/2011PASP..123..237T}
}

@INPROCEEDINGS{Seifert2003,
       author = {{Seifert}, Walter and {Appenzeller}, Immo and {Baumeister}, Harald and {Bizenberger}, Peter and {Bomans}, Dominik and {Dettmar}, Ralf-Juergen and {Grimm}, Bernard and {Herbst}, Tom and {Hofmann}, Reiner and {Juette}, Marcus and {Laun}, Werner and {Lehmitz}, Michael and {Lemke}, Roland and {Lenzen}, Rainer and {Mandel}, Holger and {Polsterer}, K. and {Rohloff}, Ralf-Rainer and {Schuetze}, A. and {Seltmann}, Andreas and {Thatte}, Niranjan A. and {Weiser}, Peter and {Xu}, Wenli},
        title = "{LUCIFER: a Multi-Mode NIR Instrument for the LBT}",
    booktitle = {Instrument Design and Performance for Optical/Infrared Ground-based Telescopes},
         year = 2003,
       editor = {{Iye}, Masanori and {Moorwood}, Alan F.~M.},
       series = {Society of Photo-Optical Instrumentation Engineers (SPIE) Conference Series},
       volume = {4841},
        month = mar,
        pages = {962-973},
          doi = {10.1117/12.459494},
       adsurl = {https://ui.adsabs.harvard.edu/abs/2003SPIE.4841..962S}
}

@ARTICLE{sipgi2022,
       author = {{Gargiulo}, A. and {Fumana}, M. and {Bisogni}, S. and {Franzetti}, P. and {Cassar{\`a}}, L.~P. and {Garilli}, B. and {Scodeggio}, M. and {Vietri}, G.},
        title = "{SIPGI: an interactive pipeline for spectroscopic data reduction}",
      journal = {\mnras},
         year = 2022,
        month = aug,
       volume = {514},
       number = {2},
        pages = {2902-2914},
          doi = {10.1093/mnras/stac1065},
archivePrefix = {arXiv},
       eprint = {2209.05441},
 primaryClass = {astro-ph.IM},
       adsurl = {https://ui.adsabs.harvard.edu/abs/2022MNRAS.514.2902G}
}

@ARTICLE{Fontana2014a,
       author = {{Fontana}, A. and {Dunlop}, J.~S. and {Paris}, D. and {Targett}, T.~A. and {Boutsia}, K. and {Castellano}, M. and {Galametz}, A. and {Grazian}, A. and {McLure}, R. and {Merlin}, E. and {Pentericci}, L. and {Wuyts}, S. and {Almaini}, O. and {Caputi}, K. and {Chary}, R.-R. and {Cirasuolo}, M. and {Conselice}, C.~J. and {Cooray}, A. and {Daddi}, E. and {Dickinson}, M. and {Faber}, S.~M. and {Fazio}, G. and {Ferguson}, H.~C. and {Giallongo}, E. and {Giavalisco}, M. and {Grogin}, N.~A. and {Hathi}, N. and {Koekemoer}, A.~M. and {Koo}, D.~C. and {Lucas}, R.~A. and {Nonino}, M. and {Rix}, H.~W. and {Renzini}, A. and {Rosario}, D. and {Santini}, P. and {Scarlata}, C. and {Sommariva}, V. and {Stark}, D.~P. and {van der Wel}, A. and {Vanzella}, E. and {Wild}, V. and {Yan}, H. and {Zibetti}, S.},
        title = "{The Hawk-I UDS and GOODS Survey (HUGS): Survey design and deep K-band number counts}",
      journal = {\aap},
         year = 2014,
        month = oct,
       volume = {570},
          eid = {A11},
        pages = {A11},
          doi = {10.1051/0004-6361/201423543},
archivePrefix = {arXiv},
       eprint = {1409.7082},
 primaryClass = {astro-ph.GA},
       adsurl = {https://ui.adsabs.harvard.edu/abs/2014A&A...570A..11F}
}

@ARTICLE{1988Ap&SS.150..291M,
       author = {{Milone}, Luis A. and {Paolantonio}, S. and {Briggi}, V. and {Mendicini}, D. and {Minniti}, E.},
        title = "{The Unusual Supernova 1987A}",
      journal = {\apss},
         year = 1988,
        month = dec,
       volume = {150},
       number = {2},
        pages = {291-297},
          doi = {10.1007/BF00641723},
       adsurl = {https://ui.adsabs.harvard.edu/abs/1988Ap&SS.150..291M}
}

@ARTICLE{1988PASA....7..401M,
       author = {{Menzies}, J.~W.},
        title = "{SN 1987A : the light curve.}",
      journal = {\pasa},
         year = 1988,
        month = jan,
       volume = {7},
       number = {4},
        pages = {401-404},
          doi = {10.1017/S1323358000022542},
       adsurl = {https://ui.adsabs.harvard.edu/abs/1988PASA....7..401M}
}

@ARTICLE{1989ARA&A..27..629A,
       author = {{Arnett}, W. David and {Bahcall}, John N. and {Kirshner}, Robert P. and {Woosley}, Stanford E.},
        title = "{Supernova 1987A.}",
      journal = {\araa},
         year = 1989,
        month = jan,
       volume = {27},
        pages = {629-700},
          doi = {10.1146/annurev.aa.27.090189.003213},
       adsurl = {https://ui.adsabs.harvard.edu/abs/1989ARA&A..27..629A}
}

@ARTICLE{2023ApJ...959..142S,
       author = {{Sit}, Tawny and {Kasliwal}, Mansi M. and {Tzanidakis}, Anastasios and {De}, Kishalay and {Fremling}, Christoffer and {Sollerman}, Jesper and {Gal-Yam}, Avishay and {Miller}, Adam A. and {Adams}, Scott and {Aloisi}, Robert and {Andreoni}, Igor and {Chu}, Matthew and {Cook}, David and {Das}, Kaustav Kashyap and {Dugas}, Alison and {Groom}, Steven L. and {Ho}, Anna Y.~Q. and {Karambelkar}, Viraj and {Neill}, James D. and {Masci}, Frank J. and {Medford}, Michael S. and {Purdum}, Josiah and {Sharma}, Yashvi and {Smith}, Roger and {Stein}, Robert and {Yan}, Lin and {Yao}, Yuhan and {Zhang}, Chaoran},
        title = "{Long-rising Type II Supernovae in the Zwicky Transient Facility Census of the Local Universe}",
      journal = {\apj},
         year = 2023,
        month = dec,
       volume = {959},
       number = {2},
          eid = {142},
        pages = {142},
          doi = {10.3847/1538-4357/ad036f},
archivePrefix = {arXiv},
       eprint = {2306.01109},
 primaryClass = {astro-ph.HE},
       adsurl = {https://ui.adsabs.harvard.edu/abs/2023ApJ...959..142S}
}

@INCOLLECTION{Jerkstrand_2017_handbook,
       author = {{Jerkstrand}, Anders},
        title = "{Spectra of Supernovae in the Nebular Phase}",
    booktitle = {Handbook of Supernovae},
         year = 2017,
       editor = {{Alsabti}, Athem W. and {Murdin}, Paul},
        pages = {795},
          doi = {10.1007/978-3-319-21846-5_29},
       adsurl = {https://ui.adsabs.harvard.edu/abs/2017hsn..book..795J}
}

@ARTICLE{Woosley_2019_Hestars,
       author = {{Woosley}, S.~E.},
        title = "{The Evolution of Massive Helium Stars, Including Mass Loss}",
      journal = {\apj},
         year = 2019,
        month = jun,
       volume = {878},
       number = {1},
          eid = {49},
        pages = {49},
          doi = {10.3847/1538-4357/ab1b41},
archivePrefix = {arXiv},
       eprint = {1901.00215},
 primaryClass = {astro-ph.SR},
       adsurl = {https://ui.adsabs.harvard.edu/abs/2019ApJ...878...49W}
}

@ARTICLE{Ertl_2020_Hestars,
       author = {{Ertl}, T. and {Woosley}, S.~E. and {Sukhbold}, Tuguldur and {Janka}, H.-T.},
        title = "{The Explosion of Helium Stars Evolved with Mass Loss}",
      journal = {\apj},
         year = 2020,
        month = feb,
       volume = {890},
       number = {1},
          eid = {51},
        pages = {51},
          doi = {10.3847/1538-4357/ab6458},
archivePrefix = {arXiv},
       eprint = {1910.01641},
 primaryClass = {astro-ph.HE},
       adsurl = {https://ui.adsabs.harvard.edu/abs/2020ApJ...890...51E}
}

@ARTICLE{Heger_2002_PISNgrid,
       author = {{Heger}, A. and {Woosley}, S.~E.},
        title = "{The Nucleosynthetic Signature of Population III}",
      journal = {\apj},
         year = 2002,
        month = mar,
       volume = {567},
       number = {1},
        pages = {532-543},
          doi = {10.1086/338487},
archivePrefix = {arXiv},
       eprint = {astro-ph/0107037},
 primaryClass = {astro-ph},
       adsurl = {https://ui.adsabs.harvard.edu/abs/2002ApJ...567..532H}
}

@ARTICLE{Jerkstrand_2016_PISN,
       author = {{Jerkstrand}, A. and {Smartt}, S.~J. and {Heger}, A.},
        title = "{Nebular spectra of pair-instability supernovae}",
      journal = {\mnras},
         year = 2016,
        month = jan,
       volume = {455},
       number = {3},
        pages = {3207-3229},
          doi = {10.1093/mnras/stv2369},
archivePrefix = {arXiv},
       eprint = {1510.02698},
 primaryClass = {astro-ph.SR},
       adsurl = {https://ui.adsabs.harvard.edu/abs/2016MNRAS.455.3207J}
}

@ARTICLE{1990AJ.....99..650S,
       author = {{Suntzeff}, Nicholas B. and {Bouchet}, Patrice},
        title = "{The Bolometric Light Curve of SN 1987A. I. Results from ESO and CTIO U to Q0 Photometry}",
      journal = {\aj},
         year = 1990,
        month = feb,
       volume = {99},
        pages = {650},
          doi = {10.1086/115358},
       adsurl = {https://ui.adsabs.harvard.edu/abs/1990AJ.....99..650S}
}

@ARTICLE{2016arXiv161205560C,
       author = {{Chambers}, K.~C. and {Magnier}, E.~A. and {Metcalfe}, N. and {Flewelling}, H.~A. and {Huber}, M.~E. and {Waters}, C.~Z. and {Denneau}, L. and {Draper}, P.~W. and {Farrow}, D. and {Finkbeiner}, D.~P. and {Holmberg}, C. and {Koppenhoefer}, J. and {Price}, P.~A. and {Rest}, A. and {Saglia}, R.~P. and {Schlafly}, E.~F. and {Smartt}, S.~J. and {Sweeney}, W. and {Wainscoat}, R.~J. and {Burgett}, W.~S. and {Chastel}, S. and {Grav}, T. and {Heasley}, J.~N. and {Hodapp}, K.~W. and {Jedicke}, R. and {Kaiser}, N. and {Kudritzki}, R.-P. and {Luppino}, G.~A. and {Lupton}, R.~H. and {Monet}, D.~G. and {Morgan}, J.~S. and {Onaka}, P.~M. and {Shiao}, B. and {Stubbs}, C.~W. and {Tonry}, J.~L. and {White}, R. and {Ba{\~n}ados}, E. and {Bell}, E.~F. and {Bender}, R. and {Bernard}, E.~J. and {Boegner}, M. and {Boffi}, F. and {Botticella}, M.~T. and {Calamida}, A. and {Casertano}, S. and {Chen}, W.-P. and {Chen}, X. and {Cole}, S. and {Deacon}, N. and {Frenk}, C. and {Fitzsimmons}, A. and {Gezari}, S. and {Gibbs}, V. and {Goessl}, C. and {Goggia}, T. and {Gourgue}, R. and {Goldman}, B. and {Grant}, P. and {Grebel}, E.~K. and {Hambly}, N.~C. and {Hasinger}, G. and {Heavens}, A.~F. and {Heckman}, T.~M. and {Henderson}, R. and {Henning}, T. and {Holman}, M. and {Hopp}, U. and {Ip}, W.-H. and {Isani}, S. and {Jackson}, M. and {Keyes}, C.~D. and {Koekemoer}, A.~M. and {Kotak}, R. and {Le}, D. and {Liska}, D. and {Long}, K.~S. and {Lucey}, J.~R. and {Liu}, M. and {Martin}, N.~F. and {Masci}, G. and {McLean}, B. and {Mindel}, E. and {Misra}, P. and {Morganson}, E. and {Murphy}, D.~N.~A. and {Obaika}, A. and {Narayan}, G. and {Nieto-Santisteban}, M.~A. and {Norberg}, P. and {Peacock}, J.~A. and {Pier}, E.~A. and {Postman}, M. and {Primak}, N. and {Rae}, C. and {Rai}, A. and {Riess}, A. and {Riffeser}, A. and {Rix}, H.~W. and {R{\"o}ser}, S. and {Russel}, R. and {Rutz}, L. and {Schilbach}, E. and {Schultz}, A.~S.~B. and {Scolnic}, D. and {Strolger}, L. and {Szalay}, A. and {Seitz}, S. and {Small}, E. and {Smith}, K.~W. and {Soderblom}, D.~R. and {Taylor}, P. and {Thomson}, R. and {Taylor}, A.~N. and {Thakar}, A.~R. and {Thiel}, J. and {Thilker}, D. and {Unger}, D. and {Urata}, Y. and {Valenti}, J. and {Wagner}, J. and {Walder}, T. and {Walter}, F. and {Watters}, S.~P. and {Werner}, S. and {Wood-Vasey}, W.~M. and {Wyse}, R.},
        title = "{The Pan-STARRS1 Surveys}",
      journal = {arXiv e-prints},
         year = 2016,
        month = dec,
          eid = {arXiv:1612.05560},
        pages = {arXiv:1612.05560},
          doi = {10.48550/arXiv.1612.05560},
archivePrefix = {arXiv},
       eprint = {1612.05560},
 primaryClass = {astro-ph.IM},
       adsurl = {https://ui.adsabs.harvard.edu/abs/2016arXiv161205560C}
}

@ARTICLE{2020ApJS..251....7F,
       author = {{Flewelling}, H.~A. and {Magnier}, E.~A. and {Chambers}, K.~C. and {Heasley}, J.~N. and {Holmberg}, C. and {Huber}, M.~E. and {Sweeney}, W. and {Waters}, C.~Z. and {Calamida}, A. and {Casertano}, S. and {Chen}, X. and {Farrow}, D. and {Hasinger}, G. and {Henderson}, R. and {Long}, K.~S. and {Metcalfe}, N. and {Narayan}, G. and {Nieto-Santisteban}, M.~A. and {Norberg}, P. and {Rest}, A. and {Saglia}, R.~P. and {Szalay}, A. and {Thakar}, A.~R. and {Tonry}, J.~L. and {Valenti}, J. and {Werner}, S. and {White}, R. and {Denneau}, L. and {Draper}, P.~W. and {Hodapp}, K.~W. and {Jedicke}, R. and {Kaiser}, N. and {Kudritzki}, R.~P. and {Price}, P.~A. and {Wainscoat}, R.~J. and {Chastel}, S. and {McLean}, B. and {Postman}, M. and {Shiao}, B.},
        title = "{The Pan-STARRS1 Database and Data Products}",
      journal = {\apjs},
         year = 2020,
        month = nov,
       volume = {251},
       number = {1},
          eid = {7},
        pages = {7},
          doi = {10.3847/1538-4365/abb82d},
archivePrefix = {arXiv},
       eprint = {1612.05243},
 primaryClass = {astro-ph.IM},
       adsurl = {https://ui.adsabs.harvard.edu/abs/2020ApJS..251....7F}
}

@ARTICLE{2019AJ....157..168D,
       author = {{Dey}, Arjun and {Schlegel}, David J. and {Lang}, Dustin and {Blum}, Robert and {Burleigh}, Kaylan and {Fan}, Xiaohui and {Findlay}, Joseph R. and {Finkbeiner}, Doug and {Herrera}, David and {Juneau}, St{\'e}phanie and {Landriau}, Martin and {Levi}, Michael and {McGreer}, Ian and {Meisner}, Aaron and {Myers}, Adam D. and {Moustakas}, John and {Nugent}, Peter and {Patej}, Anna and {Schlafly}, Edward F. and {Walker}, Alistair R. and {Valdes}, Francisco and {Weaver}, Benjamin A. and {Y{\`e}che}, Christophe and {Zou}, Hu and {Zhou}, Xu and {Abareshi}, Behzad and {Abbott}, T.~M.~C. and {Abolfathi}, Bela and {Aguilera}, C. and {Alam}, Shadab and {Allen}, Lori and {Alvarez}, A. and {Annis}, James and {Ansarinejad}, Behzad and {Aubert}, Marie and {Beechert}, Jacqueline and {Bell}, Eric F. and {BenZvi}, Segev Y. and {Beutler}, Florian and {Bielby}, Richard M. and {Bolton}, Adam S. and {Brice{\~n}o}, C{\'e}sar and {Buckley-Geer}, Elizabeth J. and {Butler}, Karen and {Calamida}, Annalisa and {Carlberg}, Raymond G. and {Carter}, Paul and {Casas}, Ricard and {Castander}, Francisco J. and {Choi}, Yumi and {Comparat}, Johan and {Cukanovaite}, Elena and {Delubac}, Timoth{\'e}e and {DeVries}, Kaitlin and {Dey}, Sharmila and {Dhungana}, Govinda and {Dickinson}, Mark and {Ding}, Zhejie and {Donaldson}, John B. and {Duan}, Yutong and {Duckworth}, Christopher J. and {Eftekharzadeh}, Sarah and {Eisenstein}, Daniel J. and {Etourneau}, Thomas and {Fagrelius}, Parker A. and {Farihi}, Jay and {Fitzpatrick}, Mike and {Font-Ribera}, Andreu and {Fulmer}, Leah and {G{\"a}nsicke}, Boris T. and {Gaztanaga}, Enrique and {George}, Koshy and {Gerdes}, David W. and {Gontcho}, Satya Gontcho A. and {Gorgoni}, Claudio and {Green}, Gregory and {Guy}, Julien and {Harmer}, Diane and {Hernandez}, M. and {Honscheid}, Klaus and {Huang}, Lijuan Wendy and {James}, David J. and {Jannuzi}, Buell T. and {Jiang}, Linhua and {Joyce}, Richard and {Karcher}, Armin and {Karkar}, Sonia and {Kehoe}, Robert and {Kneib}, Jean-Paul and {Kueter-Young}, Andrea and {Lan}, Ting-Wen and {Lauer}, Tod R. and {Le Guillou}, Laurent and {Le Van Suu}, Auguste and {Lee}, Jae Hyeon and {Lesser}, Michael and {Perreault Levasseur}, Laurence and {Li}, Ting S. and {Mann}, Justin L. and {Marshall}, Robert and {Mart{\'\i}nez-V{\'a}zquez}, C.~E. and {Martini}, Paul and {du Mas des Bourboux}, H{\'e}lion and {McManus}, Sean and {Meier}, Tobias Gabriel and {M{\'e}nard}, Brice and {Metcalfe}, Nigel and {Mu{\~n}oz-Guti{\'e}rrez}, Andrea and {Najita}, Joan and {Napier}, Kevin and {Narayan}, Gautham and {Newman}, Jeffrey A. and {Nie}, Jundan and {Nord}, Brian and {Norman}, Dara J. and {Olsen}, Knut A.~G. and {Paat}, Anthony and {Palanque-Delabrouille}, Nathalie and {Peng}, Xiyan and {Poppett}, Claire L. and {Poremba}, Megan R. and {Prakash}, Abhishek and {Rabinowitz}, David and {Raichoor}, Anand and {Rezaie}, Mehdi and {Robertson}, A.~N. and {Roe}, Natalie A. and {Ross}, Ashley J. and {Ross}, Nicholas P. and {Rudnick}, Gregory and {Safonova}, Sasha and {Saha}, Abhijit and {S{\'a}nchez}, F. Javier and {Savary}, Elodie and {Schweiker}, Heidi and {Scott}, Adam and {Seo}, Hee-Jong and {Shan}, Huanyuan and {Silva}, David R. and {Slepian}, Zachary and {Soto}, Christian and {Sprayberry}, David and {Staten}, Ryan and {Stillman}, Coley M. and {Stupak}, Robert J. and {Summers}, David L. and {Sien Tie}, Suk and {Tirado}, H. and {Vargas-Maga{\~n}a}, Mariana and {Vivas}, A. Katherina and {Wechsler}, Risa H. and {Williams}, Doug and {Yang}, Jinyi and {Yang}, Qian and {Yapici}, Tolga and {Zaritsky}, Dennis and {Zenteno}, A. and {Zhang}, Kai and {Zhang}, Tianmeng and {Zhou}, Rongpu and {Zhou}, Zhimin},
        title = "{Overview of the DESI Legacy Imaging Surveys}",
      journal = {\aj},
         year = 2019,
        month = may,
       volume = {157},
       number = {5},
          eid = {168},
        pages = {168},
          doi = {10.3847/1538-3881/ab089d},
archivePrefix = {arXiv},
       eprint = {1804.08657},
 primaryClass = {astro-ph.IM},
       adsurl = {https://ui.adsabs.harvard.edu/abs/2019AJ....157..168D}
}

@ARTICLE{2014ApJ...797....9W,
       author = {{Whalen}, Daniel J. and {Smidt}, Joseph and {Heger}, Alexander and {Hirschi}, Raphael and {Yusof}, Norhasliza and {Even}, Wesley and {Fryer}, Chris L. and {Stiavelli}, Massimo and {Chen}, Ke-Jung and {Joggerst}, Candace C.},
        title = "{Pair-instability Supernovae in the Local Universe}",
      journal = {\apj},
         year = 2014,
        month = dec,
       volume = {797},
       number = {1},
          eid = {9},
        pages = {9},
          doi = {10.1088/0004-637X/797/1/9},
archivePrefix = {arXiv},
       eprint = {1312.5360},
 primaryClass = {astro-ph.HE},
       adsurl = {https://ui.adsabs.harvard.edu/abs/2014ApJ...797....9W}
}

@ARTICLE{vanBaal_2024_grid,
       author = {{van Baal}, Bart F.~A. and {Jerkstrand}, Anders and {Wongwathanarat}, Annop and {Janka}, Hans-Thomas},
        title = "{Diagnostics of 3D explosion asymmetries of stripped-envelope supernovae by nebular line profiles}",
      journal = {\mnras},
         year = 2024,
        month = aug,
       volume = {532},
       number = {4},
        pages = {4106-4131},
          doi = {10.1093/mnras/stae1603},
archivePrefix = {arXiv},
       eprint = {2404.01763},
 primaryClass = {astro-ph.HE},
       adsurl = {https://ui.adsabs.harvard.edu/abs/2024MNRAS.532.4106V}
}

@ARTICLE{Dessart_2023_Hestars,
       author = {{Dessart}, L. and {Hillier}, D. John and {Woosley}, S.~E. and {Kuncarayakti}, H.},
        title = "{Modeling of the nebular-phase spectral evolution of stripped-envelope supernovae. New grids from 100 to 450 days}",
      journal = {\aap},
         year = 2023,
        month = sep,
       volume = {677},
          eid = {A7},
        pages = {A7},
          doi = {10.1051/0004-6361/202346626},
archivePrefix = {arXiv},
       eprint = {2306.12092},
 primaryClass = {astro-ph.SR},
       adsurl = {https://ui.adsabs.harvard.edu/abs/2023A&A...677A...7D}
}

@ARTICLE{Joggerst_2011_PISNmixing,
       author = {{Joggerst}, C.~C. and {Whalen}, Daniel J.},
        title = "{The Early Evolution of Primordial Pair-instability Supernovae}",
      journal = {\apj},
         year = 2011,
        month = feb,
       volume = {728},
       number = {2},
          eid = {129},
        pages = {129},
          doi = {10.1088/0004-637X/728/2/129},
archivePrefix = {arXiv},
       eprint = {1010.4360},
 primaryClass = {astro-ph.CO},
       adsurl = {https://ui.adsabs.harvard.edu/abs/2011ApJ...728..129J}
}

@ARTICLE{Chatzopoulos_2013_PISNmixing,
       author = {{Chatzopoulos}, E. and {Wheeler}, J. Craig and {Couch}, Sean M.},
        title = "{Multi-dimensional Simulations of Rotating Pair-instability Supernovae}",
      journal = {\apj},
         year = 2013,
        month = oct,
       volume = {776},
       number = {2},
          eid = {129},
        pages = {129},
          doi = {10.1088/0004-637X/776/2/129},
archivePrefix = {arXiv},
       eprint = {1308.4660},
 primaryClass = {astro-ph.HE},
       adsurl = {https://ui.adsabs.harvard.edu/abs/2013ApJ...776..129C}
}

@ARTICLE{Chen_2014_PISNmixing,
       author = {{Chen}, Ke-Jung and {Woosley}, Stan and {Heger}, Alexander and {Almgren}, Ann and {Whalen}, Daniel J.},
        title = "{Two-dimensional Simulations of Pulsational Pair-instability Supernovae}",
      journal = {\apj},
         year = 2014,
        month = sep,
       volume = {792},
       number = {1},
          eid = {28},
        pages = {28},
          doi = {10.1088/0004-637X/792/1/28},
archivePrefix = {arXiv},
       eprint = {1402.4134},
 primaryClass = {astro-ph.HE},
       adsurl = {https://ui.adsabs.harvard.edu/abs/2014ApJ...792...28C}
}

@ARTICLE{2016ApJ...827...90L,
       author = {{Liu}, Yu-Qian and {Modjaz}, Maryam and {Bianco}, Federica B. and {Graur}, Or},
        title = "{Analyzing the Largest Spectroscopic Data Set of Stripped Supernovae to Improve Their Identifications and Constrain Their Progenitors}",
      journal = {\apj},
         year = 2016,
        month = aug,
       volume = {827},
       number = {2},
          eid = {90},
        pages = {90},
          doi = {10.3847/0004-637X/827/2/90},
archivePrefix = {arXiv},
       eprint = {1510.08049},
 primaryClass = {astro-ph.HE},
       adsurl = {https://ui.adsabs.harvard.edu/abs/2016ApJ...827...90L}
}

@ARTICLE{2023A&A...675A..83H,
       author = {{Holmbo}, S. and {Stritzinger}, M.~D. and {Karamehmetoglu}, E. and {Burns}, C.~R. and {Morrell}, N. and {Ashall}, C. and {Hsiao}, E.~Y. and {Galbany}, L. and {Folatelli}, G. and {Phillips}, M.~M. and {Baron}, E. and {Guti{\'e}rrez}, C.~P. and {Leloudas}, G. and {M{\"u}ller-Bravo}, T.~E. and {Hoeflich}, P. and {Taddia}, F. and {Suntzeff}, N.~B.},
        title = "{The Carnegie Supernova Project I. Spectroscopic analysis of stripped-envelope supernovae}",
      journal = {\aap},
         year = 2023,
        month = jul,
       volume = {675},
          eid = {A83},
        pages = {A83},
          doi = {10.1051/0004-6361/202245334},
archivePrefix = {arXiv},
       eprint = {2302.11304},
 primaryClass = {astro-ph.HE},
       adsurl = {https://ui.adsabs.harvard.edu/abs/2023A&A...675A..83H}
}

@ARTICLE{2017ApJ...845...85L,
       author = {{Liu}, Yu-Qian and {Modjaz}, Maryam and {Bianco}, Federica B.},
        title = "{Analyzing the Largest Spectroscopic Data Set of Hydrogen-poor Super-luminous Supernovae}",
      journal = {\apj},
         year = 2017,
        month = aug,
       volume = {845},
       number = {1},
          eid = {85},
        pages = {85},
          doi = {10.3847/1538-4357/aa7f74},
archivePrefix = {arXiv},
       eprint = {1612.07321},
 primaryClass = {astro-ph.HE},
       adsurl = {https://ui.adsabs.harvard.edu/abs/2017ApJ...845...85L}
}

@software{2014zndo.....11813N,
       author = {{Newville}, Matthew and {Stensitzki}, Till and {Allen}, Daniel B. and {Ingargiola}, Antonino},
        title = "{LMFIT: Non-Linear Least-Square Minimization and Curve-Fitting for Python}",
         year = 2014,
        month = sep,
          eid = {10.5281/zenodo.11813},
          doi = {10.5281/zenodo.11813},
      version = {0.8.0},
    publisher = {Zenodo},
       adsurl = {https://ui.adsabs.harvard.edu/abs/2014zndo.....11813N}
}

@ARTICLE{2026arXiv260709209C,
       author = {{Cotter}, Aidan P. and {pearson}, William J. and {Dey}, Subhrata and {Margalef-Bentabol}, Berta and {Guzm{\'a}n-Ortega}, Alejandro and {Rodriguez-Gomez}, Vicente},
        title = "{Performance of morphological classifiers for galaxy mergers compared to current machine learning methods}",
      journal = {arXiv e-prints},
         year = 2026,
        month = jul,
          eid = {arXiv:2607.09209},
        pages = {arXiv:2607.09209},
          doi = {10.48550/arXiv.2607.09209},
archivePrefix = {arXiv},
       eprint = {2607.09209},
 primaryClass = {astro-ph.GA},
       adsurl = {https://ui.adsabs.harvard.edu/abs/2026arXiv260709209C}
}

@ARTICLE{2017NatAs...1..713T,
       author = {{Terreran}, G. and {Pumo}, M.~L. and {Chen}, T.-W. and {Moriya}, T.~J. and {Taddia}, F. and {Dessart}, L. and {Zampieri}, L. and {Smartt}, S.~J. and {Benetti}, S. and {Inserra}, C. and {Cappellaro}, E. and {Nicholl}, M. and {Fraser}, M. and {Wyrzykowski}, {\L}. and {Udalski}, A. and {Howell}, D.~A. and {McCully}, C. and {Valenti}, S. and {Dimitriadis}, G. and {Maguire}, K. and {Sullivan}, M. and {Smith}, K.~W. and {Yaron}, O. and {Young}, D.~R. and {Anderson}, J.~P. and {Della Valle}, M. and {Elias-Rosa}, N. and {Gal-Yam}, A. and {Jerkstrand}, A. and {Kankare}, E. and {Pastorello}, A. and {Sollerman}, J. and {Turatto}, M. and {Kostrzewa-Rutkowska}, Z. and {Koz{\l}owski}, S. and {Mr{\'o}z}, P. and {Pawlak}, M. and {Pietrukowicz}, P. and {Poleski}, R. and {Skowron}, D. and {Skowron}, J. and {Soszy{\'n}ski}, I. and {Szyma{\'n}ski}, M.~K. and {Ulaczyk}, K.},
        title = "{Hydrogen-rich supernovae beyond the neutrino-driven core-collapse paradigm}",
      journal = {Nature Astronomy},
         year = 2017,
        month = sep,
       volume = {1},
        pages = {713-720},
          doi = {10.1038/s41550-017-0228-8},
archivePrefix = {arXiv},
       eprint = {1709.10475},
 primaryClass = {astro-ph.SR},
       adsurl = {https://ui.adsabs.harvard.edu/abs/2017NatAs...1..713T}
}

@ARTICLE{2018MNRAS.479.3106K,
       author = {{Kozyreva}, Alexandra and {Kromer}, Markus and {Noebauer}, Ulrich M. and {Hirschi}, Raphael},
        title = "{OGLE14-073 - a promising pair-instability supernova candidate}",
      journal = {\mnras},
         year = 2018,
        month = sep,
       volume = {479},
       number = {3},
        pages = {3106-3114},
          doi = {10.1093/mnras/sty983},
archivePrefix = {arXiv},
       eprint = {1804.05791},
 primaryClass = {astro-ph.HE},
       adsurl = {https://ui.adsabs.harvard.edu/abs/2018MNRAS.479.3106K}
}

@ARTICLE{2018A&A...613A...5D,
       author = {{Dessart}, Luc and {Audit}, Edouard},
        title = "{Super-luminous Type II supernovae powered by magnetars}",
      journal = {\aap},
         year = 2018,
        month = may,
       volume = {613},
          eid = {A5},
        pages = {A5},
          doi = {10.1051/0004-6361/201732229},
archivePrefix = {arXiv},
       eprint = {1712.04492},
 primaryClass = {astro-ph.HE},
       adsurl = {https://ui.adsabs.harvard.edu/abs/2018A&A...613A...5D}
}

@ARTICLE{2018A&A...619A.145O,
       author = {{Orellana}, Mariana and {Bersten}, Melina C. and {Moriya}, Takashi J.},
        title = "{Systematic study of magnetar-powered hydrogen-rich supernovae}",
      journal = {\aap},
         year = 2018,
        month = nov,
       volume = {619},
          eid = {A145},
        pages = {A145},
          doi = {10.1051/0004-6361/201832661},
archivePrefix = {arXiv},
       eprint = {1809.06414},
 primaryClass = {astro-ph.HE},
       adsurl = {https://ui.adsabs.harvard.edu/abs/2018A&A...619A.145O}
}

@ARTICLE{2018MNRAS.475L..11M,
       author = {{Moriya}, Takashi J. and {Terreran}, Giacomo and {Blinnikov}, Sergei I.},
        title = "{OGLE-2014-SN-073 as a fallback accretion powered supernova}",
      journal = {\mnras},
         year = 2018,
        month = mar,
       volume = {475},
       number = {1},
        pages = {L11-L14},
          doi = {10.1093/mnrasl/slx200},
archivePrefix = {arXiv},
       eprint = {1712.02579},
 primaryClass = {astro-ph.HE},
       adsurl = {https://ui.adsabs.harvard.edu/abs/2018MNRAS.475L..11M}
}

@ARTICLE{2024A&A...683A.223S,
       author = {{Schulze}, Steve and {Fransson}, Claes and {Kozyreva}, Alexandra and {Chen}, Ting-Wan and {Yaron}, Ofer and {Jerkstrand}, Anders and {Gal-Yam}, Avishay and {Sollerman}, Jesper and {Yan}, Lin and {Kangas}, Tuomas and {Leloudas}, Giorgos and {Omand}, Conor M.~B. and {Smartt}, Stephen J. and {Yang}, Yi and {Nicholl}, Matt and {Sarin}, Nikhil and {Yao}, Yuhan and {Brink}, Thomas G. and {Sharon}, Amir and {Rossi}, Andrea and {Chen}, Ping and {Chen}, Zhihao and {Cikota}, Aleksandar and {De}, Kishalay and {Drake}, Andrew J. and {Filippenko}, Alexei V. and {Fremling}, Christoffer and {Fr{\'e}our}, Laurane and {Fynbo}, Johan P.~U. and {Ho}, Anna Y.~Q. and {Inserra}, Cosimo and {Irani}, Ido and {Kuncarayakti}, Hanindyo and {Lunnan}, Ragnhild and {Mazzali}, Paolo and {Ofek}, Eran O. and {Palazzi}, Eliana and {Perley}, Daniel A. and {Pursiainen}, Miika and {Rothberg}, Barry and {Shingles}, Luke J. and {Smith}, Ken and {Taggart}, Kirsty and {Tartaglia}, Leonardo and {Zheng}, WeiKang and {Anderson}, Joseph P. and {Cassara}, Letizia and {Christensen}, Eric and {George Djorgovski}, S. and {Galbany}, Llu{\'\i}s and {Gkini}, Anamaria and {Graham}, Matthew J. and {Gromadzki}, Mariusz and {Groom}, Steven L. and {Hiramatsu}, Daichi and {Andrew Howell}, D. and {Kasliwal}, Mansi M. and {McCully}, Curtis and {M{\"u}ller-Bravo}, Tom{\'a}s E. and {Paiano}, Simona and {Paraskeva}, Emmanouela and {Pessi}, Priscila J. and {Polishook}, David and {Rau}, Arne and {Rigault}, Mickael and {Rusholme}, Ben},
        title = "{1100 days in the life of the supernova 2018ibb. The best pair-instability supernova candidate, to date}",
      journal = {\aap},
         year = 2024,
        month = mar,
       volume = {683},
          eid = {A223},
        pages = {A223},
          doi = {10.1051/0004-6361/202346855},
archivePrefix = {arXiv},
       eprint = {2305.05796},
 primaryClass = {astro-ph.HE},
       adsurl = {https://ui.adsabs.harvard.edu/abs/2024A&A...683A.223S}
}

@ARTICLE{2022RNAAS...6..122P,
       author = {{Pruzhinskaya}, Maria and {Volnova}, Alina and {Kornilov}, Matwey and {Malanchev}, Konstantin and {Aleo}, Patrick D. and {Ishida}, Emille E.~O. and {Korolev}, Vladimir and {Novinskaya}, Alexandra and {Russeil}, Etienne and {Sreejith}, Sreevarsha and {Blondin}, St{\'e}phane and {Kozyreva}, Alexandra and {SNAD Team}},
        title = "{Could SNAD160 be a Pair-instability Supernova?}",
      journal = {Research Notes of the American Astronomical Society},
         year = 2022,
        month = jun,
       volume = {6},
       number = {6},
          eid = {122},
        pages = {122},
          doi = {10.3847/2515-5172/ac76cf},
       adsurl = {https://ui.adsabs.harvard.edu/abs/2022RNAAS...6..122P}
}

@ARTICLE{2026arXiv260516487H,
       author = {{Hiramatsu}, Daichi and {Berger}, Edo and {Tsuna}, Daichi and {Gomez}, Sebastian and {Kumar}, Harsh and {Blanchard}, Peter K. and {Golay}, Walter W. and {Nugent}, Anya E. and {Moriya}, Takashi J. and {Howell}, D. Andrew and {Filippenko}, Alexei V. and {Brink}, Thomas G. and {Zheng}, WeiKang and {Yang}, Yi and {Andrews}, Moira and {Bostroem}, K. Azalee and {Farah}, Joseph and {McCully}, Curtis and {Newsome}, Megan and {Padilla Gonzalez}, Estefania and {Terreran}, Giacomo},
        title = "{The pair-instability origin of supernova 2023vbw}",
      journal = {arXiv e-prints},
         year = 2026,
        month = may,
          eid = {arXiv:2605.16487},
        pages = {arXiv:2605.16487},
          doi = {10.48550/arXiv.2605.16487},
archivePrefix = {arXiv},
       eprint = {2605.16487},
 primaryClass = {astro-ph.HE},
       adsurl = {https://ui.adsabs.harvard.edu/abs/2026arXiv260516487H}
}

@ARTICLE{2014MNRAS.437..656M,
       author = {{McCrum}, M. and {Smartt}, S.~J. and {Kotak}, R. and {Rest}, A. and {Jerkstrand}, A. and {Inserra}, C. and {Rodney}, S.~A. and {Chen}, T.-W. and {Howell}, D.~A. and {Huber}, M.~E. and {Pastorello}, A. and {Tonry}, J.~L. and {Bresolin}, F. and {Kudritzki}, R.-P. and {Chornock}, R. and {Berger}, E. and {Smith}, K. and {Botticella}, M.~T. and {Foley}, R.~J. and {Fraser}, M. and {Milisavljevic}, D. and {Nicholl}, M. and {Riess}, A.~G. and {Stubbs}, C.~W. and {Valenti}, S. and {Wood-Vasey}, W.~M. and {Wright}, D. and {Young}, D.~R. and {Drout}, M. and {Czekala}, I. and {Burgett}, W.~S. and {Chambers}, K.~C. and {Draper}, P. and {Flewelling}, H. and {Hodapp}, K.~W. and {Kaiser}, N. and {Magnier}, E.~A. and {Metcalfe}, N. and {Price}, P.~A. and {Sweeney}, W. and {Wainscoat}, R.~J.},
        title = "{The superluminous supernova PS1-11ap: bridging the gap between low and high redshift}",
      journal = {\mnras},
         year = 2014,
        month = jan,
       volume = {437},
       number = {1},
        pages = {656-674},
          doi = {10.1093/mnras/stt1923},
archivePrefix = {arXiv},
       eprint = {1310.4417},
 primaryClass = {astro-ph.CO},
       adsurl = {https://ui.adsabs.harvard.edu/abs/2014MNRAS.437..656M}
}

@ARTICLE{2013Natur.502..346N,
       author = {{Nicholl}, M. and {Smartt}, S.~J. and {Jerkstrand}, A. and {Inserra}, C. and {McCrum}, M. and {Kotak}, R. and {Fraser}, M. and {Wright}, D. and {Chen}, T.-W. and {Smith}, K. and {Young}, D.~R. and {Sim}, S.~A. and {Valenti}, S. and {Howell}, D.~A. and {Bresolin}, F. and {Kudritzki}, R.~P. and {Tonry}, J.~L. and {Huber}, M.~E. and {Rest}, A. and {Pastorello}, A. and {Tomasella}, L. and {Cappellaro}, E. and {Benetti}, S. and {Mattila}, S. and {Kankare}, E. and {Kangas}, T. and {Leloudas}, G. and {Sollerman}, J. and {Taddia}, F. and {Berger}, E. and {Chornock}, R. and {Narayan}, G. and {Stubbs}, C.~W. and {Foley}, R.~J. and {Lunnan}, R. and {Soderberg}, A. and {Sanders}, N. and {Milisavljevic}, D. and {Margutti}, R. and {Kirshner}, R.~P. and {Elias-Rosa}, N. and {Morales-Garoffolo}, A. and {Taubenberger}, S. and {Botticella}, M.~T. and {Gezari}, S. and {Urata}, Y. and {Rodney}, S. and {Riess}, A.~G. and {Scolnic}, D. and {Wood-Vasey}, W.~M. and {Burgett}, W.~S. and {Chambers}, K. and {Flewelling}, H.~A. and {Magnier}, E.~A. and {Kaiser}, N. and {Metcalfe}, N. and {Morgan}, J. and {Price}, P.~A. and {Sweeney}, W. and {Waters}, C.},
        title = "{Slowly fading super-luminous supernovae that are not pair-instability explosions}",
      journal = {\nat},
         year = 2013,
        month = oct,
       volume = {502},
       number = {7471},
        pages = {346-349},
          doi = {10.1038/nature12569},
archivePrefix = {arXiv},
       eprint = {1310.4446},
 primaryClass = {astro-ph.CO},
       adsurl = {https://ui.adsabs.harvard.edu/abs/2013Natur.502..346N}
}

@ARTICLE{2015MNRAS.452.1567C,
       author = {{Chen}, T.-W. and {Smartt}, S.~J. and {Jerkstrand}, A. and {Nicholl}, M. and {Bresolin}, F. and {Kotak}, R. and {Polshaw}, J. and {Rest}, A. and {Kudritzki}, R. and {Zheng}, Z. and {Elias-Rosa}, N. and {Smith}, K. and {Inserra}, C. and {Wright}, D. and {Kankare}, E. and {Kangas}, T. and {Fraser}, M.},
        title = "{The host galaxy and late-time evolution of the superluminous supernova PTF12dam}",
      journal = {\mnras},
         year = 2015,
        month = sep,
       volume = {452},
       number = {2},
        pages = {1567-1586},
          doi = {10.1093/mnras/stv1360},
archivePrefix = {arXiv},
       eprint = {1409.7728},
 primaryClass = {astro-ph.GA},
       adsurl = {https://ui.adsabs.harvard.edu/abs/2015MNRAS.452.1567C}
}

@ARTICLE{2018ApJ...855....2Q,
       author = {{Quimby}, Robert M. and {De Cia}, Annalisa and {Gal-Yam}, Avishay and {Leloudas}, Giorgos and {Lunnan}, Ragnhild and {Perley}, Daniel A. and {Vreeswijk}, Paul M. and {Yan}, Lin and {Bloom}, Joshua S. and {Cenko}, S. Bradley and {Cooke}, Jeff and {Ellis}, Richard and {Filippenko}, Alexei V. and {Kasliwal}, Mansi M. and {Kleiser}, Io K.~W. and {Kulkarni}, Shrinivas R. and {Matheson}, Thomas and {Nugent}, Peter E. and {Pan}, Yen-Chen and {Silverman}, Jeffrey M. and {Sternberg}, Assaf and {Sullivan}, Mark and {Yaron}, Ofer},
        title = "{Spectra of Hydrogen-poor Superluminous Supernovae from the Palomar Transient Factory}",
      journal = {\apj},
         year = 2018,
        month = mar,
       volume = {855},
       number = {1},
          eid = {2},
        pages = {2},
          doi = {10.3847/1538-4357/aaac2f},
archivePrefix = {arXiv},
       eprint = {1802.07820},
 primaryClass = {astro-ph.HE},
       adsurl = {https://ui.adsabs.harvard.edu/abs/2018ApJ...855....2Q}
}

@ARTICLE{2016ApJ...831..144L,
       author = {{Lunnan}, R. and {Chornock}, R. and {Berger}, E. and {Milisavljevic}, D. and {Jones}, D.~O. and {Rest}, A. and {Fong}, W. and {Fransson}, C. and {Margutti}, R. and {Drout}, M.~R. and {Blanchard}, P.~K. and {Challis}, P. and {Cowperthwaite}, P.~S. and {Foley}, R.~J. and {Kirshner}, R.~P. and {Morrell}, N. and {Riess}, A.~G. and {Roth}, K.~C. and {Scolnic}, D. and {Smartt}, S.~J. and {Smith}, K.~W. and {Villar}, V.~A. and {Chambers}, K.~C. and {Draper}, P.~W. and {Huber}, M.~E. and {Kaiser}, N. and {Kudritzki}, R.-P. and {Magnier}, E.~A. and {Metcalfe}, N. and {Waters}, C.},
        title = "{PS1-14bj: A Hydrogen-poor Superluminous Supernova With a Long Rise and Slow Decay}",
      journal = {\apj},
         year = 2016,
        month = nov,
       volume = {831},
       number = {2},
          eid = {144},
        pages = {144},
          doi = {10.3847/0004-637X/831/2/144},
archivePrefix = {arXiv},
       eprint = {1605.05235},
 primaryClass = {astro-ph.HE},
       adsurl = {https://ui.adsabs.harvard.edu/abs/2016ApJ...831..144L}
}

@ARTICLE{2017MNRAS.468.4642I,
       author = {{Inserra}, C. and {Nicholl}, M. and {Chen}, T.-W. and {Jerkstrand}, A. and {Smartt}, S.~J. and {Kr{\"u}hler}, T. and {Anderson}, J.~P. and {Baltay}, C. and {Della Valle}, M. and {Fraser}, M. and {Gal-Yam}, A. and {Galbany}, L. and {Kankare}, E. and {Maguire}, K. and {Rabinowitz}, D. and {Smith}, K. and {Valenti}, S. and {Young}, D.~R.},
        title = "{Complexity in the light curves and spectra of slow-evolving superluminous supernovae}",
      journal = {\mnras},
         year = 2017,
        month = jul,
       volume = {468},
       number = {4},
        pages = {4642-4662},
          doi = {10.1093/mnras/stx834},
archivePrefix = {arXiv},
       eprint = {1701.00941},
 primaryClass = {astro-ph.HE},
       adsurl = {https://ui.adsabs.harvard.edu/abs/2017MNRAS.468.4642I}
}

@ARTICLE{2017ApJ...835...13J,
       author = {{Jerkstrand}, A. and {Smartt}, S.~J. and {Inserra}, C. and {Nicholl}, M. and {Chen}, T.-W. and {Kr{\"u}hler}, T. and {Sollerman}, J. and {Taubenberger}, S. and {Gal-Yam}, A. and {Kankare}, E. and {Maguire}, K. and {Fraser}, M. and {Valenti}, S. and {Sullivan}, M. and {Cartier}, R. and {Young}, D.~R.},
        title = "{Long-duration Superluminous Supernovae at Late Times}",
      journal = {\apj},
         year = 2017,
        month = jan,
       volume = {835},
       number = {1},
          eid = {13},
        pages = {13},
          doi = {10.3847/1538-4357/835/1/13},
archivePrefix = {arXiv},
       eprint = {1608.02994},
 primaryClass = {astro-ph.HE},
       adsurl = {https://ui.adsabs.harvard.edu/abs/2017ApJ...835...13J}
}

@ARTICLE{2016ApJ...826...39N,
       author = {{Nicholl}, M. and {Berger}, E. and {Smartt}, S.~J. and {Margutti}, R. and {Kamble}, A. and {Alexander}, K.~D. and {Chen}, T.-W. and {Inserra}, C. and {Arcavi}, I. and {Blanchard}, P.~K. and {Cartier}, R. and {Chambers}, K.~C. and {Childress}, M.~J. and {Chornock}, R. and {Cowperthwaite}, P.~S. and {Drout}, M. and {Flewelling}, H.~A. and {Fraser}, M. and {Gal-Yam}, A. and {Galbany}, L. and {Harmanen}, J. and {Holoien}, T.~W.-S. and {Hosseinzadeh}, G. and {Howell}, D.~A. and {Huber}, M.~E. and {Jerkstrand}, A. and {Kankare}, E. and {Kochanek}, C.~S. and {Lin}, Z.-Y. and {Lunnan}, R. and {Magnier}, E.~A. and {Maguire}, K. and {McCully}, C. and {McDonald}, M. and {Metzger}, B.~D. and {Milisavljevic}, D. and {Mitra}, A. and {Reynolds}, T. and {Saario}, J. and {Shappee}, B.~J. and {Smith}, K.~W. and {Valenti}, S. and {Villar}, V.~A. and {Waters}, C. and {Young}, D.~R.},
        title = "{SN 2015BN: A Detailed Multi-wavelength View of a Nearby Superluminous Supernova}",
      journal = {\apj},
         year = 2016,
        month = jul,
       volume = {826},
       number = {1},
          eid = {39},
        pages = {39},
          doi = {10.3847/0004-637X/826/1/39},
archivePrefix = {arXiv},
       eprint = {1603.04748},
 primaryClass = {astro-ph.SR},
       adsurl = {https://ui.adsabs.harvard.edu/abs/2016ApJ...826...39N}
}

@ARTICLE{2016ApJ...828L..18N,
       author = {{Nicholl}, M. and {Berger}, E. and {Margutti}, R. and {Chornock}, R. and {Blanchard}, P.~K. and {Jerkstrand}, A. and {Smartt}, S.~J. and {Arcavi}, I. and {Challis}, P. and {Chambers}, K.~C. and {Chen}, T.-W. and {Cowperthwaite}, P.~S. and {Gal-Yam}, A. and {Hosseinzadeh}, G. and {Howell}, D.~A. and {Inserra}, C. and {Kankare}, E. and {Magnier}, E.~A. and {Maguire}, K. and {Mazzali}, P.~A. and {McCully}, C. and {Milisavljevic}, D. and {Smith}, K.~W. and {Taubenberger}, S. and {Valenti}, S. and {Wainscoat}, R.~J. and {Yaron}, O. and {Young}, D.~R.},
        title = "{Superluminous Supernova SN 2015bn in the Nebular Phase: Evidence for the Engine-powered Explosion of a Stripped Massive Star}",
      journal = {\apjl},
         year = 2016,
        month = sep,
       volume = {828},
       number = {2},
          eid = {L18},
        pages = {L18},
          doi = {10.3847/2041-8205/828/2/L18},
archivePrefix = {arXiv},
       eprint = {1608.02995},
 primaryClass = {astro-ph.HE},
       adsurl = {https://ui.adsabs.harvard.edu/abs/2016ApJ...828L..18N}
}

@ARTICLE{2018ApJ...866L..24N,
       author = {{Nicholl}, Matt and {Blanchard}, Peter K. and {Berger}, Edo and {Alexander}, Kate D. and {Metzger}, Brian D. and {Bhirombhakdi}, Kornpob and {Chornock}, Ryan and {Coppejans}, Deanne and {Gomez}, Sebastian and {Margalit}, Ben and {Margutti}, Raffaella and {Terreran}, Giacomo},
        title = "{One Thousand Days of SN2015bn: HST Imaging Shows a Light Curve Flattening Consistent with Magnetar Predictions}",
      journal = {\apjl},
         year = 2018,
        month = oct,
       volume = {866},
       number = {2},
          eid = {L24},
        pages = {L24},
          doi = {10.3847/2041-8213/aae70d},
archivePrefix = {arXiv},
       eprint = {1809.02755},
 primaryClass = {astro-ph.HE},
       adsurl = {https://ui.adsabs.harvard.edu/abs/2018ApJ...866L..24N}
}

@ARTICLE{2009Natur.462..624G,
       author = {{Gal-Yam}, A. and {Mazzali}, P. and {Ofek}, E.~O. and {Nugent}, P.~E. and {Kulkarni}, S.~R. and {Kasliwal}, M.~M. and {Quimby}, R.~M. and {Filippenko}, A.~V. and {Cenko}, S.~B. and {Chornock}, R. and {Waldman}, R. and {Kasen}, D. and {Sullivan}, M. and {Beshore}, E.~C. and {Drake}, A.~J. and {Thomas}, R.~C. and {Bloom}, J.~S. and {Poznanski}, D. and {Miller}, A.~A. and {Foley}, R.~J. and {Silverman}, J.~M. and {Arcavi}, I. and {Ellis}, R.~S. and {Deng}, J.},
        title = "{Supernova 2007bi as a pair-instability explosion}",
      journal = {\nat},
         year = 2009,
        month = dec,
       volume = {462},
       number = {7273},
        pages = {624-627},
          doi = {10.1038/nature08579},
archivePrefix = {arXiv},
       eprint = {1001.1156},
 primaryClass = {astro-ph.CO},
       adsurl = {https://ui.adsabs.harvard.edu/abs/2009Natur.462..624G}
}

@ARTICLE{2010A&A...512A..70Y,
       author = {{Young}, D.~R. and {Smartt}, S.~J. and {Valenti}, S. and {Pastorello}, A. and {Benetti}, S. and {Benn}, C.~R. and {Bersier}, D. and {Botticella}, M.~T. and {Corradi}, R.~L.~M. and {Harutyunyan}, A.~H. and {Hrudkova}, M. and {Hunter}, I. and {Mattila}, S. and {de Mooij}, E.~J.~W. and {Navasardyan}, H. and {Snellen}, I.~A.~G. and {Tanvir}, N.~R. and {Zampieri}, L.},
        title = "{Two type Ic supernovae in low-metallicity, dwarf galaxies: diversity of explosions}",
      journal = {\aap},
         year = 2010,
        month = mar,
       volume = {512},
          eid = {A70},
        pages = {A70},
          doi = {10.1051/0004-6361/200913004},
archivePrefix = {arXiv},
       eprint = {0910.2248},
 primaryClass = {astro-ph.CO},
       adsurl = {https://ui.adsabs.harvard.edu/abs/2010A&A...512A..70Y}
}

@ARTICLE{2014A&A...565A..70K,
       author = {{Kozyreva}, A. and {Blinnikov}, S. and {Langer}, N. and {Yoon}, S.-C.},
        title = "{Observational properties of low-redshift pair instability supernovae}",
      journal = {\aap},
         year = 2014,
        month = may,
       volume = {565},
          eid = {A70},
        pages = {A70},
          doi = {10.1051/0004-6361/201423447},
archivePrefix = {arXiv},
       eprint = {1403.5212},
 primaryClass = {astro-ph.HE},
       adsurl = {https://ui.adsabs.harvard.edu/abs/2014A&A...565A..70K}
}

@ARTICLE{Dessart_2013_PISN,
       author = {{Dessart}, Luc and {Waldman}, Roni and {Livne}, Eli and {Hillier}, D. John and {Blondin}, St{\'e}phane},
        title = "{Radiative properties of pair-instability supernova explosions}",
      journal = {\mnras},
         year = 2013,
        month = feb,
       volume = {428},
       number = {4},
        pages = {3227-3251},
          doi = {10.1093/mnras/sts269},
archivePrefix = {arXiv},
       eprint = {1210.6163},
 primaryClass = {astro-ph.SR},
       adsurl = {https://ui.adsabs.harvard.edu/abs/2013MNRAS.428.3227D}
}

@ARTICLE{2022A&A...660L...9D,
       author = {{Dessart}, L. and {Hillier}, D. John},
        title = "{Modeling the signatures of interaction in Type II supernovae: UV emission, high-velocity features, broad-boxy profiles}",
      journal = {\aap},
         year = 2022,
        month = apr,
       volume = {660},
          eid = {L9},
        pages = {L9},
          doi = {10.1051/0004-6361/202243372},
archivePrefix = {arXiv},
       eprint = {2204.00446},
 primaryClass = {astro-ph.SR},
       adsurl = {https://ui.adsabs.harvard.edu/abs/2022A&A...660L...9D}
}

@ARTICLE{2020A&A...640A..56R,
       author = {{Renzo}, M. and {Farmer}, R. and {Justham}, S. and {G{\"o}tberg}, Y. and {de Mink}, S.~E. and {Zapartas}, E. and {Marchant}, P. and {Smith}, N.},
        title = "{Predictions for the hydrogen-free ejecta of pulsational pair-instability supernovae}",
      journal = {\aap},
         year = 2020,
        month = aug,
       volume = {640},
          eid = {A56},
        pages = {A56},
          doi = {10.1051/0004-6361/202037710},
archivePrefix = {arXiv},
       eprint = {2002.05077},
 primaryClass = {astro-ph.SR},
       adsurl = {https://ui.adsabs.harvard.edu/abs/2020A&A...640A..56R}
}

@ARTICLE{2023PASP..135b4503M,
       author = {{Malanchev}, Konstantin and {Kornilov}, Matwey V. and {Pruzhinskaya}, Maria V. and {Ishida}, Emille E.~O. and {Aleo}, Patrick D. and {Korolev}, Vladimir S. and {Lavrukhina}, Anastasia and {Russeil}, Etienne and {Sreejith}, Sreevarsha and {Volnova}, Alina A. and {Voloshina}, Anastasiya and {Krone-Martins}, Alberto},
        title = "{The SNAD Viewer: Everything You Want to Know about Your Favorite ZTF Object}",
      journal = {\pasp},
         year = 2023,
        month = feb,
       volume = {135},
       number = {1044},
          eid = {024503},
        pages = {024503},
          doi = {10.1088/1538-3873/acb292},
archivePrefix = {arXiv},
       eprint = {2211.07605},
 primaryClass = {astro-ph.IM},
       adsurl = {https://ui.adsabs.harvard.edu/abs/2023PASP..135b4503M}
}

@ARTICLE{2017ApJ...836..244W,
       author = {{Woosley}, S.~E.},
        title = "{Pulsational Pair-instability Supernovae}",
      journal = {\apj},
         year = 2017,
        month = feb,
       volume = {836},
       number = {2},
          eid = {244},
        pages = {244},
          doi = {10.3847/1538-4357/836/2/244},
archivePrefix = {arXiv},
       eprint = {1608.08939},
 primaryClass = {astro-ph.HE},
       adsurl = {https://ui.adsabs.harvard.edu/abs/2017ApJ...836..244W}
}

@ARTICLE{2010MNRAS.408...87M,
       author = {{Mazzali}, Paolo A. and {Maurer}, I. and {Valenti}, S. and {Kotak}, R. and {Hunter}, D.},
        title = "{The Type Ic SN 2007gr: a census of the ejecta from late-time optical-infrared spectra}",
      journal = {\mnras},
         year = 2010,
        month = oct,
       volume = {408},
       number = {1},
        pages = {87-96},
          doi = {10.1111/j.1365-2966.2010.17133.x},
archivePrefix = {arXiv},
       eprint = {1006.4259},
 primaryClass = {astro-ph.HE},
       adsurl = {https://ui.adsabs.harvard.edu/abs/2010MNRAS.408...87M}
}

@ARTICLE{2025Natur.644..634S,
       author = {{Schulze}, Steve and {Gal-Yam}, Avishay and {Dessart}, Luc and {Miller}, Adam A. and {Woosley}, Stan E. and {Yang}, Yi and {Bulla}, Mattia and {Yaron}, Ofer and {Sollerman}, Jesper and {Filippenko}, Alexei V. and {Hinds}, K.-Ryan and {Perley}, Daniel A. and {Tsuna}, Daichi and {Lunnan}, Ragnhild and {Sarin}, Nikhil and {Brennan}, Se{\'a}n J. and {Brink}, Thomas G. and {Bruch}, Rachel J. and {Chen}, Ping and {Das}, Kaustav K. and {Dhawan}, Suhail and {Fransson}, Claes and {Fremling}, Christoffer and {Gangopadhyay}, Anjasha and {Irani}, Ido and {Jerkstrand}, Anders and {Kne{\v{z}}evi{\'c}}, Nikola and {Kushnir}, Doron and {Maeda}, Keiichi and {Maguire}, Kate and {Ofek}, Eran and {Omand}, Conor M.~B. and {Qin}, Yu-Jing and {Sharma}, Yashvi and {Sit}, Tawny and {Srinivasaragavan}, Gokul P. and {Strothjohann}, Nora L. and {Takei}, Yuki and {Waxman}, Eli and {Yan}, Lin and {Yao}, Yuhan and {Zheng}, WeiKang and {Zimmerman}, Erez A. and {Bellm}, Eric C. and {Coughlin}, Michael W. and {Masci}, Frank J. and {Purdum}, Josiah and {Rigault}, Micka{\"e}l and {Wold}, Avery and {Kulkarni}, Shrinivas R.},
        title = "{Extremely stripped supernova reveals a silicon and sulfur formation site}",
      journal = {\nat},
         year = 2025,
        month = aug,
       volume = {644},
       number = {8077},
        pages = {634-639},
          doi = {10.1038/s41586-025-09375-3},
archivePrefix = {arXiv},
       eprint = {2409.02054},
 primaryClass = {astro-ph.HE},
       adsurl = {https://ui.adsabs.harvard.edu/abs/2025Natur.644..634S}
}

@ARTICLE{2017MNRAS.467..369S,
       author = {{Silverman}, Jeffrey M. and {Pickett}, Stephanie and {Wheeler}, J. Craig and {Filippenko}, Alexei V. and {Vink{\'o}}, J{\'o}zsef and {Marion}, G.~H. and {Cenko}, S. Bradley and {Chornock}, Ryan and {Clubb}, Kelsey I. and {Foley}, Ryan J. and {Graham}, Melissa L. and {Kelly}, Patrick L. and {Matheson}, Thomas and {Shields}, Joseph C.},
        title = "{After the Fall: Late-Time Spectroscopy of Type IIP Supernovae}",
      journal = {\mnras},
         year = 2017,
        month = may,
       volume = {467},
       number = {1},
        pages = {369-411},
          doi = {10.1093/mnras/stx058},
archivePrefix = {arXiv},
       eprint = {1610.07654},
 primaryClass = {astro-ph.SR},
       adsurl = {https://ui.adsabs.harvard.edu/abs/2017MNRAS.467..369S}
}

@ARTICLE{2020A&A...635A.186P,
       author = {{Prentice}, S.~J. and {Maguire}, K. and {Fl{\"o}rs}, A. and {Taubenberger}, S. and {Inserra}, C. and {Frohmaier}, C. and {Chen}, T.~W. and {Anderson}, J.~P. and {Ashall}, C. and {Clark}, P. and {Fraser}, M. and {Galbany}, L. and {Gal-Yam}, A. and {Gromadzki}, M. and {Guti{\'e}rrez}, C.~P. and {James}, P.~A. and {Jonker}, P.~G. and {Kankare}, E. and {Leloudas}, G. and {Magee}, M.~R. and {Mazzali}, P.~A. and {Nicholl}, M. and {Pursiainen}, M. and {Skillen}, K. and {Smartt}, S.~J. and {Smith}, K.~W. and {Vogl}, C. and {Young}, D.~R.},
        title = "{The rise and fall of an extraordinary Ca-rich transient. The discovery of ATLAS19dqr/SN 2019bkc}",
      journal = {\aap},
         year = 2020,
        month = mar,
       volume = {635},
          eid = {A186},
        pages = {A186},
          doi = {10.1051/0004-6361/201936515},
archivePrefix = {arXiv},
       eprint = {1909.05567},
 primaryClass = {astro-ph.HE},
       adsurl = {https://ui.adsabs.harvard.edu/abs/2020A&A...635A.186P}
}

@ARTICLE{2006NuPhA.777..424N,
       author = {{Nomoto}, Ken'ichi and {Tominaga}, Nozomu and {Umeda}, Hideyuki and {Kobayashi}, Chiaki and {Maeda}, Keiichi},
        title = "{Nucleosynthesis yields of core-collapse supernovae and hypernovae, and galactic chemical evolution}",
      journal = {\nphysa},
         year = 2006,
        month = oct,
       volume = {777},
        pages = {424-458},
          doi = {10.1016/j.nuclphysa.2006.05.008},
archivePrefix = {arXiv},
       eprint = {astro-ph/0605725},
 primaryClass = {astro-ph},
       adsurl = {https://ui.adsabs.harvard.edu/abs/2006NuPhA.777..424N}
}

@ARTICLE{1941PASP...53..224M,
       author = {{Minkowski}, R.},
        title = "{Spectra of Supernovae}",
      journal = {\pasp},
         year = 1941,
        month = aug,
       volume = {53},
       number = {314},
        pages = {224},
          doi = {10.1086/125315},
       adsurl = {https://ui.adsabs.harvard.edu/abs/1941PASP...53..224M}
}

@INCOLLECTION{2017hsn..book..195G,
       author = {{Gal-Yam}, Avishay},
        title = "{Observational and Physical Classification of Supernovae}",
    booktitle = {Handbook of Supernovae},
         year = 2017,
       editor = {{Alsabti}, Athem W. and {Murdin}, Paul},
        pages = {195},
          doi = {10.1007/978-3-319-21846-5_35},
       adsurl = {https://ui.adsabs.harvard.edu/abs/2017hsn..book..195G}
}

@ARTICLE{1977SvAL....3..215P,
       author = {{Pskovskii}, Iu. P.},
        title = "{A photometric classification of SN.}",
      journal = {Soviet Astronomy Letters},
         year = 1977,
        month = oct,
       volume = {3},
        pages = {215},
       adsurl = {https://ui.adsabs.harvard.edu/abs/1977SvAL....3..215P}
}

@ARTICLE{2024A&A...692A.208F,
       author = {{Fraga}, B.~M.~O. and {Bom}, C.~R. and {Santos}, A. and {Russeil}, E. and {Leoni}, M. and {Peloton}, J. and {Ishida}, E.~E.~O. and {M{\"o}ller}, A. and {Blondin}, S.},
        title = "{Transient classifiers for Fink: Benchmarks for LSST}",
      journal = {\aap},
         year = 2024,
        month = dec,
       volume = {692},
          eid = {A208},
        pages = {A208},
          doi = {10.1051/0004-6361/202450370},
archivePrefix = {arXiv},
       eprint = {2404.08798},
 primaryClass = {astro-ph.IM},
       adsurl = {https://ui.adsabs.harvard.edu/abs/2024A&A...692A.208F}
}

@ARTICLE{2002PASP..114..833P,
       author = {{Poznanski}, Dovi and {Gal-Yam}, Avishay and {Maoz}, Dan and {Filippenko}, Alexei V. and {Leonard}, Douglas C. and {Matheson}, Thomas},
        title = "{Not Color-Blind: Using Multiband Photometry to Classify Supernovae}",
      journal = {\pasp},
         year = 2002,
        month = aug,
       volume = {114},
       number = {798},
        pages = {833-845},
          doi = {10.1086/341741},
archivePrefix = {arXiv},
       eprint = {astro-ph/0202198},
 primaryClass = {astro-ph},
       adsurl = {https://ui.adsabs.harvard.edu/abs/2002PASP..114..833P}
}

@ARTICLE{2019ApJ...884...83V,
       author = {{Villar}, V.~A. and {Berger}, E. and {Miller}, G. and {Chornock}, R. and {Rest}, A. and {Jones}, D.~O. and {Drout}, M.~R. and {Foley}, R.~J. and {Kirshner}, R. and {Lunnan}, R. and {Magnier}, E. and {Milisavljevic}, D. and {Sanders}, N. and {Scolnic}, D.},
        title = "{Supernova Photometric Classification Pipelines Trained on Spectroscopically Classified Supernovae from the Pan-STARRS1 Medium-deep Survey}",
      journal = {\apj},
         year = 2019,
        month = oct,
       volume = {884},
       number = {1},
          eid = {83},
        pages = {83},
          doi = {10.3847/1538-4357/ab418c},
archivePrefix = {arXiv},
       eprint = {1905.07422},
 primaryClass = {astro-ph.HE},
       adsurl = {https://ui.adsabs.harvard.edu/abs/2019ApJ...884...83V}
}

@ARTICLE{2026arXiv260213036T,
       author = {{Townsend}, A. and {Nordin}, J. and {Kowalski}, M. and {Reusch}, S. and {Anderson}, J.~P. and {Bellm}, E.~C. and {Burgaz}, U. and {Chen}, T.~X. and {Chen}, T.-W. and {Dimitriadis}, G. and {Galbany}, L. and {Goobar}, A. and {Graham}, M.~J. and {Gromadzki}, M. and {Guti{\'e}rrez}, C.~P. and {Hale}, D. and {Inserra}, C. and {Kasliwal}, M. and {Kim}, Y.-L. and {Maguire}, K. and {Masci}, F.~J. and {M{\"u}ller-Bravo}, T.~E. and {Perley}, D.~A. and {Riddle}, R.~L. and {Rigault}, M. and {van Santen}, J. and {Schulze}, S. and {Smith}, M. and {Sollerman}, J. and {Yang}, S.},
        title = "{Photometric classification of supernovae detected by the Zwicky Transient Facility using noise augmentation}",
      journal = {arXiv e-prints},
         year = 2026,
        month = feb,
          eid = {arXiv:2602.13036},
        pages = {arXiv:2602.13036},
          doi = {10.48550/arXiv.2602.13036},
archivePrefix = {arXiv},
       eprint = {2602.13036},
 primaryClass = {astro-ph.IM},
       adsurl = {https://ui.adsabs.harvard.edu/abs/2026arXiv260213036T}
}

@ARTICLE{2026arXiv260414761R,
       author = {{Russeil}, E. and {Lunnan}, R. and {Peloton}, J. and {Schulze}, S. and {Pessi}, P.~J. and {Perley}, D. and {Sollerman}, J. and {Gkini}, A. and {Hu}, Y. and {Chen}, T.-W. and {Bellm}, E.~C. and {Chen}, T.~X. and {Rusholme}, B.},
        title = "{NOMAI : A real-time photometric classifier for superluminous supernovae identification. A science module for the Fink broker}",
      journal = {arXiv e-prints},
         year = 2026,
        month = apr,
          eid = {arXiv:2604.14761},
        pages = {arXiv:2604.14761},
          doi = {10.48550/arXiv.2604.14761},
archivePrefix = {arXiv},
       eprint = {2604.14761},
 primaryClass = {astro-ph.IM},
       adsurl = {https://ui.adsabs.harvard.edu/abs/2026arXiv260414761R}
}

@ARTICLE{1967PhRvL..18..379B,
       author = {{Barkat}, Z. and {Rakavy}, G. and {Sack}, N.},
        title = "{Dynamics of Supernova Explosion Resulting from Pair Formation}",
      journal = {\prl},
         year = 1967,
        month = mar,
       volume = {18},
       number = {10},
        pages = {379-381},
          doi = {10.1103/PhysRevLett.18.379},
       adsurl = {https://ui.adsabs.harvard.edu/abs/1967PhRvL..18..379B}
}

@ARTICLE{1967ApJ...148..803R,
       author = {{Rakavy}, G. and {Shaviv}, G.},
        title = "{Instabilities in Highly Evolved Stellar Models}",
      journal = {\apj},
         year = 1967,
        month = jun,
       volume = {148},
        pages = {803},
          doi = {10.1086/149204},
       adsurl = {https://ui.adsabs.harvard.edu/abs/1967ApJ...148..803R}
}

@ARTICLE{1968Ap&SS...2...96F,
       author = {{Fraley}, Gary S.},
        title = "{Supernovae Explosions Induced by Pair-Production Instability}",
      journal = {\apss},
         year = 1968,
        month = aug,
       volume = {2},
       number = {1},
        pages = {96-114},
          doi = {10.1007/BF00651498},
       adsurl = {https://ui.adsabs.harvard.edu/abs/1968Ap&SS...2...96F}
}

@ARTICLE{1984ApJ...280..825B,
       author = {{Bond}, J.~R. and {Arnett}, W.~D. and {Carr}, B.~J.},
        title = "{The evolution and fate of Very Massive Objects}",
      journal = {\apj},
         year = 1984,
        month = may,
       volume = {280},
        pages = {825-847},
          doi = {10.1086/162057},
       adsurl = {https://ui.adsabs.harvard.edu/abs/1984ApJ...280..825B}
}

@ARTICLE{2002RvMP...74.1015W,
       author = {{Woosley}, S.~E. and {Heger}, A. and {Weaver}, T.~A.},
        title = "{The evolution and explosion of massive stars}",
      journal = {Reviews of Modern Physics},
         year = 2002,
        month = nov,
       volume = {74},
       number = {4},
        pages = {1015-1071},
          doi = {10.1103/RevModPhys.74.1015},
       adsurl = {https://ui.adsabs.harvard.edu/abs/2002RvMP...74.1015W}
}

@INPROCEEDINGS{2026enap....2..680R,
       author = {{Renzo}, Mathieu and {Smith}, Nathan},
        title = "{Pair-instability evolution and explosions in massive stars}",
    booktitle = {Encyclopedia of Astrophysics, Volume 2},
         year = 2026,
       volume = {2},
        month = jan,
        pages = {680-705},
          doi = {10.1016/B978-0-443-21439-4.00019-5},
       adsurl = {https://ui.adsabs.harvard.edu/abs/2026enap....2..680R}
}

@INCOLLECTION{2017hsn..book..939K,
       author = {{Kasen}, Daniel},
        title = "{Unusual Supernovae and Alternative Power Sources}",
    booktitle = {Handbook of Supernovae},
         year = 2017,
       editor = {{Alsabti}, Athem W. and {Murdin}, Paul},
        pages = {939},
          doi = {10.1007/978-3-319-21846-5_32},
       adsurl = {https://ui.adsabs.harvard.edu/abs/2017hsn..book..939K}
}

@ARTICLE{2023NatAs...7..779L,
       author = {{Lin}, Weili and {Wang}, Xiaofeng and {Yan}, Lin and {Gal-Yam}, Avishay and {Mo}, Jun and {Brink}, Thomas G. and {Filippenko}, Alexei V. and {Xiang}, Danfeng and {Lunnan}, Ragnhild and {Zheng}, Weikang and {Brown}, Peter and {Kasliwal}, Mansi and {Fremling}, Christoffer and {Blagorodnova}, Nadejda and {Mirzaqulov}, Davron and {Ehgamberdiev}, Shuhrat A. and {Lin}, Han and {Zhang}, Kaicheng and {Zhang}, Jicheng and {Yan}, Shengyu and {Zhang}, Jujia and {Chen}, Zhihao and {Deng}, Licai and {Wang}, Kun and {Xiao}, Lin and {Wang}, Lingjun},
        title = "{A superluminous supernova lightened by collisions with pulsational pair-instability shells}",
      journal = {Nature Astronomy},
         year = 2023,
        month = jul,
       volume = {7},
        pages = {779-789},
          doi = {10.1038/s41550-023-01957-3},
archivePrefix = {arXiv},
       eprint = {2304.10416},
 primaryClass = {astro-ph.HE},
       adsurl = {https://ui.adsabs.harvard.edu/abs/2023NatAs...7..779L}
}

@ARTICLE{2011ApJ...734..102K,
       author = {{Kasen}, Daniel and {Woosley}, S.~E. and {Heger}, Alexander},
        title = "{Pair Instability Supernovae: Light Curves, Spectra, and Shock Breakout}",
      journal = {\apj},
         year = 2011,
        month = jun,
       volume = {734},
       number = {2},
          eid = {102},
        pages = {102},
          doi = {10.1088/0004-637X/734/2/102},
archivePrefix = {arXiv},
       eprint = {1101.3336},
 primaryClass = {astro-ph.HE},
       adsurl = {https://ui.adsabs.harvard.edu/abs/2011ApJ...734..102K}
}

@INPROCEEDINGS{Nomoto_2001_Hypernovae,
       author = {{Nomoto}, Ken'ichi and {Maeda}, Keiichi and {Umeda}, Hideyuki and {Nakamura}, Takayoshi},
        title = "{Hypernova Nucleosynthesis and Galactic Chemicaal Evolution}",
    booktitle = {The Influence of Binaries on Stellar Population Studies},
         year = 2001,
       editor = {{Vanbeveren}, D.},
       series = {Astrophysics and Space Science Library},
       volume = {264},
        month = jan,
        pages = {507},
          doi = {10.1007/978-94-015-9723-4_36},
archivePrefix = {arXiv},
       eprint = {astro-ph/0105127},
 primaryClass = {astro-ph},
       adsurl = {https://ui.adsabs.harvard.edu/abs/2001ASSL..264..507N}
}

@BOOK{1912vamu.book.....G,
       author = {{Gini}, C.},
        title = "{Variabilit{\`a} e mutabilit{\`a}}",
         year = 1912,
       adsurl = {https://ui.adsabs.harvard.edu/abs/1912vamu.book.....G}
}

\begin{flushleft}
\small
\textit{Affiliations}\\[6pt]
\textsuperscript{1} Astrophysics Division, National Centre for Nuclear Research, Pasteura 7, 02-093 Warsaw, Poland\\
\textsuperscript{2} The Oskar Klein Centre, Department of Astronomy, Stockholm University, AlbaNova 106 91, Stockholm, Sweden\\
\textsuperscript{3} Astronomical Observatory, University of Warsaw, Al. Ujazdowskie 4, 00-478 Warsaw, Poland\\
\textsuperscript{4} Department of Particle Physics and Astrophysics, Weizmann Institute of Science, 234 Herzl St, 76100 Rehovot, Israel\\
\textsuperscript{5} Astronomical Institute, University of Wroc{\l}aw, ul. M. Kopernika 11, 51-622 Wroc{\l}aw, Poland\\
\textsuperscript{6} Division of Physics, Mathematics, and Astronomy, California Institute of Technology, Pasadena, CA 91125, USA\\
\textsuperscript{7} Caltech Optical Observatories, California Institute of Technology, Pasadena, CA 91125, USA\\
\textsuperscript{8} Astrophysics Research Institute, Liverpool John Moores University, 146 Brownlow Hill, Liverpool L3 5RF, UK\\
\textsuperscript{9} Aryabhatta Research Institute of Observational Sciences, Manora Peak, Nainital-263001, Uttarakhand, India\\
\textsuperscript{10} IAASARS, National Observatory of Athens, Metaxa \& Vas. Pavlou St., 15236, Penteli, Athens, Greece\\
\textsuperscript{11} Center for Astrophysics and Cosmology, University of Nova Gorica, Vipavska 11c, 5270 Ajdov\v{s}\v{c}ina, Slovenia\\
\textsuperscript{12} School of Physics, Trinity College Dublin, College Green, Dublin 2, Ireland\\
\textsuperscript{13} Faculty of Physics, University of Bialystok, ul. Ciolkowskiego 1L, 15-245 Bialystok, Poland\\
\textsuperscript{14} Departament de F\'{\i}sica Qu\`antica i Astrof\'{\i}sica (FQA), Universitat de Barcelona (UB), Mart\'{\i} i Franqu\`es 1, E-08028 Barcelona, Spain\\
\textsuperscript{15} Institut de Ci\`encies del Cosmos (ICCUB), Universitat de Barcelona (UB), Mart\'{\i} i Franqu\`es 1, E-08028 Barcelona, Spain\\
\textsuperscript{16} Institut d'Estudis Espacials de Catalunya (IEEC), Edifici RDIT, Campus UPC, 08860 Castelldefels (Barcelona), Spain\\
\textsuperscript{17} Institute of Earth Systems, University of Malta, Msida, Malta\\
\textsuperscript{18} Vereniging Voor Sterrenkunde (VVS), Zeeweg 96, 8200 Brugge, Belgium\\
\textsuperscript{19} Bundesdeutsche Arbeitsgemeinschaft für Veränderliche Sterne (BAV), Munsterdamm 90, 12169 Berlin, Germany\\
\textsuperscript{20} Groupe Européen d’Observations Stellaires (GEOS), 23 Parc de Levesville, 28300 Bailleau l’Evêque, France\\
\textsuperscript{21} Adiyaman University, Astrophysics Application and Research Center, 02040, Adiyaman, Türkiye\\
\textsuperscript{22} Department of Physics \& Astronomy, University of California Los Angeles, 430 Portola Plaza, Los Angeles, CA 90095-1547, US\\
\textsuperscript{23} Astronomical Observatory of the Jagiellonian University, Orla 171, 30-244 Cracow, Poland\\
\textsuperscript{24} Department of Physics and Astronomy, Northwestern University, 2145 Sheridan Rd., Evanston, IL 60208, USA\\
\textsuperscript{25} Center for Interdisciplinary Exploration and Research in Astrophysics (CIERA), Northwestern University, 1800 Sherman Ave., Evanston, IL 60201, USA\\
\textsuperscript{26} NSF-Simons AI Institute for the Sky (SkAI), 172 E. Chestnut St., Chicago, IL 60611, USA\\
\textsuperscript{27} Vilnius University, Faculty of Physics, Institute of Theoretical Physics and Astronomy, Sauletekio av. 3, LT-10257 Vilnius, Lithuania\\
\textsuperscript{28} Astronomska revija Spika, Koprska ulica 94, 1000 Ljubljana, Slovenia\\
\textsuperscript{29} University of National Education Commission, Cracow \\
\textsuperscript{30} Silesian Planetarium, Chorzow, Poland\\
\textsuperscript{31} Finnish Centre for Astronomy with ESO (FINCA), FI-20014 University of Turku, Finland\\
\textsuperscript{32} Department of Physics and Astronomy, FI-20014 University of Turku, Finland\\
\textsuperscript{33} Department of Physics and Astronomy, Northwestern University, 2145 Sheridan Rd., Evanston, IL 60208, USA\\
\textsuperscript{34} Astronomy and Space Physics Department, Taras Shevchenko National University of Kyiv, 4 Glushkova ave, Kyiv, 03022, Ukraine\\
\textsuperscript{35} Adam Mickiewicz University, Faculty of Physics and Astronomy, Astronomical Observatory Institute, Słoneczna 36, 60-286 Poznań, Poland\\
\textsuperscript{36} INAF - Osservatorio Astronomico di Roma, via Frascati 33, Monte Porzio Catone, 00078, Rome, Italy\\
\textsuperscript{37} Janusz Gil Institute of Astronomy, University of Zielona Gora, Szafrana 2, 65-516 Zielona Gora, Poland\\
\textsuperscript{38} DIRAC Institute, Department of Astronomy, University of Washington, 3910 15th Avenue NE, Seattle, WA 98195, USA\\
\textsuperscript{39} IPAC, California Institute of Technology, 1200 E. California Blvd, Pasadena, CA 91125, USA\\
\end{flushleft}


\twocolumn

\begin{appendix} 

\section{Photometry}

The full version of each table is available as supplementary material. $UBVRI$ magnitudes are presented in the Vega system (\citealt{2023A&AGaiaCollaboration}, corresponding zero-points are derived from the Vega stellar spectrum as presented by \citealt{2012PASP..124..140B} when needed). $ocgri$ and $JHK_s$ magnitudes are reported in the AB system. UVW2, UVM2 UVW1, were originally obtained in the Vega system as provided by the \texttt{UVOTSOURCE} pipeline and and were converted to the AB system following the conversions provided in \url{https://swift.gsfc.nasa.gov/analysis/uvot_digest/zeropts.html}. We therefore report these measurements in the AB magnitude system.

The unbinned BHTOM observations are available at \url{https://bhtom.space/public/targets/SN2024afyu}. The facilities involved in the follow up campaign are: Las Cumbres Observatory 40-cm and 1-m telescopes located at Cerro Tololo Interamerican Observatory, Haleakala Observatory, McDonald Observatory, South African Astronomical Observatory, Siding Spring Observatory and Teide Observatory; Lisnyky observational station of Taras Shevchenko National University of Kyiv's AZT-8 70-cm telescope; Adiyaman University Observatory 60-cm telescope; GoChile 40-cm Ritchey-Chretien telescope at El Sauce Observatory; Kryoneri Observatory of the National Observatory of Athens 1.2-meter telescope; private Seestar S30 and S50 telescopes; Vilnius University's Moletai Observatory 35-cm and 80-cm telescopes; Warsaw University Observatory Ostrowik station 60-cm telescope; Planetarium Śląskie 70-cm telescope in Chorzów; Rapid Eye Mount 60-cm telescope at La Silla Observatory operated by the Italian National Institute for Astrophysics; Remote Observatory Atacama Desert 40-cm ODK telescope; Szkolne Obserwatorium Astronomiczne in Bolencina ZWO ASI2600MM Monochrome Astronomy Camera; Montsec Astronomical Observatory 80-cm Joan Oró Telescope; University of Zielona Góra 50-cm telescope at Deep Sky Chile in the Rio Hurtado Valley; Uniwersytetu w Białymstoku Astronomical Observatory 60-cm Telescope; Znith Astronomy Observatory 20-cm Telescope in Malta; and Jagiellonian University Astronomical Observatory 50-cm telescopes.

\begin{table}[h!]
  \centering
  \caption{Portion of ZTF, SEDM and ATLAS photometry. Magnitudes with no associated error bars are upper limits.}
  \label{tab:ZTFSEDMATLASphot}
\begin{tabular}{llll}
\hline
MJD & mag & mag\_err & filter \\
\hline
60368.51 & 20.11 & -- & r \\
60383.36 & 20.04 & -- & o \\
60390.64 & 20.05 & -- & o \\
\hline
\end{tabular}
\end{table}

\begin{table}[h!]
  \centering
  \caption{Portion of BHTOM photometry used in this work.}
  \label{tab:BHTOMphot}
\begin{tabular}{llll}
\hline
MJD & mag & mag\_err & filter \\
\hline
60727.50 & 13.88 & 0.13 & I \\                      
60727.50 & 14.28 & 0.02 & sdssi \\                  
60727.50 & 14.34 & 0.11 & R \\                      
\hline
\end{tabular}
\end{table}

\begin{table}[h!]
  \centering
  \caption{Portion of LBT NIR photometry.}
  \label{tab:LBTNIRphot}
\begin{tabular}{llll}
\hline
MJD & mag & mag\_err & filter \\
\hline
61102.00 & 15.04 & 0.39 & J \\
61102.00 & 14.00 & 0.91 & H \\
61102.00 & 13.81 & 0.46 & Ks \\
\hline
\end{tabular}
\end{table}

\section{Spectra}
\label{app:spec-reduction}

Spectral reduction is performed as follows.
SEDM spectra were reduced using the \texttt{pysedm} fully automated integral field spectrograph pipeline \citep{2019A&A...627A.115R, 2022PASP..134b4505K}. 
ALFOSC and Binospec spectra were reduced using the \texttt{PypeIt} python spectroscopic data reduction pipeline \citep{2020JOSS....5.2308P}. 
The NGPS spectrum was reduced using the quicklook reduction tool
(Fremling et al., in prep). 
The KAST spectrum was obtained using the 300/7500 grating for the red side, and the 600/4310 grism for the blue side. A slit width of 1.0\farcs\ was used in the first 430 seconds of the observation to compensate for high altitude clouds, after which it was opened to 1.5\farcs\ for the remainder of the observation. The \texttt{UCSC} spectral pipeline\footnote{\url{https://github.com/msiebert1/UCSC_spectral_pipeline}.} \citep{2020ApJ...900L..27S} was used for reduction.
The LRIS spectrum was reduced using the \texttt{LPIPE} pipeline \citep{2019PASP..131h4503P}.
NIRES spectra were reduced using a modified version of the IDL based reduction package \texttt{Spextool} \citep{2004PASP..116..362C}, using \texttt{xtellcor} \citep{2003PASP..115..389V}  to correct for telluric features.
LUCI spectra were obtained using ABBA offsets and the zJ and HK filters providing a wavelength coverage of 0.90--1.25  $\mu m$ and 1.47--2.35  $\mu m$. Telluric features were removed using observed standard stars. The two dimensional spectra was reduced using the  Spectroscopic Interactive Pipeline and Graphical Interface \citep[SIPGI;][]{sipgi2022} and then combined to extract the one-dimensional spectrum of each source. 
   
\begin{table}[h!]
  \setlength{\tabcolsep}{3pt}
  \caption{Log of spectroscopic observations.}
  \label{tab:speclog}
\begin{tabular}{lll|lll}
\hline
MJD & Telescope & Instrument &   MJD & Telescope & Instrument\\\hline
60739.45  & P60     & SEDM     &    60849.07  & NOT     & ALFOSC   \\         
60743.43  & P60     & SEDM     &    60862.38  & P60     & SEDM     \\         
60753.50  & P60     & SEDM     &    60864.35  & P60     & SEDM     \\         
60756.24  & NOT     & ALFOSC   &    60869.26  & P60     & SEDM     \\         
60756.52  & P200    & NGPS     &    60880.93  & NOT     & ALFOSC   \\         
60760.44  & P60     & SEDM     &    60881.27  & P60     & SEDM     \\         
60771.39  & P60     & SEDM     &    60889.17  & P60     & SEDM     \\         
60773.41  & P60     & SEDM     &    60906.28  & P60     & SEDM     \\         
60776.00  & Keck-II & NIRES    &    60922.00  & Keck-II & NIRES    \\         
60778.38  & P60     & SEDM     &    60922.91  & NOT     & ALFOSC   \\         
60779.38  & P60     & SEDM     &    60944.16  & P60     & SEDM     \\         
60792.21  & NOT     & ALFOSC   &    60955.84  & NOT     & ALFOSC   \\         
60803.02  & NOT     & ALFOSC   &    60960.00  & Keck-II & NIRES    \\         
60809.00  & Keck-II & NIRES    &    61081.25  & NOT     & ALFOSC   \\         
60811.41  & MMT     & Binospec &    61093.51  & P60     & SEDM     \\         
60816.30  & P60     & SEDM     &    61101.47  & LBT     & LUCI     \\         
60817.37  & P60     & SEDM     &    61103.51  & P60     & SEDM     \\         
60821.42  & P200    & NGPS     &    61104.51  & P60     & SEDM     \\         
60822.31  & P60     & SEDM     &    61111.49  & P60     & SEDM     \\         
60824.39  & P60     & SEDM     &    61116.47  & LBT     & LUCI     \\         
60827.40  & Shane   & KAST     &    61125.00  & Keck-II & NIRES    \\         
60834.35  & P60     & SEDM     &    61128.09  & NOT     & ALFOSC   \\         
60837.39  & P60     & SEDM     &    61168.11  & NOT     & ALFOSC   \\         
60848.43  & P60     & SEDM     &    61200.00  & Keck-I  & LRIS     \\
\hline
\end{tabular}
\end{table}

\section{Explosion epoch estimation}

\begin{figure}[h!]
   \includegraphics[width=\linewidth]{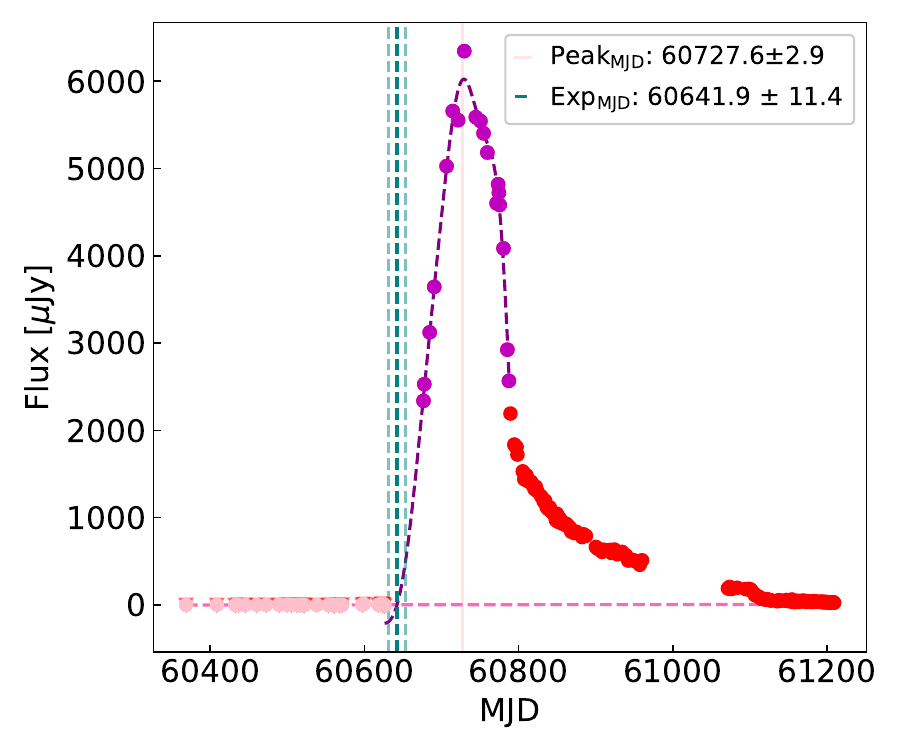}
      \caption{Explosion epoch estimation. ZTF $r$-band points corresponding to the baseline are marked in pink. Points included in the GP interpolation are marked in purple. The horizontal dashed line marks the considered baseline. Vertical dashed green lines mark the explosion epoch, the darker line corresponds to the adopted value and the lighter ones to the associated error. Light pink vertical line marks the peak epoch.
              }
         \label{fig:exp_peak_epoch}
   \end{figure}

\pagebreak

\section{\texttt{SYNAPPS} identification of the feature near 6300\AA}
\label{app:synapps}

Given the several spectral matches to SN~Ib (see above), we investigate the possible presence of hydrogen in the spectra using the open-source spectral-synthesis code \texttt{SYNAPPS} \citep{Thomas2011}. The code assumes spherical symmetry, homologous expansion, and a sharp photosphere emitting a blackbody continuum. We considered observations of SN 2024afyu at $\sim$28 and 120 days normalized with respect to a black body fit. 

The spectral modeling of the line profiles with \texttt{SYNAPPS} reproduce the major absorption features in the observed spectra, supporting the identification of the dominant ions included in the model. Note that the code is not developed to reproduce emission features. The considered ions are: H, He, Si, C, S, O, and Ca. Particular attention was given to the absorption feature near 6300~\AA, which could potentially be associated with H, Si, C or Ca. We therefore explored models with each of these species individually, models with different combinations of them, and models in which one or more of the candidate ions were excluded.

Modeling was performed providing a range of velocities from 5000 km/sec to 20000 km/sec for the considered elements, considering it a free parameter. 
\texttt{SYNAPPS} then searches the resulting multidimensional parameter space (including temperature, optical depths and velocities) using its parallel optimization framework. Figure~\ref{fig:synapps_spec} presents the individual ion contributions to the synthetic spectra. In both cases, carbon produces only a negligible contribution in the region around the absorption feature near 6300~\AA, disfavouring a C-dominated interpretation. Instead, the feature is reproduced primarily by a blend of Si and H, with H providing the dominant contribution to the absorption trough. Ca could also be contributing to the broadness of the feature. The fitting suggests that the observed feature is more plausibly associated with H$\alpha$, potentially blended with \ion{Si}{ii}$\lambda6355$ and/or \ion{Ca}{i}$\lambda$6572, rather than with carbon. However, the identification remains model-dependent because of line blending and the many simplifying assumptions of \texttt{SYNAPPS}.

\begin{figure*}[h!]
   \centering
   \includegraphics[width=\textwidth]{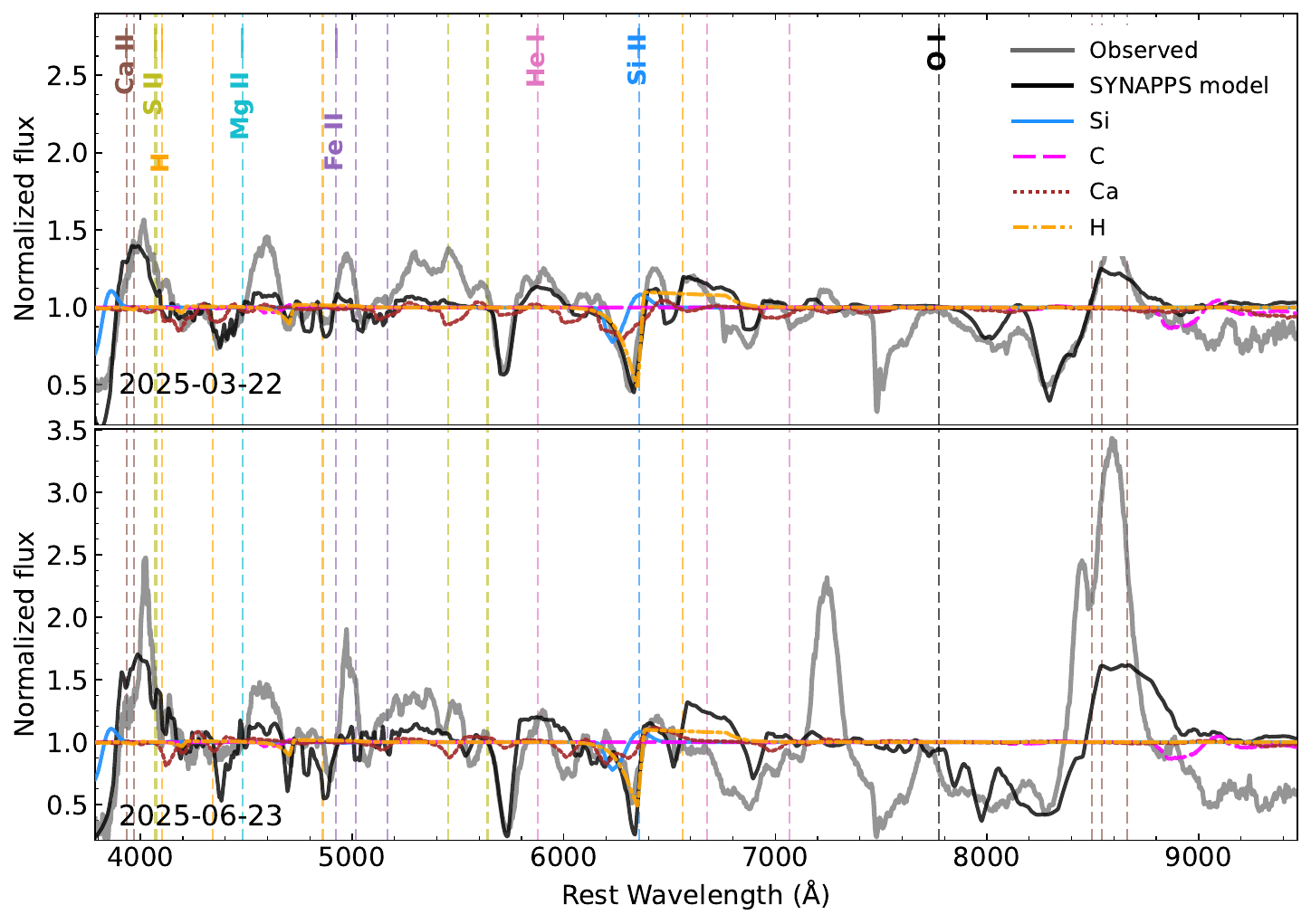}
      \caption{SN 2024afyu (observed with ALFOSC on 2025 March 22, 28.4 rest frame days past peak; and on 2025 June 23, 120.5 rest frame days past peak) against \texttt{SYNAPPS} synthetic spectra. Each panel shows the observed spectrum overplotted with a model including individual contribution of ions in different phases. Spectra are normalised by the best-fit blackbody continuum. Vertical dashed lines mark the rest-frame positions of prominent spectral features; only ions present in the respective model are labeled per panel.}
         \label{fig:synapps_spec}
\end{figure*}

\section{Host SED, morphology and metallicity}
\label{app:host}

Following \citet{Schulze2021a}, we retrieved the science-ready images from the Galaxy Evolution Explorer (GALEX) general release 6/7 \citep{Martin2005a}, the DESI Imaging Legacy Surveys \citep{2019AJ....157..168D}, the Two Micron All Sky Survey \citep[2MASS;][]{2006AJSkrutskie}, and re-processed images from the Wide-field Infrared Survey Explorer \citep[WISE;][]{2010AJ....140.1868W} from the unWISE archive \citep{Lang2014a}. The unWISE images include data from the NEOWISE Reactivation mission \citep{Mainzer2014a, Meisner2017a} through Year 6/7. We measured the host brightness using \texttt{LAMBDAR} \citep[Lambda Adaptive Multi-Band Deblending Algorithm in R;][]{Wright2016a} and utilised tools developed by \citet{Schulze2021a}. All measurements are summarised in Table \ref{tab:host:photometry}.

\begin{table}
    \centering
    \caption{Host galaxy photometry. All measurements are in the AB system and are not corrected for reddening.}
    \begin{tabular}{ccc}
         \hline
         Instrument & Filter & Apparent  \\
         & & magnitude (mag)\\
         \hline
\textit{GALEX} & $FUV$ &$ 18.70 \pm 0.14$\\
\textit{GALEX} & $NUV$ &$ 17.82 \pm 0.05$\\
LS &$ g $&$ 15.71 \pm 0.02$ \\
LS &$ r $&$ 15.19 \pm 0.04$ \\
LS &$ i $&$ 14.97 \pm 0.05$ \\
LS &$ z $&$ 14.73 \pm 0.07$ \\
2MASS &$ J $&$ 14.65 \pm 0.05$\\
2MASS &$ H $&$ 14.89 \pm 0.12$\\
2MASS &$ K $&$ 14.55 \pm 0.08$\\
\textit{WISE} &$W1$ &$15.24 \pm 0.05 $\\
\textit{WISE} &$W2$ &$15.68 \pm 0.05 $\\
         \hline
    \end{tabular}
    \label{tab:host:photometry}
\end{table}

We model the observed host-galaxy spectral energy distribution (SED) with the software package \texttt{Prospector} version 1.4 \citep{Johnson2021a}. \texttt{Prospector} uses the \texttt{Flexible Stellar Population Synthesis} (\texttt{FSPS}) code \citep{Conroy2009a} to generate the underlying physical model and \texttt{python-fsps} \citep{ForemanMackey2014a} to interface with \texttt{FSPS} in \texttt{python}. The \texttt{FSPS} code also accounts for the contribution from diffuse gas based on the \texttt{Cloudy} models of \citet{Byler2017a}. We use the dynamic nested-sampling package \texttt{dynesty} \citep{Speagle2020a} to sample the posterior probability distribution. We assume a Chabrier initial mass function \citep{Chabrier2003a}, a parametric star-formation history that rises linearly at early times and declines exponentially at late times [$t\times\exp(-t/t_{1/e})$], and attenuation following the prescription of \citet{Calzetti2000a}. The priors are identical to those adopted by \citet{Schulze2021a}. Figure \ref{fig:host:sed} shows the SED fit. 

\begin{figure}
\includegraphics[width=\columnwidth]{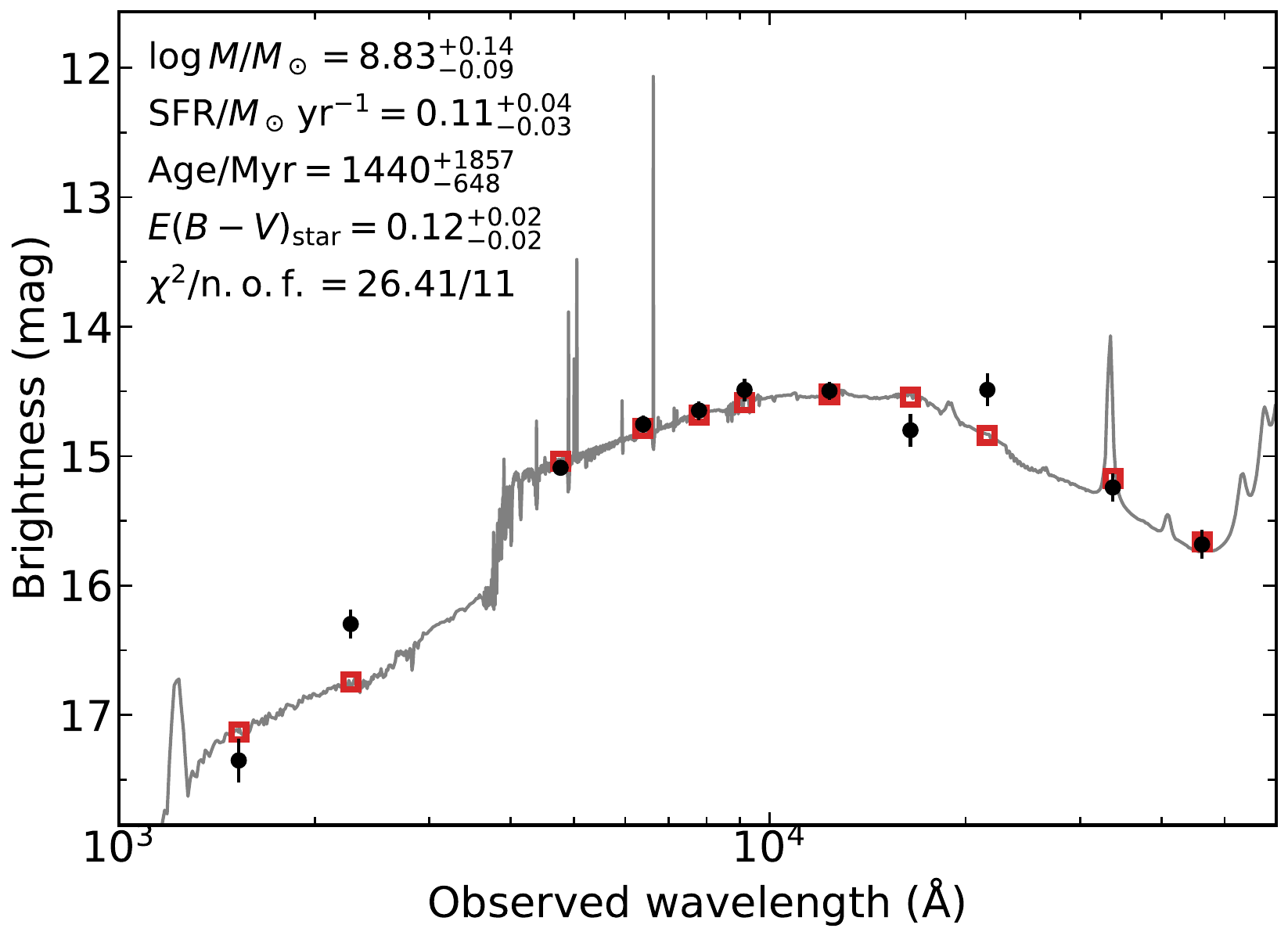}
\caption{Host SED from 1000 to 60000~\AA\ (black dots). The solid line shows the best-fitting model of the SED. The red squares represent the model-predicted magnitudes. The fitting parameters are shown in the upper-left corner. ``n.o.f.'' stands for number of filters.}
\label{fig:host:sed}
\end{figure}

The host shows an irregular and clumpy morphology, which may indicate a disturbed system or a recent merger event. To test this possibility quantitatively, we measured the Gini coefficient ($G$\footnote{The Gini coefficient \citep{1912vamu.book.....G} is typically used in economics to describe wealth inequalities. It has been adopted by \cite{2003ApJ...588..218A} to measure light distribution in galaxies and has since become a widely used non-parametric morphological indicator.}) , the second-order moment of the brightest 20\% of the galaxy light ($M_{20}$) \citep{Lotz2004, Lotz2008}, and the concentration index ($C$) \citep{Conselice2003}. We then applied the merger criteria defined by \cite{2026arXiv260709209C}:

\begin{equation}
\begin{aligned}
S(G, M_{20}) &= G + 0.267M_{20} - 0.143,\\
S(G, C) &= G - 0.162C + 0.149.
\end{aligned}
\label{eq:merger_criteria}
\end{equation}
\vspace{0.1pt}

Table~\ref{tab:hoststats} shows the merger criteria values obtained considering image cutouts of the host obtained from the Pan-STARRS1 Image Access service\footnote{\url{https://ps1images.stsci.edu/}.} \citep{2020ApJS..251....7F}, and the Legacy Survey DR9\footnote{\url{https://www.legacysurvey.org/dr9/}.} \citep{2019AJ....157..168D}. According to \cite{2026arXiv260709209C}, galaxy samples with $S(G,M_{20})>0$ contain approximately 63.5--69.5\% mergers, while samples with $S(G,C)>0$ contain approximately 68.7--72.3\% mergers. Except for the $S(G,M_{20})$ values corresponding to the Legacy Survery $g$ band cutout, the galaxy has both $S(G,M_{20})>0$ and $S(G,C)>0$ in all analysed photometric bands. Thus, the host is consistent with being a merger system.

\begin{table}
\centering
\caption{Morphological statistics measured from Pan-STARRS1 and Legacy Survey DR9 image cutouts.}
\label{tab:hoststats}
\begin{tabular}{lcc|lcc}
\hline
Band & $S(G,M_{20})$ & $S(G,C)$ & Band & $S(G,M_{20})$ & $S(G,C)$ \\
\hline
\multicolumn{3}{l|}{\textbf{Pan-STARRS1}} & 
\multicolumn{3}{l}{\textbf{Legacy Survey DR9}}\\\hline
$g$ & 0.260 & 0.373 & $g$ & -0.053 & 0.256   \\
$r$ & 0.037 & 0.369 & $r$ & 0.353    & 0.880 \\
$i$ & 0.196 & 0.334 & $z$ & 0.182    & 0.385 \\
$y$ & 0.239 & 0.390 & & & \\
$z$ & 0.299 & 0.359 & & & \\
\hline
\end{tabular}
\end{table}

To put SN~2024afyu's host in the context of a general SN host galaxy population, we compare it with hosts of ordinary Type~II and IIb SNe from the Palomar Transient Factory\footnote{The host galaxies of Type~II and IIb SNe from the Palomar Transient Factory have very similar properties \citep{Schulze2021a} and, for simplicity, are treated as a single sample.} study of \citet{Schulze2021a} as well as previous PISN candidates (reported in Sec.~\ref{sec:PISNinterpretation}). The host photometry of the PISN candidates was taken from \citet{2016ApJ...831..144L}, \citet{Perley2016a}, \citet{2017NatAs...1..713T}, \citet{Schulze2018a, 2024A&A...683A.223S}, \citet{Hiramatsu2024a} and refitted with the same host galaxy model. Fig.~\ref{fig:host:mass_vs_sfr} places all objects in the $M_\star$-SFR plane. 

\begin{figure*}
\centering
\includegraphics[width=\textwidth]{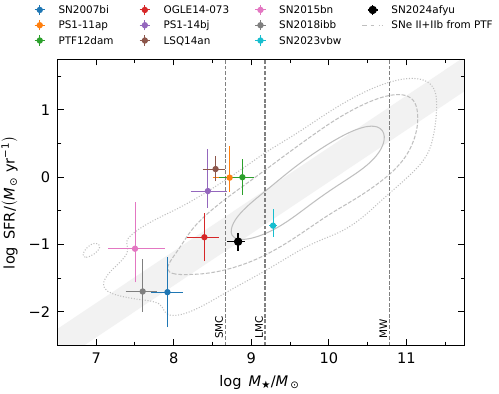}
\caption{Star formation rate against stellar mass for the host of SN\,2024afyu (black circle) and for the hosts of previously proposed PISN candidates (coloured circles). Error bars give 1$\sigma$ uncertainties. Grey contours enclose 66 (solid), 90 (dashed) and 95 (dotted) per cent of the PTF SNe II and IIb host sample. Vertical dashed lines mark the stellar masses of the SMC, the LMC and the Milky Way. SN~2024afyu's host lies on the main sequence of star-forming galaxies (grey band), and within the 66\% contour of the PTF SNe II and IIb host sample, but its mass is fairly large compared to previous PISN candidates.}
\label{fig:host:mass_vs_sfr}
\end{figure*}

Using our latest spectrum, that shows narrow emission from H~II regions along the line of sight, we use the so-called O3N2 diagnostic together with the parametrization of \citet{Curti2017a}, to infer a gas-phase metallicity of $0.55\pm0.02$ solar, where the uncertainty is statistical only. Line fluxes are summarized in Table~\ref{tab:host:emission_lines}.

\begin{table}
    \centering
    \caption{Emission line fluxes measured from the LRIS spectrum observed at $\sim$468 days. The flux measurements are corrected for MW extinction.}
    \begin{tabular}{cc}
         \hline
         Species & Flux  \\
         & ($10^{-15}\,\rm erg\,s^{-1}\,cm^{-2}$) \\
         \hline
         H$\beta$ & $2.68\pm0.20$\\
         {[}\ion{O}{iii}{]}\,$\lambda$4960 & $2.30\pm0.28$\\
         {[}\ion{O}{iii}{]}\,$\lambda$5008 & $7.37\pm0.60$\\
         H$\alpha$ & $6.14\pm0.20$\\
         {[}\ion{N}{ii}{]}\,$\lambda$6584 & $0.72\pm0.11$\\
         \hline
    \end{tabular}
    \label{tab:host:emission_lines}
\end{table}

\section{Spectroscopic measurements}
\label{app:spec-measurements}

The expansion velocities and their associated uncertainties were determined by fitting a Gaussian profile to the core of each absorption feature using the \texttt{lmfit} package \citep{2014zndo.....11813N}. To account for the uncertainty introduced by the choice of fitting region, we repeated the fit using several different wavelength intervals to define the absorption core. The standard deviation of the obtained velocities was taken as the corresponding uncertainty.
The pEW measurements were obtained in an analogous manner, but consider the entirety of the line, instead of only the core. We repeatedly selected slightly different wavelength boundaries to define the pseudo-continuum and adopted the standard deviation of all measurements as the error bar.

\begin{figure*}
   \centering
   \includegraphics[scale=0.5]{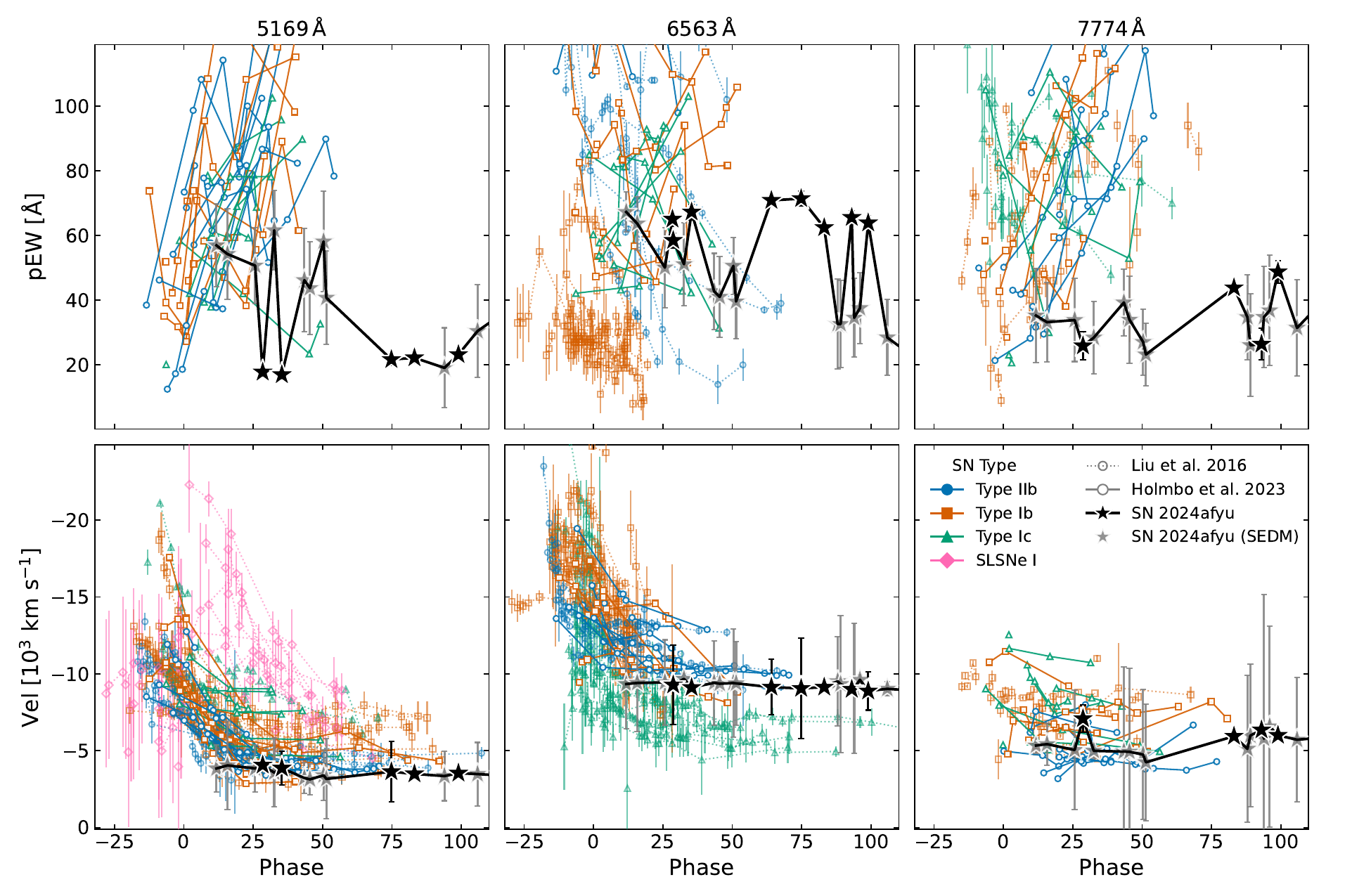}
      \caption{Measured pEWs (top) and velocities (bottom) of SN~2024afyu's absorption features bluewards to 5160\AA\ (left), 6563\AA\ (center) and 7774\AA\ (right) compared to the measurements of the corresponding features presented by \cite{2016ApJ...827...90L} and \cite{2023A&A...675A..83H} for SESN and by \cite{2017ApJ...845...85L} for SLSNe~I. SN~2024afyu measurements are presented in black for the higher resolution spectra and in grey for the lower resolution SEDM spectra. The phases are in rest frame days with respect to $V$ band peak, except for the sample of \cite{2023A&A...675A..83H} that is presented in rest frame days with respect to $B$ band peak.}
         \label{fig:pEWandVEL}
   \end{figure*}
      
\end{appendix}
\end{document}